\documentclass{bmcart}

\usepackage[utf8]{inputenc} 

\usepackage{amsfonts}
\usepackage{amsmath}
\usepackage{arydshln}
\usepackage{booktabs}
\usepackage{graphicx}
\usepackage{multirow}
\usepackage{lscape}
\usepackage{soul}
\usepackage{url}

\startlocaldefs
\endlocaldefs

\begin{document}

\begin{frontmatter}

\begin{fmbox}
\dochead{Research}


\title{Keep Your Friends Close, and Your Enemies Closer: Structural Properties of Negative Relationships on Twitter}


\author[
   addressref={cnr,sns},                   
   corref={cnr,sns},                       
   noteref={},                          
   email={jack.tacchi@sns.it}   
]{\inits{J}\fnm{Jack} \snm{Tacchi}}
\author[
   addressref={cnr},
]{\inits{C}\fnm{Chiara} \snm{Boldrini}}
\author[
   addressref={cnr},
]{\inits{A}\fnm{Andrea} \snm{Passarella}}
\author[
   addressref={cnr},
]{\inits{M}\fnm{Marco} \snm{Conti}}


\address[id=cnr]{
  \orgname{Consiglio Nazionale delle Ricerche}, 
  \street{Via Giuseppe Moruzzi},                     %
  \postcode{56124}                                
  \city{Pisa},                              
  \cny{Italy}                                    
}
\address[id=sns]{%
  \orgname{Scuola Normale Superiore},
  \street{Piazza dei Cavalieri},
  \postcode{56126}
  \city{Pisa},
  \cny{Italy}
}


\begin{artnotes}
\end{artnotes}

\end{fmbox}


\begin{abstractbox}

\begin{abstract} 

The Ego Network Model (ENM) is a model for the structural organisation of relationships, rooted in evolutionary anthropology, that is found ubiquitously in social contexts. It takes the perspective of a single user (Ego) and organises their contacts (Alters) into a series of (typically 5) concentric circles of decreasing intimacy and increasing size. Alters are sorted based on their tie strength to the Ego, however, this is difficult to measure directly. Traditionally, the interaction frequency has been used as a proxy but this misses the qualitative aspects of connections, such as signs (i.e. polarity), which have been shown to provide extremely useful information. However, the sign of an online social relationship is usually an implicit piece of information, which needs to be estimated by interaction data from Online Social Networks (OSNs), making sign prediction in OSNs a research challenge in and of itself. This work aims to bring the ENM into the signed networks domain by investigating the interplay of signed connections with the ENM. This paper delivers 2 main contributions. Firstly, a new and data-efficient method of signing relationships between individuals using sentiment analysis and, secondly, we provide an in-depth look at the properties of Signed Ego Networks (SENs), using 9 Twitter datasets of various categories of users. We find that negative connections are generally over-represented in the active part of the Ego Networks, suggesting that Twitter greatly over-emphasises negative relationships with respect to ``offline" social networks. Further, users who use social networks for professional reasons have an even greater share of negative connections. Despite this, we also found weak signs that less negative users tend to allocate more cognitive effort to \emph{individual} relationships and thus have smaller ego networks on average. All in all, our results indicate that, even though \emph{structurally} ENMs are known to be similar in both offline and online social networks, they generally tend to nurture more negative feelings and relationships in the latter.

\end{abstract}




\end{abstractbox}
%

\end{frontmatter}



\section{Introduction}
\label{sec:intro}

\begin{figure}
    \centering
    \includegraphics[scale=0.25]{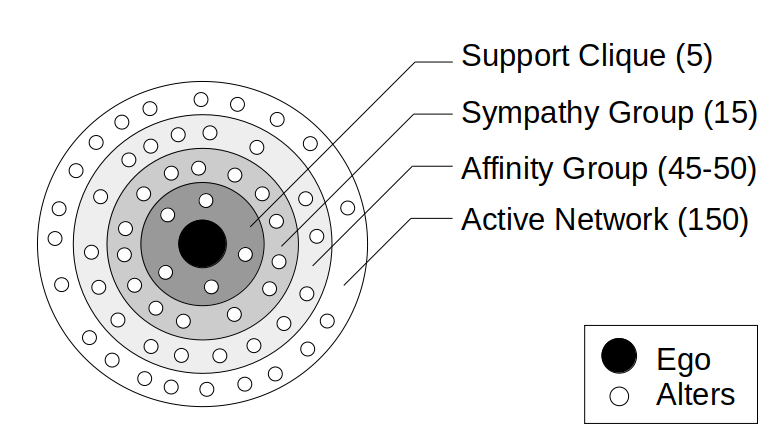}
    \caption{The Ego Network Model, with the names and expected sizes of each subgroup for social networks of humans.}
    \label{Ego_Network_Model}
\end{figure}

Online social networks (OSN) can be seen as a social microscope to investigate the properties of our social interactions in the online world. The increasing global connectivity underscores the significance of understanding social networks and the interactions that occur within them. 
Social network analysis has extensively employed graph-based models to study the structural characteristics of relationships. One such representation, the Ego Network Model (ENM), is rooted in evolutionary anthropology research on how humans structure their social networks~\cite{Dunbar_1995}. The ENM model is centred around a single user, the \emph{Ego}, and portrays all their immediate connections, named \emph{Alters}, based on their relationship strength to the Ego. This results in a series of concentric circles with increasing size but decreasing intimacy, as illustrated in Fig.\ref{Ego_Network_Model}. The number and sizes of the circles are generally consistent, with an average of around 5, 15, 50, and 150 Alters~\cite{Zhou_2005}. The size ratio between them is also quite consistent, with a value close to 3~\cite{Hill_2003}. Note that an ENM only contains meaningful relationships, i.e. those the Ego spends some time nurturing regularly. 

The importance of the ENM is due in large part to its omnipresence in social networks. Indeed, its structure is prevalent across an extremely diverse range of social communities; including traditional hunter-gatherer groups, small-scale horticultural societies, ancient Roman armies and modern-day military units~\cite{Dunbar_1993}. The ENM is so prevalent that it can even be observed in many non-human primate species, although with smaller group sizes~\cite{Dunbar_1998}. The Social Brain Hypothesis proposed by Dunbar explains this pervasiveness, positing that primates have a cognitive limit that restricts the size and complexity of social groups they can maintain. For humans, this limit is approximately 150, also known as Dunbar's number. When the limit is exceeded, social groups tend to become unstable and fragment into smaller, more manageable groups~\cite{Dunbar_1992}. Although one might assume that the ease of online communication would require less cognitive effort and therefore allow for larger social networks to be maintained, the ENM structure remains largely consistent in online contexts. The only notable difference is the occasional presence of an additional innermost circle, with an average size of around 1.5 Alters~\cite{Dunbar_2015}. While this has been postulated for offline networks as well, quantities of data sufficient enough to confirm its existence in offline contexts have never been available.

Furthermore, because each individual in a social network can be viewed as an Ego, the entire network itself can be thought of as a collection of interconnected Ego Networks. Thus, observing a network from the perspectives of the individual Egos can reveal insights that are only visible at a microscopic scale, yet have far-reaching consequences across the entire network. Indeed, the structural properties of the ENM have been shown to influence a number of social behaviours, such as collaboration and information diffusion~\cite{Sutcliffe_2012}. 

Despite its ability to provide many insights, the ENM does have some notable limitations. One such drawback is how the tie strength between Egos and Alters is measured, which has traditionally been done by measuring their frequency of interactions. While this has been shown to be a good proxy measure for the strength of a relationship~\cite{Gilbert_2009}, not all relationships can be differentiated merely by their strength. For example, an individual with a supportive coworker and an angry neighbour will have two very different relationships: even though the interaction frequencies may be very similar, the former relationship will be far more positive than the latter. One way to include some of the important qualitative information that is being lost is to use a signed representation of the network, known as a \emph{signed network}. Each connection in a signed network has a polarity (+/-) indicating either a positive or negative link. The former denotes friendship, trust, and similarity, whilst the latter is associated with hatred and distrust. Positive and negative relationships play different roles in a network and can be leveraged to improve network-related tasks, such as community detection~\cite{Esmailian_2015} and opinion dynamics~\cite{Shi_2016}. Negative links are more informative than positive ones because, among other things, they are usually located along social divisions in a network, such as between two communities, and they can therefore reveal important information about the structure of the overall network~\cite{Leskovec_2010_b_Predicting}. Thus, the inclusion of signs may improve our understanding of the ENM and social networks in general. However, the sign of an OSN relationship is an implicit piece of information, which typically needs to be estimated by interaction data, making sign prediction in OSNs a research challenge in and of itself.

\subsection{Contributions}

In this work, we set out to extend the Ego Network Model with information about the signs of relationships. To this aim, we propose a novel method, grounded in quantitative results from psychology~\cite{Gottman_1995}, of inferring \emph{signed relationships} in unsigned network data (which are typically used to build ego networks), allowing an unsigned network to be converted into a signed one. This method (i) requires only text-based interactions to sign a relationship (hence, it can be applied to any network in which users interact principally via text, i.e. in the vast majority of popular OSNs), (ii) is designed for the short texts typical of OSNs interactions, (iii) requires only data about the interactions over the links we want to sign (hence scales linearly with them). Note that, while signing individual interactions between users simply boils down to attaching a sentiment to the interaction (typically with a sentiment classifier), signing relationships is more nuanced, as it implies deciding on an overall sentiment that captures the whole relation, and, for this sign to reflect human perception, we decided to ground our approach in psychology. This methodology is then shown to be robust to the chosen sentiment classifier for individual interactions and produces results that are consistent with Structural Balance Theory~\cite{Leskovec_2010_a_Signed}.

The second original contribution is the analysis of Signed Ego Networks (SENs), i.e. Ego Networks where edges have a polarity. This was done by obtaining unsigned Ego Networks, for 9 Twitter datasets, and applying the aforementioned method of generating signs to them. The unsigned and signed versions of the networks are analysed, including the distribution of signed links across the various circles of the SEN. The main findings are that: (i) Twitter users engage in much more negative relationships than expected in the Active Networks (illustrated in Fig.~\ref{Ego_Network_Model}), (ii) specialised users (e.g. journalists) do so to an even higher extent, (iii) negative relationships are particularly present in the intimate EN layers of specialised users, and (iv) there is evidence for a potential weak effect of negativity leading to a slightly higher-than-average number of distinct connections, but fewer interactions in each relationship. All in all, the results confirm the popular notion that higher engagement in online social interactions results in being exposed to increasingly negative relationships and sentiments. They also extend beyond this with the surprising revelation that negative relationships tend to be proportionally more present in the social circles of the Ego Networks closer to the Ego.

Some preliminary results on the Signed Ego Network Model (SENM) were first presented in~\cite{Tacchi_2022}. These were then expanded on in~\cite{Tacchi_2023}, where the generalisability of the SENM was observed across several cultures and types of communities. The main extensions of this current work are the following. First, the robustness of the method of signing relationships is tested using 4 different sentiment analysis models for labelling individual interactions (Section~\ref{sec:comparison_NLP_models}). The results show that the proportions of positive and negative relationships were similar for all 4 of the models. Furthermore, the models agreed on the signs of around 70-80\% of the relationships and when the models did disagree, the disagreements tended to be very close to the threshold used for signing the relationships (i.e. when the models disagreed, they tended to only disagree slightly). Next, the method of signing relationships is further validated via triad analysis (Section~\ref{sec:triad_analysis}). Specifically, repeated analysis of the signed triads produced by each of the 4 models shows that the distribution of signs produced by this method fitted expectations of known psychological effects in social networks (i.e. Structural Balance Theory). These distributions are also extremely and significantly different from what would be obtained by chance. Finally, we have included an analysis of the impact of negative social relationships on the cognitive effort of the Ego (Section~\ref{sec:negativity_metrics_results}).

\section{Background}
\subsection{Ego Network Model}
\label{sec:egonets_background}
As previously mentioned, the ENM is centred around an individual Ego, who is surrounded by their Alters, organised in a series of concentric circles. The ENM stems from the anthropological Social Brain Hypothesis~\cite{Dunbar_1998}, which posits that the social capabilities of primates are constrained by the sizes of their neocortices. Based on the size of our own neocortex, the maximum social group size that can be maintained by a human is estimated to be around 150 (the famous Dunbar's number). Note that these 150 contacts with whom a person engages do not include acquaintances, rather they are exclusively relationships that are regularly nurtured. Traditionally, this has been defined as a minimum interaction frequency of at least once a year; for example, exchanging annual holiday wishes. These relationships constitute the so-called \emph{active} part of the Ego Network.

Of course, the frequency and importance of the interactions generated by each relationship varies significantly from Alter to Alter. Indeed, by arranging the Alters based on their tie strength to the Ego, the aforementioned concentric structure will typically emerge~\cite{Hill_2003,Zhou_2005}, with each subsequent circle containing the Alters of the previous ones (thus, the size of the active part of the Ego Network is equivalent to its outermost circle). Both the number of circles (approximately 4 or 5) and their sizes -- $1.5, 5, 15, 50, 150$ -- are fairly regular, in offline and online social networks~\cite{Dunbar_2015}.

As the tie strength between Ego and Alter directly determines which circle the Alters are placed into, this is obviously a core concept of the ENM. Tie strength was defined by Granovetter as the equally weighted combination of 4 elements in a relationship: the time spent maintaining it, its emotional intensity, its level of intimacy and the reciprocal services it generates~\cite{Granovetter_1973}. This definition can be a crucial consideration for understanding how various users interact socially. For example, individuals who engage in OSNs for professional purposes may devote more time to social platforms, thereby generating more reciprocal services and investing greater amounts of time in maintaining relationships. Indeed, it has previously been suggested that journalists are likely to be more cognitively engaged with Twitter than other types of users~\cite{Toprak_2021_b_Region-based}. While the time spent maintaining a relationship is just one of the tie strength dimensions described by Granovetter, it has largely been the sole focus of the related literature on Ego Networks due to its widespread availability and ease of computation (using the number of interactions as its proxy). Therefore, the objective of this work is to advance the state of the art by exploring the hitherto underrepresented qualitative aspects of tie strength, in addition to the traditional metric of the time spent maintaining them.

\subsection{Signed Networks}
\label{sec:signed_networks_background}

In contrast to unsigned networks, whose connections are either binary (i.e. a connection between two users either exists or doesn't) or weighted connections (usually based on tie strength), signed networks feature connections that can be further distinguished as either positive or negative (sometimes referred to as the \emph{polarity} of edges~\cite{Tang_2016}). Positive links indicate positive relationships and are used to infer trust and homogeneity~\cite{Maniu_2011}. On the other hand, negative links indicate negative relationships, distrust, and dissimilarities. Therefore, signed networks contain additional information that can be leveraged to enhance the performance of many tasks, such as community detection~\cite{Traag_2009} and information diffusion~\cite{Ferrara_2015}.

Previous research on networks with publicly available signed connections has revealed that negative connections are significantly less prevalent than positive connections, accounting for approximately 15.0\% to 22.6\% of the total connections within a network~\cite{Leskovec_2010_a_Signed}. In these networks, the users' awareness of link polarity may intensify social pressure and effects such as social capital~\cite{Coleman_1988}, whereby relationships between individuals who have many relationships in common are more likely to be positive due to social pressure from the surrounding community to get along. Conversely, even if an unsigned network contains implicit positive and negative relationships, the lack of explicitly visible negative links results in lower social pressure. Therefore, we can anticipate that networks without explicit signed relationships will have a higher proportion of negative relations than those with explicitly signed ones. We will investigate this hypothesis further in section~\ref{sec:results}.

Despite the added advantages of signed networks, they are rarely the focus of research because the vast majority of popular social platforms do not allow users to create explicitly negative links. This makes it very difficult to obtain signed network data in sufficient enough quantities for in-depth analysis. Nevertheless, some exceptions do exist, most notably Slashdot and Epinions, which have provided two of the most widely used benchmark datasets for signed networks~\cite{Tang_2016}. Unfortunately, these datasets do not provide information on interaction frequencies and therefore cannot be used for Ego Network analysis. ENM studies typically use Twitter data (due to their public nature and easy access via the Twitter API) but Twitter does not provide explicit relationship signs between users. However, just as with real-world relationships, relationships that take place online usually contain implicit information about their polarity, which can potentially be gleaned from the interactions they produce~\cite{Maniu_2011}.

Several approaches have been developed to predict the signs of unsigned networks. However, most of these focus on the structural aspects of the surrounding network in order to deduce the sign of a connection (e.g. by leveraging topological notions like the clustering coefficient~\cite{Javari_2014}), which is an indirect way of extracting signs, without looking directly at how people communicate with each other. Classification algorithms, trained on preexisting datasets with known signs, have also been used to compute the signs of novel networks~\cite{Ye_2013}. All these techniques have taken a top-down perspective, viewing the network's features as a whole and inferring signs based on the structure of the connections. However, if the inverse approach is taken, viewing the problem from the bottom up, then it is possible to take into consideration the more tacit aspects of connections that have largely gone uninvestigated, as we discuss below.

The basic building blocks that form a relationship are the interactions and exchanges between users and their corresponding sentiments. Sentiment analysis for individual exchanges is extremely well established~\cite{Liu_2012}. This allows signs to be obtained for these singular interactions with an extremely high degree of confidence. However, methods for extending the signs of these bottom-level interactions to whole series of interactions, or relationships, have not received anywhere near the same level of scientific interest. One study~\cite{Hassan_2012} that has previously examined this problem trained a Support Vector Machine (SVM) on a manually-annotated dataset of relationships in discussion forums. The SVM took in 4 user features and 3 interaction features and achieved an accuracy of 0.835 on a subset of annotated data. Unfortunately, this approach cannot be directly replicated for Twitter interactions due to their very short and unstructured nature compared to discussion forums. In addition, there is a lack of publicly available ground truth data for Twitter relationships. In response to these problems, we propose an alternative approach that is specifically designed for dealing with short texts and can leverage models that have been established within the previous literature in order to obtain the sentiment of individual interaction.

\subsection{Structural Balance Theory}
\label{sec:balance_theory}
\begin{figure}
    \centering
    \includegraphics[scale=0.35]{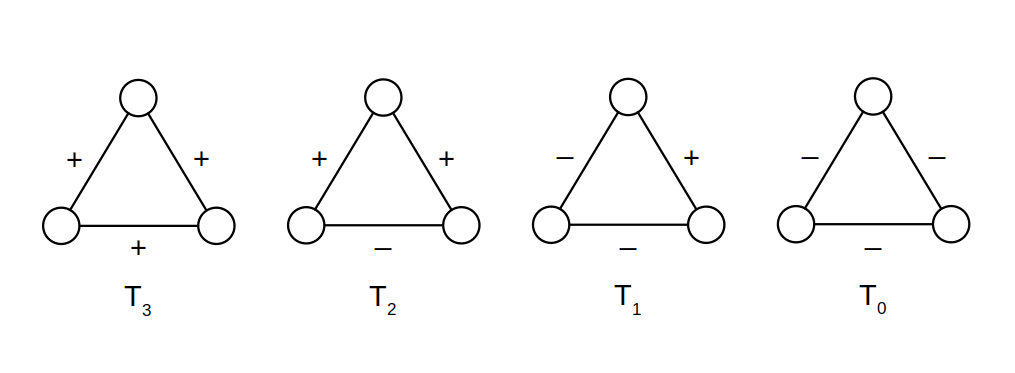}
    \caption{All four possible signed triads, as per Structural Balance Theory. The subscript number following the “T" corresponds to the number of positive connections for that triad.}
    \label{fig:triad_types}
\end{figure}

Signed networks are known to conform to certain properties and configurations. A theory that lays out such a set of informative expectations is Structural Balance Theory~\cite{Heider_1946,Cartwright_1956}; a psychological theory, which postulates that certain configurations of signed triads (i.e. groups of three individuals who are all interconnected by signed edges) should be more common than others when observed across a social network\footnote{The standard nomenclature for these triads is a capital letter “T" followed by the number of positive connections in the triad in subscript: T\textsubscript{3}, T\textsubscript{2}, T\textsubscript{1} and T\textsubscript{0}.}. This is because connections are not independent but rather influenced by the other connections in the surrounding network. With regards to signed triads, those with odd numbers of positive connections, i.e. one and three, are considered plausible, or “balanced" (see T\textsubscript{3} and T\textsubscript{1} in Figure~\ref{fig:triad_types}), while those with even numbers of positive connections, i.e. two or zero, are considered implausible, or “unbalanced" (see T\textsubscript{2} and T\textsubscript{0} in Figure~\ref{fig:triad_types}). This is because these latter configurations correspond to socially problematic situations: the first, where one individual has two friends who are enemies, and the second, where all three individuals are hostile to one another and none of them decide to pair up against the third. However, a more lenient variant of this theory, commonly known as Weak Structural Balance Theory, argues that it should not be unexpected to have a situation in which three enemies refuse to team up (T\textsubscript{0}) or for two friends to have a common enemy (T\textsubscript{1}). Therefore, one should only expect triads with exactly two positive connections (T\textsubscript{2}) to be underrepresented and only triads with three positive connections (T\textsubscript{3}) to be overrepresented, with no expectations for T\textsubscript{1} or T\textsubscript{0} \cite{Davis_1967}. 

Given the expectations of Structural Balance Theory, it is possible to validate the predicted signs of a network by analysing the resulting triads~\cite{Hassan_2012} and comparing them to the expected numbers of each triad if the signs were distributed at random. This is indeed the approach we use to validate our method for signing relationships. Previously, it has been found that the expectations of the weaker version of Structural Balance Theory tend to fit online datasets better than those of the original theory~\cite{Leskovec_2010_a_Signed}, so this is the version we use in our analysis. The exact methodology used for this is given in Subsection~\ref{sec:triad_analysis_method}.

\section{Methodology}
\label{sec:methodology}
This section outlines the methodology for obtaining Signed Ego Networks, assuming that the input data is taken from Twitter (Twitter being the de-facto standard for data in the relevant literature~\cite{Arnaboldi_2017,Dunbar_2015,Toprak_2021_a_Harnessing,Toprak_2021_b_Region-based}). Our methodology comprises three steps: first, we attach a sign to each relationship based on the signs of individual interactions (Sections~\ref{sec:signing_relationships} and~\ref{sec:choice_of_models}); then, we validate the obtained relationship signs against Structural Balance Theory (Section~\ref{sec:triad_analysis_method}); finally, we enrich the standard Ego Network Model by transposing the sign information onto it (Section~\ref{sec:computing_signed_egonets}). Afterwards, in Section~\ref{sec:negativity_metrics}, we discuss how to measure the burden of negative relationships on overall social cognitive capacity.

In order to construct Ego Networks, it is necessary to acquire Tweets that involve direct communications between Twitter users. These communications occur when users explicitly reply to another's post (Replies), mention another user using the ``@" symbol (Mentions) or share another user's Tweet (Retweets). This latter case is sometimes accompanied by an additional piece of text made by the sharing user (Quote Retweets). Each of these directed Tweets corresponds to an interaction between an Ego and an Alter. While some of these interactions may involve the wider network beyond the specific Alter, they nonetheless reflect a cognitive involvement of the Ego towards the Alter, which is the most critical characteristic for mapping an interaction to a specific social relationship~\cite{Dunbar_1998}.

\subsection{Signing Relationships}
\label{sec:signing_relationships}

As anticipated in the introduction, in this work we take a bottom-up approach to sign extraction, inferring signs from the sentiment of individual interactions. Indeed, the effects of positive and negative exchanges have been studied in a variety of contexts. One such observation that is particularly relevant here is that a ratio of around 1 negative interaction for every 5 positive interactions, or roughly 17\%, appears to be an important tipping point for numerous different types of relationships. Once this threshold is crossed, marriages become significantly less likely to last~\cite{Gottman_1995} and, for parent-child relationships, children are more likely to underperform at school and have developmental problems~\cite{Hart_1995}. 

This ratio, which we will refer to as the \emph{golden interaction threshold}, is leveraged for our proposed method for signing relationships, which culminates in a binary classification (positive or negative) for each Ego-Alter pair. More precisely, our method consists of 2 main steps:

\emph{Step 1: label single interactions--} First, sentiment analysis is carried out to obtain a positive, neutral or negative label for each text-based communication Tweet made by an Ego towards one of their Alters\footnote{Thus, the labels are directional, meaning that if two users in a given dataset are both Egos and have each other as Alters, the signs of their relationship are not guaranteed to be the same in both directions}. The models used for the sentiment analysis of single interactions are discussed in Section~\ref{sec:choice_of_models}.

The sentiment analysis was done for Replies, Mentions and Quote Retweets. Regular Retweets are instead always classified as neutral because they were not originally written by the Ego and, therefore, do not reflect the same level of cognitive effort. Returning to Granovetter's definition, these regular Retweets can be regarded instead as a reciprocal service generated by a relationship because they correspond to an Ego's desire to share the content of an Alter. In addition, automatically assigning a neutral sentiment to regular Retweets reduces their relative impact on the overall sign of a relationship without completely ignoring it. This is also consistent with the lower relative cognitive and temporal costs required for clicking the Retweet button compared to composing a Quote Retweet, Reply or Mention. Neutral interactions are treated the same as positive interactions at the moment of signing the relationships. This is because the time spent on a relationship is directly correlated to its strength, as per Granovetter's definition. Therefore, any active effort made by an individual to communicate with another should, intuitively, be considered positively unless there is reason to think otherwise.

\emph{Step 2: label relationships--} Next, a sign is computed for each relationship based on the ratio of negative interactions produced by the relationship. Specifically, by applying the golden interaction ratio~\cite{Gottman_1995} as a threshold, we determine relationships exhibiting greater than 17\% negative interactions as negative, otherwise, the relationship is classified as positive. According to the psychological literature, the former scenario would indicate an unstable relationship, while the latter corresponds to a stable one.

The use of a threshold for determining the relationship signs in the described manner may be inappropriate for relationships that have very few interactions; namely, fewer than 6, given the 1:5 interaction ratio. This point is addressed in Section~\ref{sec:results_circles}, where we observe the numbers of interactions at each level of the ENM.

\subsection{Choice of sentiment classifier for individual interactions}
\label{sec:choice_of_models}
To check how susceptible the relationship signs are to the choice of model used to label the individual interactions, 4 sentiment analysis models were selected to be compared. Recently, there has been a strong shift towards the use of transformer-based methods for Natural Language Processing (NLP). This is largely due to transformers' robustness and improved ability to process the sequential aspects of language. Reflecting this shift in focus, the models chosen for this study consist of a more traditional, lexicon- and rule-based model and 3 transformer-based models.

All the models were used to obtain relationship signs for the largest of the datasets used in this paper (that being the Snowball dataset, see Section~\ref{sec:datasets}). The numbers of each label predicted by the 4 models, as well as how often they agreed with each other can be seen in Subsection~\ref{sec:comparison_NLP_models}.

\subsubsection{VADER}
The first model is Valence Aware Dictionary and sEntiment Reasoner (VADER), a well-established sentiment analysis tool developed specifically for use with social media data~\cite{Hutto_2014}. VADER provides a compound sentiment score between -1 and 1 for a given text. This score can be converted into a positive label if it is above 0.05, negative if it is below -0.05 or neutral if it is between these values~\cite{Hutto_2014}. VADER was compared to 7 state-of-practice alternatives, as well as individual human annotators, using a test set of 4,200 Tweets. It obtained an F1 score of 0.99, outperforming all other models and humans~\cite{Hutto_2014}.

\subsubsection{BERTweet}
The first BERT-based model used in this paper is BERTweet~\cite{Nguyen_2020}, a version of BERT~\cite{Devlin_2018} that has been purposefully optimised for Twitter data. Specifically, it was fine-tuned for the task of sentiment classification using a corpus of 850 million English Tweets collected between January 2012 and March 2020. BERTweet was tested using the SemEval 2017 (Task 4) corpus~\cite{Rosenthal_2019}, a common benchmark dataset for sentiment classification, which contains around 50,000 English Tweets; BERTweet achieved an F1 score of 0.73~\cite{Nguyen_2020}.

\subsubsection{XLM-T}
The next model is XLM-T \cite{Barbieri_2021}, a fine-tuned version of XLM-RoBERTa \cite{Conneau_2019}. This latter model is a general NLP model that was trained on 2.5TB of CommonCrawl data, containing 100 languages, which had been filtered following pre-established guidelines based on perplexity \cite{Wenzek_2019}. The former was then further trained specifically for sentiment classification using 198 million Tweets from over 60 languages. XLM-T's performance varies from language to language, but attained a mean F1 score of 0.69 when tested across monolingual datasets for 8 languages (Arabic, English, French, German, Hindi, Italian, Portuguese and Spanish). The F1 scores for 7 of these languages were between 0.69 and 0.78, however, Hindi only reached 0.56, highlighting the model's difficulty when dealing with certain languages. The English F1 score, 0.71, was obtained using a subset of 3,033 Tweets from the SemEval 2017 dataset, thus, this model's performance seems to be similar to that of BERTweet.

\subsubsection{BERT-C}
The final model is a downstream version of BERTweet, also fine-tuned for sentiment classification, this time on a classified dataset. This model was released by HuggingFace~\cite{HuggingFace_2022} and it is referred to here as the BERT Classified (BERT-C) model. Although we have no prior metrics for estimating the performance of this model, it is assumed that it will have a performance comparable to that of the original BERTweet model.

\subsection{Triad Analysis}
\label{sec:triad_analysis_method}

As previously mentioned (in Subsection~\ref{sec:balance_theory}), signed connections in a social network are known to follow certain patterns, predicted by Structural Balance Theory. Thus, in this work, we leverage these expected patterns to validate the relationship signs obtained with our method. In order to form the triads, an interconnected network of users is required. 

This is different from the standard data used for computing Ego Networks, where only the interactions between the Ego and the Alters are of interest. For triad analysis, we also need Alter-Alter interactions. The Snowball dataset described in Section~\ref{sec:datasets} satisfies this requirement. Thus, each edge of the graph is assigned a sign with the methodology described in Section~\ref{sec:signing_relationships}. The final step entails counting the triad types in the resulting signed graph. This makes it possible to obtain an idea of how under- or overrepresented each triad is and, thus, whether or not the predictions match the expectations of Structural Balance Theory.

In order to rule out that the same sign distribution could have been produced at random from the same background distribution of positive and negatives, we compare the triad counts in the signed graph above with those obtained after shuffling the signs~\cite{Leskovec_2010_a_Signed}. For statistical reliability, the random shuffling was repeated 10 times and the final results use the mean values. 
The further away the quantities observed in the real signed graph are from the random ones, the more ``surprise'' there is and the lower the likelihood of the predictions occurring due to random chance. Here, \emph{surprise} is defined as the number of standard deviations by which the observed number of Triad \textit{i} differs from that of the randomly shuffled network with the same proportion of positive and negative signs.

The precise formula (taken from \cite{Leskovec_2010_a_Signed}) used for calculating the level of surprise $s(T_i)$ for the observed number of Triad $i$ is given in Equation~\ref{eq:surprise}. 
\begin{equation}
    s(T_i) = \frac{T_i - \mathbb{E}[T_i]}{\sqrt{\Delta p_0(T_i)(1 - p_0(T_i))}}
    \label{eq:surprise}
\end{equation}
Here, $\Delta$ is the total number of triads in the dataset, $p_0(T_i)$ is the fraction of $T_i$ triads to be expected in the network given a random distribution of signs, and $\mathbb{E}[T_i]$ is the expected number of triads $T_i$ in the randomly shuffled model. $s(T_i)$ effectively measures the number of standard deviations by which the actual quantity of $T_i$ triads differs from the expected number under the randomly shuffled model. The denominator in Equation~\ref{eq:surprise} corresponds to the standard deviation of a binomial distribution where the success probability is $p_0(T_i)$ and the number of trials are $\Delta$.


\subsection{Computation of Signed Ego Networks}
\label{sec:computing_signed_egonets}
The computation of the Ego Networks is achieved by first computing the frequency of interaction between each Ego-Alter pair and then clustering the Alters based on these frequencies. This method is well-established and has previously been done using a variety of different clustering algorithms; including k-means~\cite{MacQueen_1967}, DBSCAN~\cite{Ester_1996} and MeanShift~\cite{Fukunaga_1975}. MeanShift is used for this paper as it is one of the most commonly used algorithms and it also automatically finds the optimum number of clusters (corresponding to the number of circles in the Ego Network, into which the Alters are organised). The signs of the Ego Network relationships are computed separately, in the manner previously described. These signs are then matched to each Ego-Alter relationship in the Ego Networks, resulting in Signed Ego Networks.

\subsection{Negativity Metrics}
\label{sec:negativity_metrics}

Given the obvious differences in the effects that positive and negative interactions can have on a relationship, an additional investigation was conducted to examine whether interactions and relationships of differing sentiments exert different amounts of cognitive effort. Given that negative information is generally harder and more time-consuming for humans to process~\cite{Baumeister_2001}, one would expect negative relationships to be more cognitively demanding than positive ones. Therefore, the hypothesis we tested is whether greater numbers of negative relationships are associated with smaller active Ego Networks. For this analysis, the mean active Ego Network sizes of users with an optimum number of circles equal to 5 were compared. The users' levels of negativity were measured using 3 different metrics. Before introducing their formal definitions, let us denote with $\mathcal{A}_i$ the set of Alters in the active Ego network of Ego $i$. Considering the signs of the relationships with the Alters, we can also split $\mathcal{A}_i$ into $\mathcal{A}_i^+$ and $\mathcal{A}_i^-$, for Alters whose relationship with the Ego $i$ is positive and negative, respectively. Further, we denote with $n_{ij}^+$ and $n_{ij}^-$  the number of positive and negative interactions between Ego $i$ and Alter $j$. We denote their sum as $n_{ij}$. Leveraging this notation, the first negativity metric $l_1 $ corresponds to the proportion of negative relationships, i.e. the number of negative relationships that each Ego had, divided by their total number of relationships:
\begin{equation}
    l_1(i) = \frac{|\mathcal{A}_i^-|}{|\mathcal{A}_i|}.
    \label{eq:negativity_metric1}
\end{equation}
The second negativity metric measures the proportion of negative interactions, even if they belong to positive relationships, i.e. the number of negative interactions for each Ego divided by their total number of interactions:
\begin{equation}
    l_2(i) = \frac{\sum_{j \in \mathcal{A}_i} n_{i,j}^-}{\sum_{j \in \mathcal{A}_i} n_{i,j}}.
    \label{eq:negativity_metric2}
\end{equation}
Finally, the third negativity metric follows the proportion of interactions that belong to negative relationships, even if the interaction itself is positive, i.e. the number of each Ego's interactions that correspond to a negative relationship divided by their total number of interactions:
\begin{equation}
    l_3(i) = \frac{\sum_{j \in \mathcal{A}_i^-} n_{i,j}}{\sum_{j \in \mathcal{A}_i} n_{i,j}}.
    \label{eq:negativity_metric3}
\end{equation}
When compared against the Ego Network size, the first of these metrics directly investigates the cognitive effects of maintaining negative relationships regardless of how often we interact with said negative contacts. The latter two metrics take a more fine-grained look at the role of interactions. Indeed, the second metric gauges whether negative interactions, rather than relationships, have a different impact on cognitive effort, even if the negative interaction is with someone we have a positive relationship with. The third metric checks whether interacting with negative relationships elicits a different level of cognitive effort, even if some of the interactions are positive.

The values of the metrics are defined between 0 and 1 (inclusive) and the Egos in each dataset were grouped into bins based on their negativity values for each of the 3 negativity metrics. This ensures that all the bins of a given dataset contain similar numbers of Egos, although it does mean that the bin boundaries change between dataset and metric.
The Egos' negativities were then compared to the sizes of their Ego Networks (the results are discussed in Subsection~\ref{sec:negativity_metrics_results}).

\section{Datasets}
\label{sec:datasets}
All of the data used in this paper were collected from Twitter using the official Twitter Developer API. Twitter has long been a reliable source of Ego Network data due to its vast and active userbase as well as providing mostly public data. At the time of collection, the standard Twitter API allowed the most recent 3,200 public Tweets created by a given user to be collected. These Tweets are referred to collectively as the user's Timeline. Although this may not correspond to all the Tweets a user has created, this has been shown to be a significant quantity of information to generate meaningful Ego Networks (e.g.~\cite{Arnaboldi_2017,Arnaboldi_2015,Dunbar_2015}).

In total, 9 datasets were used. These were collected from previous works and represent a mixture of specialised users, who use Twitter mainly for professional reasons, and generic users, who use the platform primarily for social reasons. The distinction between these two types of users is important as they have been observed to exhibit differing behaviours in certain online contexts~\cite{Toprak_2021_b_Region-based}. Information describing these datasets in terms of the numbers of Egos, Alters, relationships and interactions they contain can be seen in Table~\ref{descriptive_full} and Table~\ref{descriptive_active}, the former containing all collected users and the latter containing only the users that remained after the preprocessing steps detailed in section ~\ref{subsub:preprocessing}.

\begin{table*}[htbp]
    \centering
    \caption{Number of Egos, Alters, relationships and interactions in the full Ego Networks, before removing unengaged users (as described in section~\ref{subsub:preprocessing})}
    \label{descriptive_full}
    \begin{tabular}{@{}lrrrr@{}}
        \toprule
        \textbf{Dataset} & \textbf{  Egos  } & \textbf{  Alters  } & \textbf{ Relationships } & \textbf{ Interactions }\\
        \midrule
        American Journalists & 1,714 & 505,023 & 1,479,764 & 4,677,736\\
        Australian Journalists & 957 & 185,245 & 709,764 & 2,466,111\\
        British Journalists & 512 & 209,402 & 469,863 & 1,397,996\\
        NYT Journalists & 678 & 173,620 & 521,917 & 1,493,199\\
        Science Writers & 497 & 182,240 & 463,624 & 1,350,799\\
        British MPs & 584 & 157,053 & 343,366 & 1,277,010\\
        \hdashline
        Monday Motivation & 6,946 & 1,151,899 & 2,291,692 & 9,449,775\\
        UK Users & 3,512 & 12,088,975 & 2,507,634 & 9,931,908\\
        Snowball & 12,200 & 4,065,930 & 9,636,070 & 77,088,560\\ 
        \bottomrule
    \end{tabular}
\end{table*}

\begin{table*}[htbp]
    \centering
    \caption{Number of Egos, Alters, relationships and interactions in the active networks of each dataset, after removing unengaged users (as described in section~\ref{subsub:preprocessing})} 
    \label{descriptive_active}
    \begin{tabular}{@{}lrrrr@{}}
        \toprule
        \textbf{Dataset} & \textbf{  Egos  } & \textbf{  Alters  } & \textbf{ Relationships } & \textbf{ Interactions }\\
        \midrule
        American Journalists & 1,037 & 68,792 & 143,390 & 1,639,623\\
        Australian Journalists & 520 & 26,561 & 75,455 & 937,764\\
        British Journalists & 281 & 24,614 & 41,524 & 434,477\\
        NYT Journalists & 558 & 23,327 & 59,922 & 561,563\\
        Science Writers & 241 & 18,531 & 35,185 & 381,340\\
        British MPs & 440 & 27,538 & 76,857 & 323,765\\
        \hdashline
        Monday Motivation & 1,461 & 78,906 & 158,374 & 894,648\\
        UK Users & 921 & 84,993 & 111,426 & 1,474,882\\
        Snowball & 4,049 & 366,168 & 574,585 & 8,593,290\\
        \bottomrule
    \end{tabular}
\end{table*}

\subsection{Specialised Users}
\label{subsub:datasets_specialised}
\hfill \\
\emph{Journalists--}
The first set of specialised users contains data from journalists. This set consists of 3 datasets that were originally collected during a previous study, which observed the Ego Networks of journalists from 17 different countries across the globe~\cite{Toprak_2021_b_Region-based}. Unfortunately, many of these datasets contained, either entirely or in large part, non-English Tweets. The sentiment analysis of non-English tweets would introduce an additional level of complexity (since the vast majority of tools are trained and optimised for the English language) without contributing to the scope of the paper. Therefore, only data from anglophone countries were included in the present study; specifically: the United States of America, Australia and the United Kingdom. The American and Australian datasets were collected in May 2018 and the British dataset was collected in January 2018, using existing lists of Twitter journalists (validated in~\cite{Boldrini_2018}).

In addition to these, another set of journalist data was taken from a different study~\cite{Ollivier_2022}. This dataset was collected from a list of New York Times journalists, created by the New York Times itself. All the users from this list were downloaded in February 2018. This dataset will be referred to as NYT Journalists.

\emph{Science Writers--}
The next dataset of specialised users contains science writers. Again, these are users who use Twitter for professional means, albeit to a potentially different extent compared to journalists. This dataset was collected using a curated list of science writers, created by a writer at Scientific American, Jennifer Frazer. Its Timelines were gathered in June 2018, as part of a previous study~\cite{Ollivier_2022}.

\emph{British Members of Parliament (MPs)--}
The final specialised dataset was collected during the preliminary investigation of SENMs \cite{Tacchi_2022}. This one includes the Timelines of members of the British Parliament, taken from a publicly available list provided by UKinbound~\cite{UK_inbound_2020}. These Timelines were collected in March 2022. At the time of collection of this dataset, Twitter allowed academics to retrieve full user timelines (i.e. not just the first 3,200 Tweets), however, for the sake of comparison with previous work, we limited our analysis to include only the first 3,200 Tweets for each user.

\subsection{Generic Users}
\label{subsub:datasets_generic}
\hfill \\
\emph{Monday Motivation--}
The first generic dataset consisted of users who tweeted in English using the hashtag \#MondayMotivation on 16\textsuperscript{th} January 2020. The Timelines of these users were then collected in January 2020, during a previous study~\cite{Ollivier_2022}.

\emph{UK Users--}
The second generic dataset came from a random sample of all users who tweeted in English from the United Kingdom on February 11\textsuperscript{th} 2020. These users' Timelines  were collected in February 2020, as part of a previous study~\cite{Ollivier_2022}. 

\emph{Snowball--}
The final dataset, taken from a cross-cultural analysis of SENMs \cite{Tacchi_2023} (in which it was referred to as Baseline) consists of a collection of interconnected Ego Networks, collected using a snowball sampling methodology. Specifically, an initial set of 31 interconnected seed users were selected, pseudorandomly to ensure a degree of interconnectivity between the seeds, from another preexisting dataset, which itself was collected using a snowball sampling starting from Barack Obama~\cite{Arnaboldi_2013}. The timelines of these users were then collected, followed by those of their Alters and then of their Alters' Alters. This means that Egos have common Alters and can be themselves Alters for other Egos, which is an important distinction as it is a requirement for carrying out Structural Balance analysis (see Subsection~\ref{sec:triad_analysis_method}). The Timelines for the Snowball dataset were collected between April and May 2022. As with the British MPs dataset, the full timelines of each user were accessible at the time of collection, however, they were limited to 3,200 Tweets per user during our analyses to ensure comparability with the other datasets.

\subsection{Preprocessing} 
\label{subsub:preprocessing}
The first step of preprocessing was required to remove any undesired types of users from the data, namely by filtering out any user accounts that are not owned by individual humans. This is an important consideration as, for example, bots and other types of automated accounts will not have any cognitive constraints. As the specialised user datasets were gathered from verified lists of Twitter users, this step was only necessary for the generic datasets: Monday Motivation, UK Users and Snowball. A Support Vector Machine (SVM)~\cite{Cortes_1995} was trained on a set of 500 Twitter users that were manually classified as either “people" or “other". This classifier and the training set are established in ENM research~\cite{Arnaboldi_2013} and an accuracy of 81.3\% was achieved using k-fold cross-validation (with k=5). Any user accounts that were labelled as “other" by the SVM were removed by the original authors of each dataset.

Next, before conducting any analyses on the ENMs, it was necessary to filter out inactive and irregular users for all the datasets. This is because such users are unlikely to be engaged enough with Twitter to have fully developed Ego Networks on the platform. For this, Egos were removed if their timeline consisted of fewer than 2,000 Tweets total, spanned a period of fewer than 6 months (from the first to the last Tweet in their Timeline) or if they tweeted less than once every 3 days for more than 50\% of the months that they were active. The main rationale behind these choices is to keep only Twitter users that are active and engage regularly with Twitter. These filtration parameters are in line with those of previous work on Ego Networks~\cite{Arnaboldi_2015,Toprak_2021_b_Region-based}, to which we refer for further details.

\section{Results}
\label{sec:results}
In this section, we report our experimental findings. First, we conduct 2 tests: to investigate the impact of the choice of sentiment analysis model on the interactions and relationships labels (Section~\ref{sec:comparison_NLP_models}) and to support the validity of said labels (Section~\ref{sec:triad_analysis}). Next, we investigate the properties of the Signed Ego Networks of the 9 selected datasets extracted according to the methodology discussed in Section~\ref{sec:computing_signed_egonets}. Recalling from Section~\ref{sec:egonets_background} that an Ego Network is composed of an active and inactive part, we study how negative relationships are distributed in the full vs active network in Section~\ref{sec:results_full_active}. Then, in Section~\ref{sec:results_specialised_generic}, we discuss the differences between specialised and generic users and, in Section~\ref{sec:results_circles}, analyse how positive and negative relationships are distributed across the Ego Network social circles. Finally, in Section~\ref{sec:negativity_metrics_results} we investigate the effects of negativity on cognitive effort by observing the correlations between users' Ego Network sizes and their level of negativity, using the 3 negativity metrics defined in Subsection~\ref{sec:negativity_metrics}.

\subsection{Sensitivity of Signing Method to Sentiment Classifier}
\label{sec:comparison_NLP_models}

In Section~\ref{sec:signing_relationships}, we have introduced our method for signing social relationships from unsigned social network data. It comprises two steps: labelling of individual interactions (using a state-of-the-art sentiment classifier) and labelling of relationships applying the psychology-grounded golden interaction ratio. Here, we investigate the sensitivity of the proposed relationship signing method to the choice of sentiment classifier, selected among the ones discussed in Section~\ref{sec:choice_of_models}.
The Snowball dataset was chosen as the focus of this comparison as it is the largest dataset in this paper; it is also the only dataset that can be used for the Triad Analysis in the next section.

\begin{figure}
    \centering
    \includegraphics[scale=0.275]{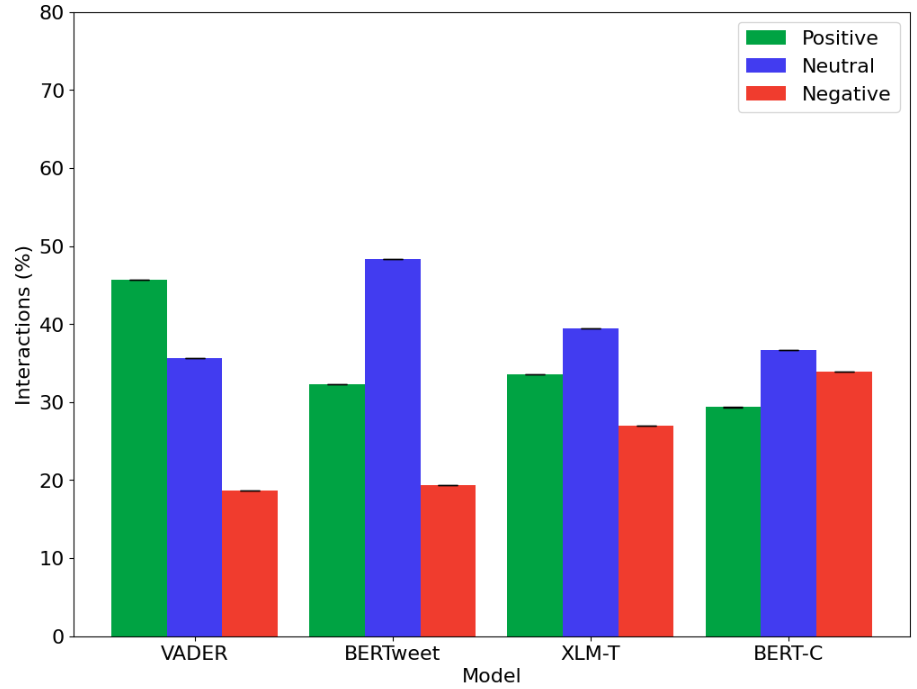}
    \caption{Percentages of positive, neutral and negative interaction labels estimated by each model (95\% confidence intervals)}
    \label{fig:interaction_bar_chart}
\end{figure}

\begin{figure}
    \centering
    \includegraphics[scale=0.275]{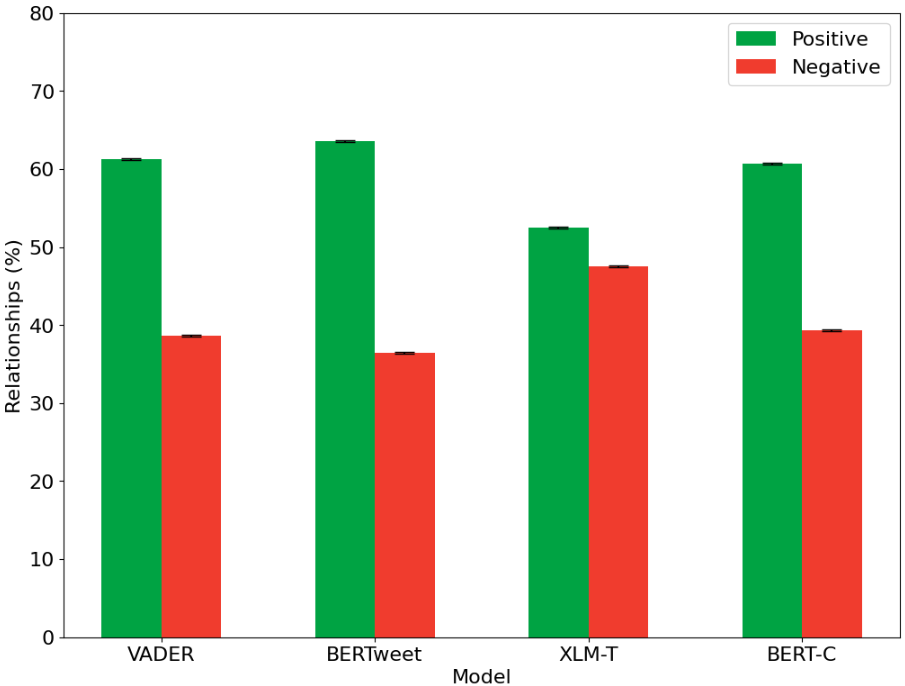}
    \caption{Percentages of positive and negative relationship labels estimated by each model (95\% confidence intervals)}
    \label{fig:relationshiop_bar_chart}
\end{figure}

We first compare the sentiment classifiers on the task of labelling single interactions. For the interaction labels (Figure~\ref{fig:interaction_bar_chart}), the models show a fair degree of variability, with around 30 to 45\% for positive, 35 to 50\% for neutral and 20 to 30\% for negative. However, when looking at the relationship labels (Figure~\ref{fig:relationshiop_bar_chart}), there is a very tight percentage range for 3 of the models (VADER, BERTweet and BERT-C): between 60.71\% and 63.53\% positive (39.29\% and 36.47\% negative). By contrast, XLM-T, while still not far from the others\footnote{Note that the difference between XLM-T and the other models could be partly due to XLM-T being a multilingual model.}, leans towards almost equal numbers of positive and negative relationships (52.48\% positive to 47.52\% negative).  

Overall, these observations suggest that even though the models may have significant variations in their predicted labels for interactions, these differences shrink when it comes to labelling relationships. As we verify at the end of this section, given the use of a threshold for signing relationships, this finding is due to the models disagreeing on interactions in relationships that are either very positive or very negative (i.e. where the signs of a few interactions could change without changing the sign of the relationship). Thus, the golden interaction threshold approach of signing relationships appears to achieve very similar results with three of the models used for signing the individual interactions and reasonably close results for the fourth. Effectively, this robustness is due to the threshold-based nature of the relationship signing method, which can tolerate a certain degree of disagreement.

Note, as an additional remark, that the percentages in Figure~\ref{fig:relationshiop_bar_chart} are more negative than the aforementioned observations of previous research (between 15.0\% and 22.6\% negative~\cite{Leskovec_2010_a_Signed}). However, as mentioned in Subsection~\ref{sec:signed_networks_background}, those results were observed in networks with publicly visible signed links, meaning that the number of negative links could have been suppressed due to the effects of Social Capital ~\cite{Coleman_1988}. Thus, it is expected that datasets without explicit signs that are disclosed to the users (as is the case for all datasets used in this paper) would be more negative than these previous findings.

Next, the level of agreement between each of the models was calculated using the proportion of predicted labels that matched exactly with the corresponding labels predicted by the other models. This was done to verify that the models are not just displaying similar amounts of negative relationships but are actually agreeing on the signs of specific interactions and relationships. A matrix displaying these proportions for both the individual interactions labels and the relationships labels can be seen in Table~\ref{nlp_comparison_interactions} and Table~\ref{nlp_comparison_relationships}, respectively. For the relationships, only those with 6 or more interactions are included as this is the minimum length required for the relationship signs to be considered reliable. This is due to the 1:5 golden interaction ratio used for signing the relationships (see Section~\ref{sec:signing_relationships}). 
Indeed, for interactions, the models display somewhat high levels of agreement; ranging from 0.56 to 0.73. What's more, when looking at relationships, the models tend to agree much more; between 0.66 and 0.84.
While the models' strong agreements do not explicitly give an indication of their performance for the task of signing relationships, it does further illustrate that the relationship labels obtained are reasonably independent of models. Thus, the method of signing relationships proposed in this paper can work irrespective of the choice of model used to analyse the sentiments of individual interactions.

\begin{table}[t]
    \centering
    \caption{The proportions of interactions that each pair of sentiment analysis models agree upon.} 
    \label{nlp_comparison_interactions}
    \begin{tabular}{@{}lrrrrr@{}}
        \toprule
        & \textbf{VADER} & \textbf{BERTweet} & \textbf{XLM-T} & \textbf{BERT-C}\\ 
        \midrule
        VADER & - & 0.64 & 0.60 & 0.56\\
        BERTweet & 0.64 & - & 0.73 & 0.60\\
        XLM-T & 0.60 & 0.73 & - & 0.64\\
        BERT-C & 0.56 & 0.60 & 0.64 & - &\\
        \bottomrule
    \end{tabular}
\end{table}

\begin{table}[t]
    \centering
    \caption{The proportions of relationships that each pair of sentiment analysis models agree upon. Only relationships with at least 6 interactions are included.}
    \label{nlp_comparison_relationships}
    \begin{tabular}{@{}lrrrrr@{}}
        \toprule
        & \textbf{VADER} & \textbf{BERTweet} & \textbf{XLM-T} & \textbf{BERT-C}\\
        \midrule
        VADER & - & 0.79 & 0.76 & 0.66\\
        BERTweet & 0.79 & - & 0.84 & 0.69\\
        XLM-T & 0.76 & 0.84 & - & 0.76 &\\
        BERT-C & 0.66 & 0.69 & 0.76 & - &\\
        \bottomrule
    \end{tabular}
\end{table}

\begin{figure}
    \centering
    \includegraphics[scale=0.7]{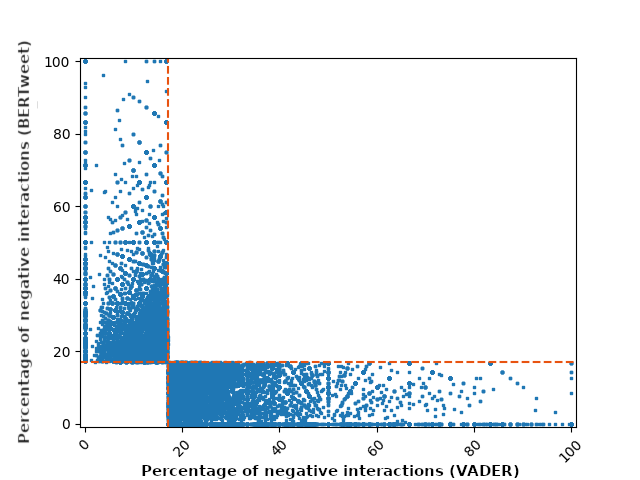}
    \caption{Example disagreement scatter plot. Each point corresponds to a relationship where two target models (here, VADER and BERTweet) disagree. The x-coordinate of the point corresponds to the percentage of negative interactions in the relationship according to VADER, and the y-coordinate to the percentage of negative interactions in the relationship according to BERTweet. Only relationships with at least 6 interactions are included.}
    \label{fig:threshold_disagreements_example}
\end{figure}

In order to gain a better understanding of the degree to which the models disagree with each other, we then investigated the percentage of negative interactions in the relationships that pairs of models disagreed on (i.e. the percentages that are used in combination with the golden interaction ratio to determine a relationship's sign). Again, only relationships with at least 6 interactions are included. By plotting these negativity percentages for pairs of models, it is possible to visualise where the models are disagreeing, as in the example Figure~\ref{fig:threshold_disagreements_example}\footnote{Observing the graphs, one may take note of the horizontal lines at the 0.0 mark on the y-axis. This corresponds to the case in which one model considers the relationship to be entirely positive but the other model still marks it as negative. While these strong disagreements are somewhat surprising, the majority of these occur before the 33\% mark along the x-axes, i.e. close to the threshold, so most of them still correspond to relatively slight disagreements. What's more, the average number of interactions corresponding to these strong disagreements is 12.15, compared to 27.69 for all disagreements, meaning that strong disagreements are much more likely to happen for relationships with fewer interactions.}. However, given the fractional nature of these values, there are many points that overlap with one another. To combat this, and to gain a more precise, numerical perspective, we then look at where the quantiles of these disagreements are. Specifically, for each relationship marked as positive when using model X (meaning that the corresponding fraction of negative interactions is below 0.17) and as negative when using model Y (meaning that the fraction $\gamma_Y$ of negative interactions is above 0.17), we compute the distribution of $\gamma_Y$. If our hypothesis is correct, we expect $\gamma_Y$ to be concentrated in the area close to 0.17. The exact values of the quantiles corresponding to the distribution of disagreements in the bottom-right area\footnote{As some of the information in these plots is duplicated, for example, the comparison between model A and model B would be the mirror of the comparison between model B and model A, only the lower half of these plots have been included.} of the example Figure~\ref{fig:threshold_disagreements_example} are displayed in Table~\ref{disagreement_quantiles}, along with those of the other combinations of models. The associated figures can be found in Appendix~\ref{appendix:model_disagreement_plots}. These numbers show that the vast majority of disagreements are indeed happening in the area immediately above the 0.17 golden interaction ratio. This suggests that, even when the models do disagree, they usually don't disagree by very much. Even the model that disagrees the most strongly with the others, BERT-C, has its third quantiles, i.e. 75\% of its disagreements, under and around 40, which corresponds to approximately only 30\% of the disagreement range $(17,100)$.

\begin{table}[t]
    \centering
    \caption{Disagreement quantiles. The model giving a positive label is on the top and the model giving a negative label is on the left.} 
    \label{disagreement_quantiles}
    \begin{tabular}{@{}llrrrr@{}}
        \toprule
        & & \textbf{VADER} & \textbf{BERTweet} & \textbf{XLM-T} & \textbf{BERT-C}\\
        \midrule
        \multirow{4}{*}{Q1} & VADER & - & 20.83 & 20.83 & 22.22\\
        & BERTweet & 21.43 & - & 20.00 & 22.22\\
        & XLM-T & 23.08 & 22.22 & - & 22.73\\
        & BERT-C & 25.00 & 23.73 & 22.22 & -\\
        \midrule
        \multirow{4}{*}{Q2} & VADER & - & 25.00 & 25.00 & 27.78\\
        & BERTweet & 26.09 & - & 25.00 & 28.57\\
        & XLM-T & 28.57 & 27.27 & - & 28.57\\
        & BERT-C & 33.33 & 30.52 & 28.57 & -\\
        \midrule
        \multirow{4}{*}{Q3} & VADER & - & 33.33 & 33.33 & 33.33\\
        & BERTweet & 33.33 & - & 28.57 & 33.33\\
        & XLM-T & 37.50 & 33.33 & - & 35.71\\
        & BERT-C & 42.86 & 40.00 & 37.50 & -\\
        \bottomrule
    \end{tabular}
\end{table}

\subsection{Validation via Triad Analysis}
\label{sec:triad_analysis}

The results in the previous section have shown that the signing method is sufficiently robust to the choice of classifier but they do not tell us anything about the soundness of the obtained signs. In order to validate the assigned signs, we leverage triad analysis as discussed in Section~\ref{sec:triad_analysis_method}. Recall that there are four types of triads (as illustrated in Figure~\ref{fig:triad_types}, depending on the number $i$ of positive edges in them, with $T_i$ denoting triads with $i$ positive edges).
As a triad requires interconnected users, most of the datasets included in this work are unsuitable for this analysis, as they contain data from a series of largely disconnected users. The one exception to this is the Snowball dataset, which, due to its snowball collection methodology, contains interconnected users. Therefore, the analysis of the signed triads was only conducted for the Snowball dataset. Fortunately, this is the largest dataset included in this study and is therefore the most likely to produce reliable results.

Four sets of signed triads were obtained using each of the four sentiment analysis classifiers. These were then compared against the signed triads extracted from their corresponding null models where the signs are randomly shuffled, as explained in Section~\ref{sec:triad_analysis_method}. The triad counts and proportions, as well as the mean expectations and surprise levels (calculated using Equations~\ref{eq:surprise}), can be seen in Table~\ref{triad_table}. The main focus for this analysis is the surprise (rightmost column), which indicates the number of standard deviations by which the predicted number of each triad differs from that of the randomly shuffled version. According to the weaker version of Structural Balance Theory, triad 3 should be overrepresented and triad 2 should be underrepresented, and this is indeed the case for all 4 of the models. This qualitatively confirms that the patterns of the extracted signs are compatible with what is observed in explicitly signed human social networks. Additionally, the surprisingly abundant $T_0$ provides an initial glimpse at the higher prevalence of negative relationships on Twitter, which we explore further in the subsequent sections.

Before moving on, it is important to note that, quantitatively, this triad analysis does not provide a means of comparison between the models. In other words, the magnitude of the surprise in the expected direction (e.g., $T_3$ being overrepresented) is not a measure of how good the model is (because there is no such numerical notion of ``correct amount of surprise'').

\begin{table*}[t]
    \centering
    \caption{Results of the triad analysis, with the counts and proportions of the observed triads from each model, along with the expected proportions (for a random distribution of signs) and the level of surprise (as described in Subsection~\ref{sec:triad_analysis_method}).} 
    \label{triad_table}
    \begin{tabular}{@{}lcrrrr@{}}
        \toprule
        \textbf{Model} & \textbf{Triad $T_i$} & \textbf{Counts} & \textbf{Proportions} & \textbf{Expectation} & \textbf{Surprise}\\
        \midrule
        
        \multirow{4}{*}{VADER}
         & $T_3$ & 16,734 & 0.267 & 0.212 & 33.4\\
         & $T_2$ & 19,018 & 0.303 & 0.431 & -64.1\\
         & $T_1$ & 16,934 & 0.270 & 0.287 & -12.0\\
         & $T_0$ & 10,020 & 0.160 & 0.064 & 94.9\\
        \hline

        \multirow{4}{*}{BERTweet}
         & $T_3$ & 21,439 & 0.342 & 0.232 & 65.5\\
         & $T_2$ & 15,771 & 0.252 & 0.437 & -93.8\\
         & $T_1$ & 15,057 & 0.240 & 0.274 & -18.8\\
         & $T_0$ & 10,439 & 0.166 & 0.057 & 117.7\\
        \hline

        \multirow{4}{*}{XLM-T}
         & $T_3$ & 15,873 & 0.253 & 0.122 & 100.1\\
         & $T_2$ & 12,715 & 0.203 & 0.372 & -87.7\\
         & $T_1$ & 15,946 & 0.254 & 0.377 & -63.6\\
         & $T_0$ & 18,172 & 0.290 & 0.128 & 120.9\\
        \hline

        \multirow{4}{*}{BERT-C}
         & $T_3$ & 20,683 & 0.330 & 0.222 & 64.8\\
         & $T_2$ & 15,623 & 0.249 & 0.435 & -93.8\\
         & $T_1$ & 15,366 & 0.245 & 0.281 & -20.2\\
         & $T_0$ & 11,034 & 0.176 & 0.062 & 119.2\\
        \bottomrule
    \end{tabular}
\end{table*}

In the interest of time, all subsequent analyses were conducted using only the signs of a single model. As all the models met the expectations of Structural Balance Theory, they are all equally appropriate. However, given that VADER is well-established and known to annotate individual Tweets more accurately than individual humans~\cite{Hutto_2014}, this was the model that was selected.

\subsection{Negative Relationships in Full and Active Networks}
\label{sec:results_full_active}

\begin{figure}
    \centering
    \includegraphics[scale=0.355]{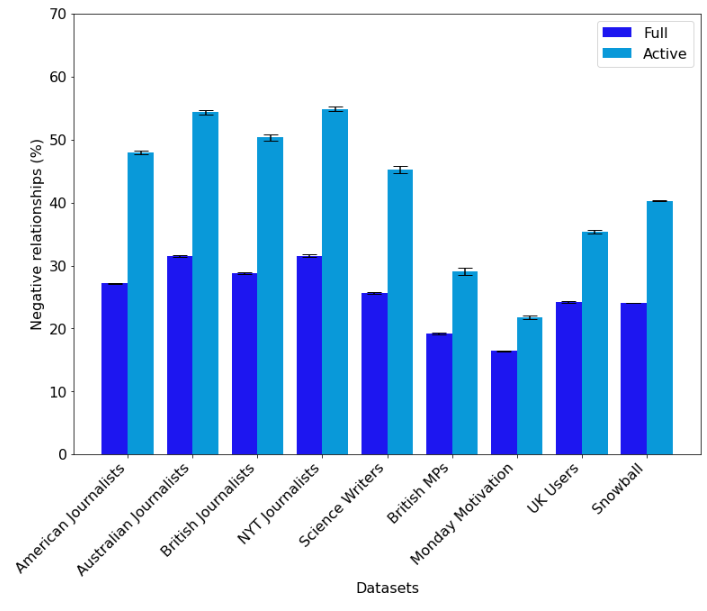}
    \caption{Percentages of relationships that are negative for the full and active networks of each dataset (95\% confidence intervals)}
    \label{fig:full_active_bar_chart}
\end{figure}

We now investigate how the signs are distributed inside the Ego Networks. The percentages of negative relationships in the full and active Ego Networks were compared for each of the 9 datasets. Recall from Subsection~\ref{sec:egonets_background} that the active Ego Network is defined as the set of Alters with whom the Ego engages meaningfully (at least one interaction a year, as per the anthropological definition). These percentages are displayed in Figure~\ref{fig:full_active_bar_chart}.

For the full networks, the datasets display levels of negativity within and slightly above the previously observed range of 15.0\% to 22.6\%~\cite{Leskovec_2010_a_Signed} (mentioned in Subsection~\ref{sec:signed_networks_background}). Specifically, the full Ego Network negativities all fall between 16.45\% (Monday Motivation) and 31.58\% (NYT Journalists). Given that the signs of the links in the current datasets are not explicitly visible to the users, and that, therefore, social pressure towards having positive links will likely be reduced, these observations are very much in line with a priori expectations. 

By contrast, the active networks show significantly higher, albeit more varied, levels of negativity, between 21.83\% (Monday Motivation) and 54.89\% (NYT Journalists). This increase in negativity from the full to active networks suggests that individuals have proportionally greater numbers of negative relationships amongst close contacts with whom they engage frequently than amongst acquaintances. Messages containing or eliciting negative emotions have previously been shown to elicit stronger responses~\cite{Baumeister_2001} and to spread faster~\cite{Rozin_2001} than positive ones. Therefore, one explanation for the higher negativities of the active networks could be that, because the users of the active networks are communicating more frequently, any negative content that enters a user's Ego Network is more likely to be dispersed along the more active connections. Therefore, the connections of the active networks may display higher negatives because they have an elevated risk of being exposed to and spreading negativity. Thus, the more engaged an individual is, the seemingly greater the likelihood their relationships have of being negative.

In addition, although the increase in negativity from full to active network is most pronounced for the journalist datasets and science writers, this change is observable for all 9 of the included datasets. Therefore, rather than being a unique feature of any specific community, it appears that increased negativity is an inevitable byproduct of engaging with Twitter. Investigating whether this phenomenon is observable for other social platforms, as well as how the effects differ, could be an interesting avenue for future research.

\subsection{Negative Relationships of Specialised and Generic Users}
\label{sec:results_specialised_generic}
After observing the full and active networks, the negativities of the specialised and generic users were compared. As can be seen in Figure~\ref{fig:full_active_bar_chart}, most of the specialised users display higher percentages of negative relationships, compared to the generic users. However, this difference is fairly small for the full networks, with Snowball and the generic UK Users dataset actually containing more negative relationships (24.05\% and 24.22\% respectively) than the British MPs (19.24\%) and nearly as much as the Science Writers (25.62\%). By comparison, the difference for the active networks is much starker. With the only exception of the British MPs (whose change in negativity better matches those of the generic datasets), the least negative specialised dataset, Science Writers (45.23\%), was nearly 5 percentage points more negative than the most negative generic dataset, Snowball (40.31\%).

The greater negativities of specialised users also support the hypothesis that more engaged users are more likely to have a greater number of negative relationships, mentioned in the previous subsection. 

\subsection{Circle-by-Circle Analysis of the ENM}
\label{sec:results_circles}

\begin{figure}
    \centering
    \includegraphics[scale=0.35]{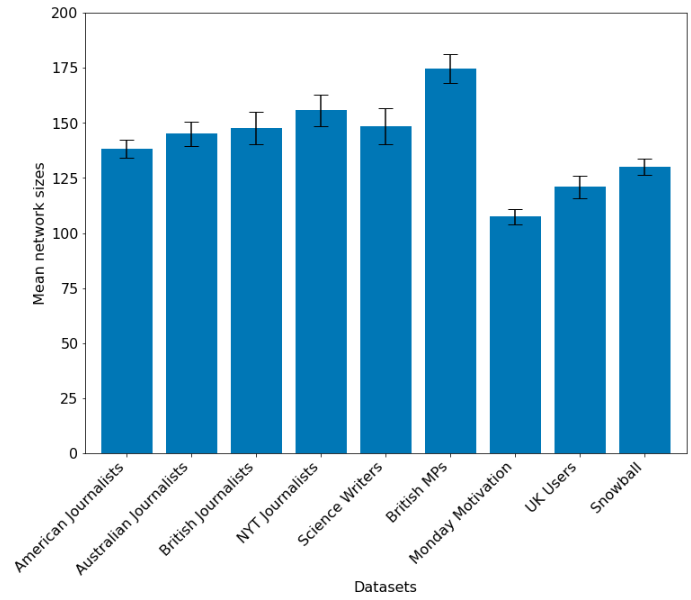}
    \caption{Mean active Ego Network sizes of users with 5 circles in each dataset (95\% confidence intervals)}
    \label{fig:egonetwork_sizes_bar_chart}
\end{figure}

\begin{figure}
    \centering
    \includegraphics[scale=0.35]{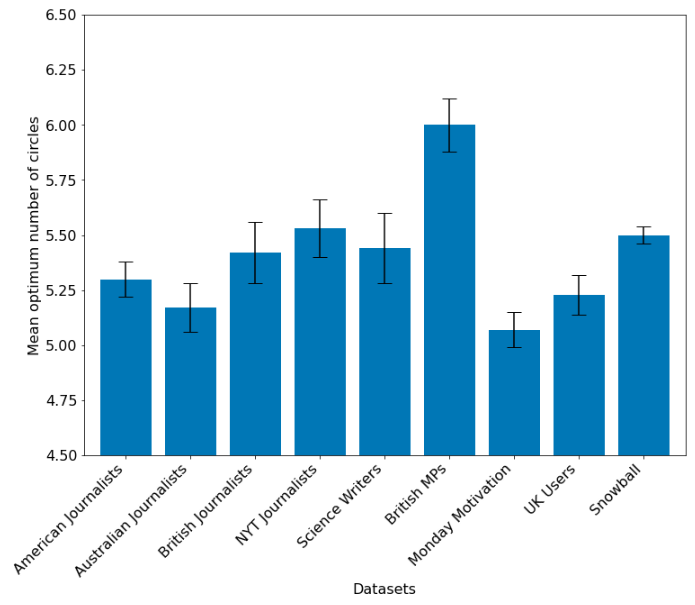}
    \caption{Mean number of circles for each dataset (95\% confidence intervals)}
    \label{fig:optimum_circle_sizes_bar_chart}
\end{figure}

As previously mentioned, the ENM is concentric, meaning that each of its circles contains all the Alters of the circles that come before it. In this section, we briefly analyze the circle sizes and the scaling ratios in the ENMs of our datasets, before proceeding with the SENM discussion\footnote{This is necessary in order to ensure that the data used for the SENM analysis is compatible with the general models of Ego Networks that emerged in the related literature.}.
%
%
It is important to note that the size of Ego Networks tends to vary slightly from Ego to Ego due to various social differences between individuals, as can be seen in Figure~\ref{fig:egonetwork_sizes_bar_chart}. 
Because of these common variations, and in order to standardise the results of any analysis performed on the circles, it is standard practice to focus on Egos who have a common number of circles~\cite{Boldrini_2018,Toprak_2021_b_Region-based}. Usually, the chosen number of circles is 5 as it is the most common number for OSN data~\cite{Toprak_2021_a_Harnessing} and, as can be seen in Figure~\ref{fig:optimum_circle_sizes_bar_chart}, 5 is the closest whole number for all except 2 of the datasets, the exceptions being NYT Journalists and British MPs (with mean circle numbers of 5.53 and 6.00 respectively). Further, the mode of all of the datasets is 5, except for NYT journalists and British MPs (which were both 6), so there is, indeed, a concentration of values around 5. 
Therefore, only Egos with 5 circles were considered for the subsequent circle-by-circle analyses.

The first part of the circle-by-circle analysis is to examine the mean sizes of the circles. As expected from previous studies, the sizes are close to those of Dunbar's expected values: i.e. 1.5, 5, 15, 50, 150 (with the typical scaling factor of roughly 3)~\cite{Dunbar_2015}, the exact numbers can be seen in Table~\ref{circle_all} (along with the remaining number of Egos after considering only those with 5 circles for each dataset). Note that the difference between the numbers in column ``Circle 5" of Table~\ref{circle_all} and the values displayed in Figure~\ref{fig:egonetwork_sizes_bar_chart} is due to the fact that, in the former case, we only consider egos with five circles, while all Egos are included in the latter. Some of the datasets (such as NYT Journalists, UK Users and Monday Motivation) become somewhat distant for the expected numbers in the outermost circle, however, this has also been observed in previous research~\cite{Arnaboldi_2017,Toprak_2021_b_Region-based}. What's more, the increasing scale of roughly 3 is clearly visible in Table~\ref{tab:circle_scaling}, with 3 being the closest whole number to every single one of the ratios between subsequent circles as well as for the overall means of each dataset.

\begin{table*}[htbp]
    \centering
    \caption{Mean circle sizes and number of Egos with 5 circles}
    \label{circle_all}
    \begin{tabular}{@{}lrrrrrr@{}}
        \toprule
        \textbf{Dataset} & \textbf{Circle 1} & \textbf{Circle 2} & \textbf{Circle 3} & \textbf{Circle 4} & \textbf{Circle 5} & \textbf{\# Egos w/ 5 circles}\\
        \midrule
        American Journalists & 1.61 & 5.33 & 15.01 & 41.78 & 127.28 & 300\\
        Australian Journalists & 1.41 & 4.76 & 13.61 & 40.22 & 134.71 & 146\\
        British Journalists & 1.83 & 6.27 & 16.87 & 48.07 & 142.52 & 86\\
        NYT Journalists & 1.65 & 5.43 & 14.76 & 40.16 & 114.68 & 97\\
        Science Writers & 1.70 & 5.81 & 16.40 & 44.29 & 124.86 & 67\\
        British MPs & 1.98 & 6.67 & 18.09 & 49.00 & 146.79 & 103\\
        \hdashline
        Monday Motivation & 1.72 & 5.26 & 13.22 & 33.58 & 103.71 & 421\\
        UK Users & 1.84 & 5.96 & 15.72 & 39.32 & 114.66 & 224\\
        Snowball & 1.78 & 6.16 & 16.86 & 44.19 & 125.91 & 1,160\\
        \bottomrule
    \end{tabular}
\end{table*}

\begin{table*}[htbp]
    \centering
    \caption{Scaling ratios between circle sizes}
    \label{tab:circle_scaling}
    \begin{tabular}{@{}lrrrrr@{}}
        \toprule
        \textbf{Dataset} & \textbf{Circle 1-2} & \textbf{Circle 2-3} & \textbf{Circle 3-4} & \textbf{Circle 4-5} & \textbf{Mean}\\
        \midrule
        American Journalists & 3.30 & 2.82 & 2.78 & 3.05 & 2.99\\
        Australian Journalists & 3.37 & 2.86 & 2.96 & 3.35 & 3.13\\
        British Journalists & 3.43 & 2.69 & 2.85 & 2.96 & 2.98\\
        NYT Journalists & 3.30 & 2.72 & 2.72 & 2.86 & 2.90\\
        Science Writers & 3.41 & 2.82 & 2.70 & 2.82 & 2.94\\
        British MPs & 3.37 & 2.71 & 2.71 & 3.00 & 2.95\\
        \hdashline
        Monday Motivation & 3.07 & 2.51 & 2.54 & 3.09 & 2.80\\
        UK Users & 3.24 & 2.64 & 2.50 & 2.92 & 2.82\\
        Snowball & 3.46 & 2.74 & 2.62 & 2.85 & 2.92\\
        \bottomrule
    \end{tabular}
\end{table*}

Next, before considering the relationship signs at each level of the SENM, we gauge the appropriacy of the threshold method of signing relationships (described in~\ref{sec:signing_relationships}) for our data. Indeed, given that psychological research has found the golden interaction ratio to be 1:5, we consider that a relationship requires a minimum of 6 interactions in order to be signed reliably. Therefore, we investigate the Egos' mean numbers of interactions per Alter at each level of the ENM, in order to verify that we have enough data to properly apply the threshold. The results, summarised in Table~\ref{num_interactions}, show that circles 1 to 4 have mean numbers of interactions that are equal to or greater than the required 6. Indeed, only the outermost circle tends to have numbers that are lower than necessary. This means that for circles 1 to 4, there is enough data, on average, to properly estimate the signs. 

\begin{table*}[htbp]
    \centering
    \caption{Mean number of interactions per Alter at each level of the ENM} 
    \label{num_interactions}
    \begin{tabular}{@{}lrrrrr@{}}
        \toprule
        \textbf{Dataset} & \textbf{Circle 1} & \textbf{Circle 2} & \textbf{Circle 3} & \textbf{Circle 4} & \textbf{Circle 5}\\
        \midrule
        American Journalists & 59.09 & 23.66 & 12.69 & 6.69 & 3.18\\
        Australian Journalists & 83.90 & 21.80 & 12.56 & 6.33 & 3.02\\
        British Journalists & 49.14 & 20.10 & 12.26 & 6.60 &  3.13\\
        NYT Journalists & 50.30 & 22.48 & 11.88 & 6.00 & 2.56\\
        Science Writers & 59.91 & 25.59 & 14.91 & 7.11 &  3.03\\
        British MPs & 106.59 & 50.95 & 27.07 & 13.14 & 5.11\\
        \hdashline
        Monday Motivation & 105.80 & 46.44 & 26.06 & 12.25 & 3.77\\
        UK Users & 86.53 & 44.34 & 22.53 & 10.81 & 3.72\\
        Snowball & 174.19 & 68.69 & 34.95 & 16.71 & 6.48\\
        \bottomrule
    \end{tabular}
\end{table*}

Beyond validating the application of the golden interaction ratio to relationships in circles 1 to 4, Table~\ref{num_interactions} also shows that journalists tend to interact about half as much per Alter compared to generic users. While this finding is initially counter-intuitive (given that journalists are generally considered to be more engaged with Twitter), a follow-up examination of the different types of interactions sent from the Egos revealed that this is actually in line with the findings of previous works. Essentially, specialised users, such as journalists, tend to generate more Mentions and Retweets, and fewer Replies, than generic users. Based on the conclusions of previous work~\cite{Toprak_2022}, this suggests that specialised users generally spread their cognitive effort across slightly more distinct connections than generic users, while generic users tend to spend slightly more cognition on each individual relationship. This is supported by the slightly higher active ego network sizes of the specialised users in Table~\ref{circle_all} (see Circle 5 column).
While this investigation is important for properly understanding the results of Table~\ref{num_interactions}, its findings are only tangentially related to the main focus of this paper. Consequently, the full details have been placed in Appendix~\ref{appendix:number_interactions_per_alter_investigation}.

\subsection{Circle-by-Circle Analysis of the SENM}

Next, moving on to the analysis of the SENM, we observe the mean numbers and percentages of negative relationships for each circle, these can be seen in Table~\ref{tab:circle_negative}. The proportions of negative relationships are found to be disproportionately higher at the innermost circles of the ENM, especially for specialized users, decreasing steadily towards the outer layers. The negative percentages of all journalist datasets are above 61\% at the innermost circle and are below 55\% at the outermost. This is very surprising as the inner sections of the ENM should be associated with an individual's most trusted and similar connections. Indeed, one of the four components from Granovetter's definition of tie strength is reciprocal services~\cite{Granovetter_1973}, and reciprocity is thought to be very closely related to trust~\cite{Ostrom_2003}. What makes these findings even more surprising is that the aforementioned effect of social capital, which creates a bias towards maintaining positive connections, would be strongest in the innermost circles, where individuals are expected to be the most tightly knit.

\begin{table*}[htbp]
    \centering
    \caption{Mean number and percentage of negative relationships at each level of the Signed Ego Network (for Egos with 5 circles). In bold, the most negative circle of each dataset.}
    \label{tab:circle_negative}
    \begin{tabular}{@{}lrrrrrr@{}}
        \toprule
        \textbf{Dataset} & \textbf{Circle 1} & \textbf{Circle 2} & \textbf{Circle 3} & \textbf{Circle 4} & \textbf{Circle 5} & \textbf{Difference}\textsuperscript{b}\\
        \midrule
        \multirow{2}{*}{American Journalists} & \textbf{0.99} & 3.15 & 8.53 & 22.28 & 60.23 &\\
        & \textbf{61.37\%} & 59.13\% & 56.85\% & 53.33\% & 47.32\% & -14.05\\
        \multirow{2}{*}{Australian Journalists} & \textbf{1.09} & 3.34 & 9.08 & 25.27 & 73.03 &\\
        & \textbf{77.30\%} & 70.14\% & 66.74\% & 62.82\% & 54.21\% & -23.09\\
        \multirow{2}{*}{British Journalists} & \textbf{1.16} & 3.63 & 9.76 & 27.02 & 70.94 &\\
        & \textbf{63.33\%} & 57.98\% & 57.85\% & 56.22\% & 49.77\% & -13.56 \\
        \multirow{2}{*}{NYT Journalists} & 1.11 & \textbf{3.73} & 9.90 & 24.73 & 60.43 &\\
        & 67.21\% & \textbf{68.66\%} & 67.05\% & 61.59\% & 52.70\% & -14.51\\
        \multirow{2}{*}{Science Writers} & 0.82 & \textbf{2.90} & 7.97 & 21.31 & 55.87 &\\
        & 48.39\% & \textbf{49.91\%} & 48.59\% & 48.11\% & 44.75\% & -3.64\\
        \multirow{2}{*}{British MPs} & \textbf{0.58} & 1.88 & 5.08 & 13.09 & 31.31 &\\
        & \textbf{29.41\%} & 28.24\% & 28.07\% & 26.71\% & 21.33\% & -8.08\\
        \hdashline
        \multirow{2}{*}{Monday Motivation} & 0.30 & 0.97 & \textbf{2.46} & 5.79 & 14.09 &\\
        & 17.72\% & 18.37\% & \textbf{18.59\%} & 17.25\% & 13.58\% & -4.14\\
        \multirow{2}{*}{UK Users} & \textbf{0.64} & 2.00 & 5.27 & 13.14 & 37.63 &\\
        & \textbf{34.75\%} & 33.46\% & 33.54\% & 33.41\% & 32.81\% & -1.94\\
        \multirow{2}{*}{Snowball} & 0.71 & \textbf{2.56} & 6.93 & 17.80 & 48.99 &\\
        & 40.17\% & \textbf{41.54\%} & 41.12\% & 40.29\% & 38.91\% & -1.26\\
        \bottomrule
        \multicolumn{7}{l}{$^{\mathrm{b}}$ Difference between circle 1 and circle 5 in percentage points.}
    \end{tabular}
\end{table*}

Despite these observed differences in the proportions of negative relations across the circles,  an observable ratio similar to that of the circle sizes appears to be fairly consistent, as can be seen in Table~\ref{negative_circle_scaling}. The mean value of this negativity ratio is marginally lower than that of the circle sizes, however, it is still roughly equal to 3. Looking at the mean column, this ratio appears to be roughly 2.8.

\begin{table*}[htbp]
    \centering
    \caption{Scaling ratios of negative relationships counts between circle sizes}
    \label{negative_circle_scaling}
    \begin{tabular}{@{}lrrrrrr@{}}
        \toprule
        \textbf{Dataset} & \textbf{Circle 1-2} & \textbf{Circle 2-3} & \textbf{Circle 3-4} & \textbf{Circle 4-5} & \textbf{Mean}\\
        \midrule
        American Journalists & 3.18 & 2.71 & 2.61 & 2.70 & 2.80\\
        Australian Journalists & 3.06 & 2.72 & 2.78 & 2.89 & 2.86\\
        British Journalists & 3.14 & 2.68 & 2.77 & 2.63 & 2.80\\
        NYT Journalists & 3.37 & 2.65 & 2.50 & 2.44 & 2.74\\
        Science Writers & 3.52 & 2.75 & 2.67 & 2.62 & 2.89\\
        British MPs & 3.23 & 2.70 & 2.58 & 2.39 & 2.72\\
        \hdashline
        Monday Motivation & 3.18 & 2.54 & 2.36 & 2.43 & 2.63\\
        UK Users & 3.12 & 2.64 & 2.49 & 2.86 & 2.78\\
        Snowball & 3.58 & 2.71 & 2.57 & 2.75 & 2.90\\
        \bottomrule
    \end{tabular}
\end{table*}

In Table~\ref{tab:circle_negative}, we can compare the proportions of negative relationships between the different types of users. Once again, there appears to be a divide between specialised and generic users. This difference becomes even more noticeable when the journalists are compared to the non-journalists. Indeed, the variations in negativity across the circles appear to be much greater for journalists than for any of the other datasets. The most stable journalist dataset (British Journalists) drops by 13.56 percentage points from circle 1 to circle 5. By contrast, the biggest variation for the non-journalists is 8.08 percentage points (British MPs).

Again, these observations lend support to the notion that increased levels of engagement with Twitter lead to increased levels of negativity. Egos engage the most with their innermost circles and this is where the strongest concentration of negative relationships is found. What's more, the difference between the negativity at this innermost level and that of the outer level is greatest for the most engaged category of users (journalists). Otherwise said, the most negativity is found at the highest levels of engagement and this is true at every level of the Ego Networks as well as between different types of users. This could also explain why the $T_0$ triads in Section~\ref{sec:triad_analysis} were so prevalent.

\subsection{Negativity Metrics}
\label{sec:negativity_metrics_results}

As discussed in Subsection~\ref{sec:negativity_metrics}, a final analysis was carried out to investigate whether maintaining negative relationships is more cognitively demanding than maintaining positive ones. For this, 3 different metrics were computed: the proportion of negative relationships, the proportion of negative interactions and the cognitive effort spent on negative relationships (details in Section~\ref{sec:negativity_metrics}). These metrics were then compared to the sizes of the users' active Ego Networks and the number of users' interactions: both statistically and graphically.

For the statistical comparisons, Pearson's R was used. Our hypothesis is that an increase in negativity may correspond to an increase in cognitive effort, hence to smaller active Ego Networks and fewer interactions (this latter hypothesis is based on our observations in Section~\ref{sec:results_circles}, which showed that specialised users, who show higher negativity levels, tend to display roughly half the number of interactions as generic users). Thus, a 1-tailed analysis was employed. The results showed no significant correlations for any of the datasets for either the active Ego Network sizes (\emph{p}\textgreater.523 for all cases) or the number of interactions (\emph{p}\textgreater.531 for all cases). This suggests that negativity does not decrease the size of Ego Networks, on average. 

\begin{figure*}
    \centering
    \includegraphics[scale=0.39]{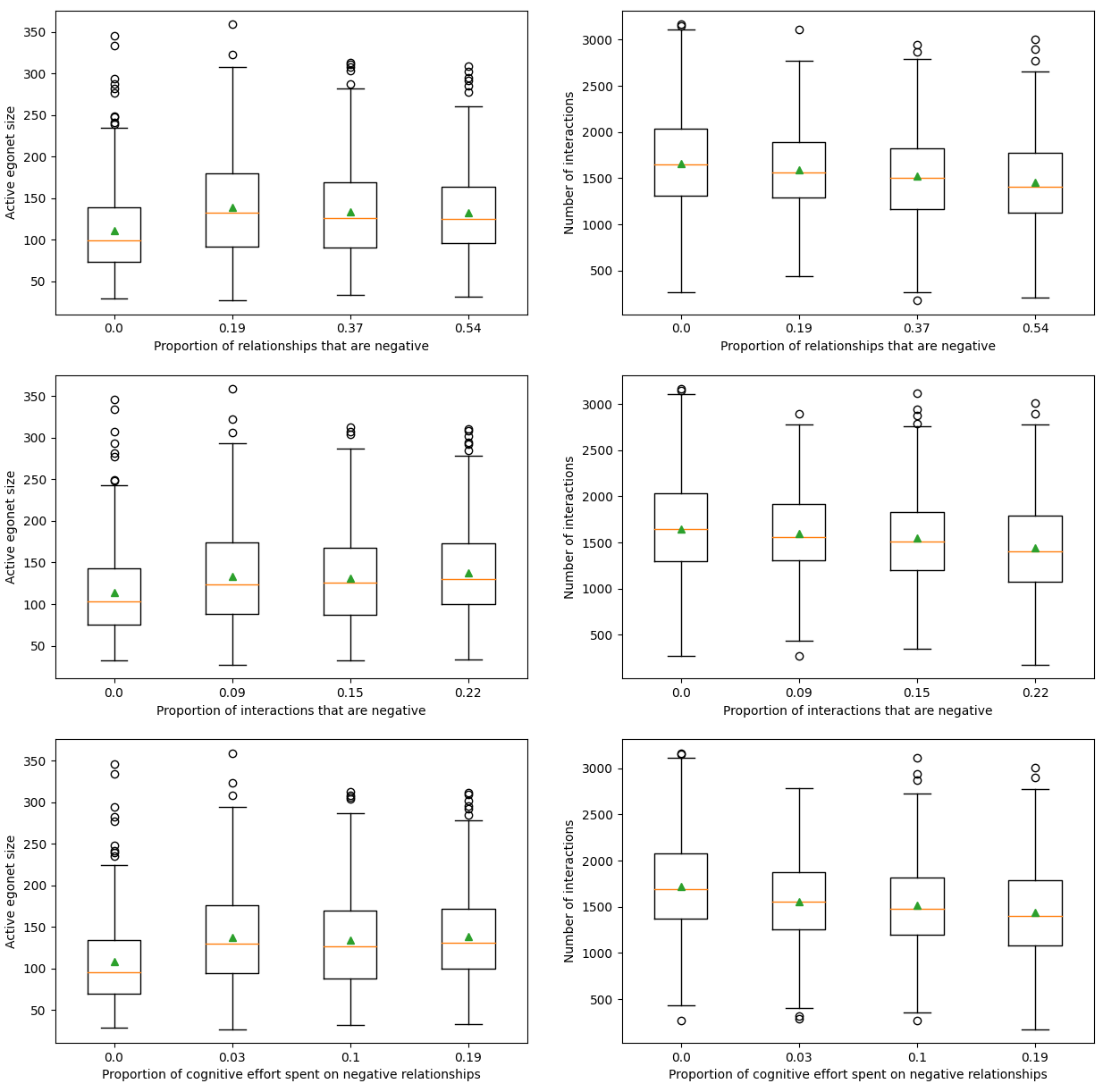}
    \caption{Boxplots for active Ego Network size (left column) and number of interactions (right column) against the 3 negativity metrics (top, middle and bottom) for the Snowball dataset. For each group of binned Egos, the boxplots display mean (orange line), median (green triangle), first to third quartile (box), 1.5 times the interquartile range beyond the box (whiskers) and outliers (black circles).}
    \label{fig:snowball_boxplots_egonets}
\end{figure*}

Next, binned boxplots were made to visualise the interplay between negativity and cognitive effort, for different classes of Ego negativity. We binned the Egos into quantiles with respect to the negativity metrics (as described in Subsection~\ref{sec:negativity_metrics}), and then analysed the distributions of the active Ego Network sizes and the number of interactions in each bin. The corresponding boxplots for the 2 largest datasets in terms of Egos, Snowball and Monday Motivation, can be seen in Figures~\ref{fig:snowball_boxplots_egonets} and \ref{fig:monday_motivation_boxplots_egonets}. The complete set of boxplots is available in Appendix~\ref{appendix:negativity_metrics}. For the majority of the datasets, the means, medians, boxes and whiskers of the boxplots are fairly flat across the bins (as expected given the non-significant correlations). However, the Snowball dataset shows a smaller active ego network for the first quantile and numbers of interactions that steadily decrease from the first to fourth quantile. This 2 observations are seen for all 3 of the negativity metrics.

\begin{figure*}
    \centering
    \includegraphics[scale=0.39]{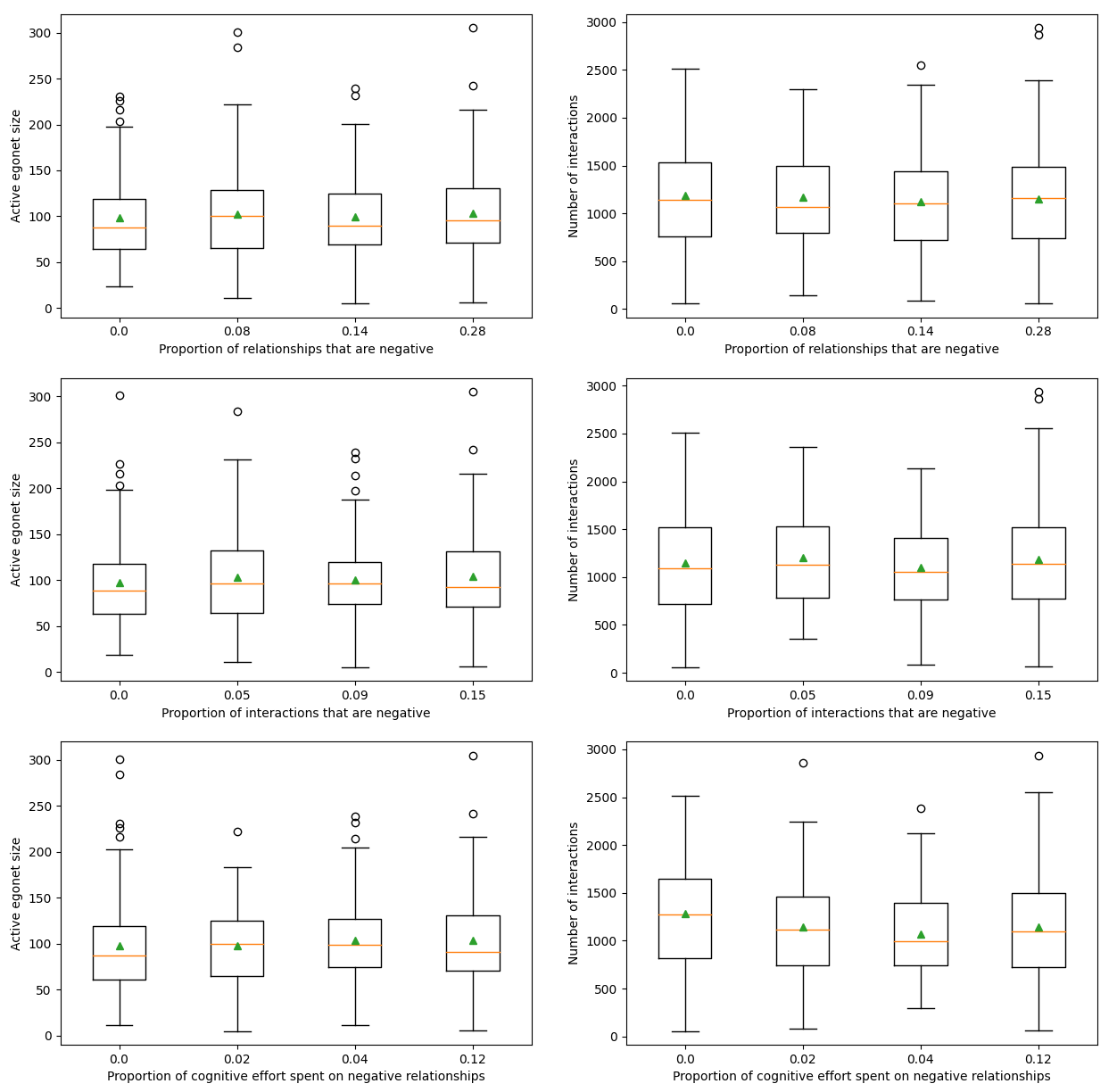}
    \caption{Boxplots for active Ego Network size (left column) and number of interactions (right column) against the 3 negativity metrics (top, middle and bottom) for the Monday Motivation dataset. For each group of binned Egos, the boxplots display mean (orange line), median (green triangle), first to third quartile (box), 1.5 times the interquartile range beyond the box (whiskers) and outliers (black circles).}
    \label{fig:monday_motivation_boxplots_egonets}
\end{figure*}

We followed up on these observations by conducting t-tests between pairs of bins for each dataset, with the null hypothesis being that there should not be any differences between them. This was done for both the Ego network sizes and the number of interactions. The resulting p-values are displayed in Tables~\ref{p_values_ego_networks} and~\ref{p_values_num_interactions} respectively and the t-scores are available in Appendix~\ref{appendix:negativity_metrics_t_scores}. The only dataset that displays consistently significant values is Snowball, which shows significant differences for all comparisons involving the first bin, for the Ego network sizes, and all comparisons involving the last bin for the number of interactions. The Snowball results would suggest that users with many positive relationships are likely to have slightly smaller Ego networks (i.e. fewer connections) and those with many negative relationships are likely to have more overall interactions.
Given that the Snowball dataset is significantly larger than the others this may be a relatively weak effect that is only statistically significant when observing a very large sample size.

These two results together would suggest that more positive users tend to have fewer connections and interact less frequently overall but more intimately with the connections they do have (at least on the Twitter platform). While more negative users have more, yet less intimate, connections with whom they interact less frequently compared to the positive users, they still end up interacting the most overall. In other words, these results suggest that nurturing positive relationships in online social networks is more cognitively engaging, resulting in smaller ego networks for more positive users. However, while these results seem very promising, given some of the limitations of the negativity metrics analysis (i.e. the observations were only found to be significant for Snowball dataset), it would be pertinent to further investigate the interplay between these effects.

\begin{table*}[htbp]
    \centering
    \caption{The p-values from the pairwise comparisons between bins for Ego network sizes and negativity. Statistically significant values ($<0.05$) are displayed in bold.}
    \label{p_values_ego_networks}
        \begin{tabular}{@{}llcccccc@{}}
        \toprule
        & & \multicolumn{6}{c}{\textbf{Bin pairs}}\\
        & \textbf{Dataset} & \textbf{1-2} & \textbf{1-3} & \textbf{1-4} & \textbf{2-3} & \textbf{2-4} & \textbf{3-4}\\
        \midrule
        \multirow{9}{*}{Metric 1}
        & American Journalists & 0.743 & 0.152 & \textbf{0.029} & 0.294 & 0.076 & 0.431\\
        & Australian Journalists & 0.235 & 0.966 & 0.414 & 0.143 & \textbf{0.024} & 0.333\\
        & British Journalists & 0.162 & 0.354 & 0.381 & \textbf{0.016} & 0.643 & 0.072\\
        & NYT Journalists & 0.431 & 0.764 & \textbf{0.002} & 0.548 & \textbf{0.001} & \textbf{0.000}\\
        & Science Writers & 0.850 & 0.384 & 0.773 & 0.447 & 0.913 & 0.514\\
        & British MPs & 0.127 & 0.189 & 0.904 & 0.610 & 0.151 & 0.230\\
        & Monday Motivation & 0.598 & 0.892 & 0.447 & 0.680 & 0.846 & 0.519\\
        & UK Users & 0.321 & 0.548 & 0.577 & 0.669 & 0.645 & 0.970\\
        & Snowball & \textbf{0.000} & \textbf{0.000} & \textbf{0.000} & 0.221 & 0.146 & 0.829\\

        \midrule
        \multirow{9}{*}{Metric 2}
        & American Journalists & 0.230 & 0.582 & 0.164 & 0.513 & 0.917 & 0.420\\
        & Australian Journalists & 0.944 & 0.825 & 0.384 & 0.712 & 0.315 & 0.155\\
        & British Journalists & 0.872 & 0.584 & 0.806 & 0.728 & 0.721 & 0.495\\
        & NYT Journalists & 0.289 & 0.375 & 0.236 & 0.787 & 0.967 & 0.770\\
        & Science Writers & 0.242 & 0.383 & 0.959 & 0.712 & 0.232 & 0.378\\
        & British MPs & 0.341 & 0.192 & 0.977 & 0.939 & 0.321 & 0.171\\
        & Monday Motivation & 0.388 & 0.633 & 0.314 & 0.658 & 0.881 & 0.550\\
        & UK Users & 0.326 & 0.089 & 0.387 & 0.558 & 0.875 & 0.426\\
        & Snowball & \textbf{0.000} & \textbf{0.000} & \textbf{0.000} & 0.667 & 0.439 & 0.212\\

        \midrule
        \multirow{9}{*}{Metric 3}
        & American Journalists & 0.323 & 0.791 & 0.274 & 0.511 & 0.995 & 0.473\\
        & Australian Journalists & 0.143 & 0.502 & 0.780 & 0.337 & 0.176 & 0.649\\
        & British Journalists & 0.180 & 0.295 & 0.793 & 0.747 & 0.199 & 0.298\\
        & NYT Journalists & \textbf{0.045} & 0.319 & 0.445 & 0.263 & 0.054 & 0.636\\
        & Science Writers & 0.142 & \textbf{0.020} & 0.768 & 0.483 & 0.207 & \textbf{0.031}\\
        & British MPs & 0.169 & 0.101 & 0.540 & 0.895 & 0.377 & 0.321\\
        & Monday Motivation & 0.997 & 0.364 & 0.390 & 0.280 & 0.315 & 0.998\\
        & UK Users & \textbf{0.036} & 0.452 & 0.426 & \textbf{0.002} & 0.176 & 0.101\\
        & Snowball & \textbf{0.000} & \textbf{0.000} & \textbf{0.000} & 0.421 & 0.978 & 0.391\\
        \bottomrule
    \end{tabular}
\end{table*}

\begin{table*}[htbp]
    \centering
    \caption{The p-values from the pairwise comparisons between bins for number of interactions and negativity. Statistically significant values ($<0.05$) are displayed in bold.}
    \label{p_values_num_interactions}
    \begin{tabular}{@{}llcccccc@{}}
        \toprule
        & & \multicolumn{6}{c}{\textbf{Bin pairs}}\\
        & \textbf{Dataset} & \textbf{1-2} & \textbf{1-3} & \textbf{1-4} & \textbf{2-3} & \textbf{2-4} & \textbf{3-4}\\
        \midrule
        \multirow{9}{*}{Metric 1}
        & American Journalists & 0.175 & 0.910 & 0.414 & 0.148 & 0.560 & 0.359\\
        & Australian Journalists & 0.508 & 0.705 & 0.324 & 0.773 & 0.659 & 0.499\\
        & British Journalists & 0.257 & 0.902 & 0.692 & 0.200 & 0.450 & 0.595\\
        & NYT Journalists & 0.703 & 0.556 & 0.173 & 0.820 & 0.272 & 0.372\\
        & Science Writers & 0.659 & 0.719 & 0.318 & 0.930 & 0.617 & 0.545\\
        & British MPs & 0.726 & 0.239 & 0.385 & 0.373 & 0.579 & 0.718\\
        & Monday Motivation & 0.810 & 0.335 & 0.611 & 0.457 & 0.781 & 0.649\\
        & UK Users & 0.366 & 0.503 & \textbf{0.012} & 0.771 & 0.099 & \textbf{0.042}\\
        & Snowball & 0.126 & \textbf{0.003} & \textbf{0.000} & 0.082 & \textbf{0.001} & 0.110\\

        \midrule
        \multirow{9}{*}{Metric 2}
        & American Journalists & 0.469 & 0.993 & 0.911 & 0.454 & 0.401 & 0.916\\
        & Australian Journalists & 0.567 & 0.606 & 0.100 & 0.929 & 0.223 & 0.184\\
        & British Journalists & 0.873 & 0.783 & 0.607 & 0.893 & 0.714 & 0.843\\
        & NYT Journalists & 0.271 & 0.087 & 0.597 & 0.528 & 0.491 & 0.163\\
        & Science Writers & 0.482 & 0.441 & 0.699 & 0.932 & 0.744 & 0.689\\
        & British MPs & 0.166 & \textbf{0.030} & 0.905 & 0.446 & 0.287 & 0.080\\
        & Monday Motivation & 0.504 & 0.475 & 0.668 & 0.127 & 0.821 & 0.226\\
        & UK Users & 0.462 & 0.993 & 0.466 & 0.448 & 0.129 & 0.440\\
        & Snowball & 0.270 & \textbf{0.022} & \textbf{0.000} & 0.166 & \textbf{0.000} & \textbf{0.015}\\

        \midrule
        \multirow{9}{*}{Metric 3}
        & American Journalists & 0.114 & 0.089 & 0.661 & 0.950 & 0.269 & 0.228\\
        & Australian Journalists & 0.551 & 0.433 & 0.180 & 0.867 & 0.063 & \textbf{0.041}\\
        & British Journalists & 0.422 & 0.643 & 0.350 & 0.780 & 0.097 & 0.194\\
        & NYT Journalists & 0.091 & 0.075 & 0.442 & 0.862 & 0.270 & 0.221\\
        & Science Writers & 0.760 & 0.380 & 0.855 & 0.442 & 0.569 & 0.240\\
        & British MPs & 0.368 & 0.099 & 0.806 & 0.437 & 0.565 & 0.196\\
        & Monday Motivation & 0.080 & \textbf{0.003} & 0.073 & 0.241 & 0.949 & 0.276\\
        & UK Users & 0.284 & 0.286 & 0.470 & \textbf{0.021} & 0.710 & 0.054\\
        & Snowball & \textbf{0.000} & \textbf{0.000} & \textbf{0.000} & 0.445 & \textbf{0.010} & 0.060\\
        \bottomrule
    \end{tabular}
\end{table*}

\section{Conclusion}
\label{sec:conclusion}

The present study introduces a novel method for the inferral of signs in unsigned networks, which leveraged text-based communications among individual pairs of users. The proposed method is founded on solid theoretical underpinnings and enables the application of signed network techniques to non-signed networks in future research, even in situations where data about the global network topology is scarce or unavailable (hence, topology-based tools cannot be applied). The method was shown to be robust to the choice of the underlying sentiment classifier and to reproduce a sign distribution that matches the expectation of the well-known Structural Balance Theory.
To demonstrate its effectiveness, this approach was used to generate signed relationships and Ego Networks across 9 distinct datasets. The resulting signed networks were then systematically examined and compared against their unsigned counterparts.
This concluded in 4 main findings: (i) somewhat unexpectedly, percentages of negative relationships tend to be higher for active networks than for full networks and this is more pronounced for specialised users than for generic users; (ii) specialised users display a higher propensity towards having negative relationships than generic users; (iii) very surprisingly, negative relationships are found disproportionately more at the more intimate levels of the ENM; (iv) having and maintaining negative relationships appears to have a weak detrimental effect on the number of interactions an individual creates and a weak incremental effect on the distinct number of individuals one interacts with.
On top of these core findings, a consolidated signed version of the ENM is also established, with a scaling ratio of negative relationships that decreases slightly from the inner circles to the outer circles for most types of users and which has an overall value that is slightly lower than that of the original model's circle sizes (i.e. roughly 2.8). 

The overall message is that OSNs, while generating \emph{structurally} similar Ego Networks with respect to offline relationships (i.e. not mediated by social platforms), tend to drastically overemphasise negativity, leading to unexpectedly high percentages of negative relationships. On the other hand, our results also provide weak signs of a more positive use of online social platforms, as users who allocated more cognitive efforts to individual relationships tend to enjoy more positive relationships than the average.

These contributions enable several avenues for further research. For instance, investigating the observed effects in other OSNs such as Reddit or Mastodon or examining the interplay between ``positive connections that share negative content'' and ``actually negative relationships'', which greatly increases our understanding of what it means to have and interact with negative relationships as well as how sharing negative content online can affect the polarity of communications over time.


\begin{backmatter}

\section*{Availability of data and materials}
The datasets generated as part of the current study are available on Zenodo for the following dataset:

Snowball (Baseline) at \url{https://zenodo.org/record/7717006#.ZLGhyNKUdkg}

British MPs at \url{https://zenodo.org/record/6420845#.ZLGiANKUdkg}

For the remaining datasets, please contact the authors of the original papers in which they were collected.

\section*{Competing interests}

The authors declare that they have no competing interests.

\section*{Funding}

This work was partially supported by SoBigData.it. SoBigData.it receives funding from European Union – NextGenerationEU – National Recovery and Resilience Plan (Piano Nazionale di Ripresa e Resilienza, PNRR) – Project: “SoBigData.it – Strengthening the Italian RI for Social Mining and Big Data Analytics” – Prot. IR0000013 – Avviso n. 3264 del 28/12/2021. 

C. Boldrini was also supported by PNRR - M4C2 - Investimento 1.4, Centro Nazionale CN00000013 - "ICSC - National Centre for HPC, Big Data and Quantum Computing" - Spoke 6, funded by the European Commission under the NextGeneration EU programme.

A. Passarella and M. Conti were also supported by the PNRR - M4C2 - Investimento 1.3,
Partenariato Esteso PE00000013 - "FAIR", funded by the European Commission under the NextGeneration EU programme.

\section*{Author's contributions}
Designed the study: JT CB AP MC. Collected and processed the data: JT. Analyzed the data: JT CB AP MC. Wrote the paper: JT CB AP. All authors read and approved the final manuscript.

\section*{Acknowledgements}
Not applicable.

\bibliographystyle{bmc-mathphys}
\bibliography{biblio.bib}


\begin{thebibliography}{54}
\ifx \bisbn   \undefined \def \bisbn  #1{ISBN #1}\fi
\ifx \binits  \undefined \def \binits#1{#1}\fi
\ifx \bauthor  \undefined \def \bauthor#1{#1}\fi
\ifx \batitle  \undefined \def \batitle#1{#1}\fi
\ifx \bjtitle  \undefined \def \bjtitle#1{#1}\fi
\ifx \bvolume  \undefined \def \bvolume#1{\textbf{#1}}\fi
\ifx \byear  \undefined \def \byear#1{#1}\fi
\ifx \bissue  \undefined \def \bissue#1{#1}\fi
\ifx \bfpage  \undefined \def \bfpage#1{#1}\fi
\ifx \blpage  \undefined \def \blpage #1{#1}\fi
\ifx \burl  \undefined \def \burl#1{\textsf{#1}}\fi
\ifx \doiurl  \undefined \def \doiurl#1{\textsf{#1}}\fi
\ifx \betal  \undefined \def \betal{\textit{et al.}}\fi
\ifx \binstitute  \undefined \def \binstitute#1{#1}\fi
\ifx \binstitutionaled  \undefined \def \binstitutionaled#1{#1}\fi
\ifx \bctitle  \undefined \def \bctitle#1{#1}\fi
\ifx \beditor  \undefined \def \beditor#1{#1}\fi
\ifx \bpublisher  \undefined \def \bpublisher#1{#1}\fi
\ifx \bbtitle  \undefined \def \bbtitle#1{#1}\fi
\ifx \bedition  \undefined \def \bedition#1{#1}\fi
\ifx \bseriesno  \undefined \def \bseriesno#1{#1}\fi
\ifx \blocation  \undefined \def \blocation#1{#1}\fi
\ifx \bsertitle  \undefined \def \bsertitle#1{#1}\fi
\ifx \bsnm \undefined \def \bsnm#1{#1}\fi
\ifx \bsuffix \undefined \def \bsuffix#1{#1}\fi
\ifx \bparticle \undefined \def \bparticle#1{#1}\fi
\ifx \barticle \undefined \def \barticle#1{#1}\fi
\ifx \bconfdate \undefined \def \bconfdate #1{#1}\fi
\ifx \botherref \undefined \def \botherref #1{#1}\fi
\ifx \url \undefined \def \url#1{\textsf{#1}}\fi
\ifx \bchapter \undefined \def \bchapter#1{#1}\fi
\ifx \bbook \undefined \def \bbook#1{#1}\fi
\ifx \bcomment \undefined \def \bcomment#1{#1}\fi
\ifx \oauthor \undefined \def \oauthor#1{#1}\fi
\ifx \citeauthoryear \undefined \def \citeauthoryear#1{#1}\fi
\ifx \endbibitem  \undefined \def \endbibitem {}\fi
\ifx \bconflocation  \undefined \def \bconflocation#1{#1}\fi
\ifx \arxivurl  \undefined \def \arxivurl#1{\textsf{#1}}\fi
\csname PreBibitemsHook\endcsname

\bibitem{Dunbar_1995}
\begin{barticle}
\bauthor{\bsnm{Dunbar}, \binits{R.I.}},
\bauthor{\bsnm{Spoors}, \binits{M.}}:
\batitle{Social networks, support cliques, and kinship}.
\bjtitle{Human nature}
\bvolume{6}(\bissue{3}),
\bfpage{273}--\blpage{290}
(\byear{1995})
\end{barticle}
\endbibitem

\bibitem{Zhou_2005}
\begin{barticle}
\bauthor{\bsnm{Zhou}, \binits{W.-X.}},
\bauthor{\bsnm{Sornette}, \binits{D.}},
\bauthor{\bsnm{Hill}, \binits{R.A.}},
\bauthor{\bsnm{Dunbar}, \binits{R.I.}}:
\batitle{Discrete hierarchical organization of social group sizes}.
\bjtitle{Proceedings of the Royal Society B: Biological Sciences}
\bvolume{272}(\bissue{1561}),
\bfpage{439}--\blpage{444}
(\byear{2005})
\end{barticle}
\endbibitem

\bibitem{Hill_2003}
\begin{barticle}
\bauthor{\bsnm{Hill}, \binits{R.A.}},
\bauthor{\bsnm{Dunbar}, \binits{R.I.}}:
\batitle{Social network size in humans}.
\bjtitle{Human nature}
\bvolume{14}(\bissue{1}),
\bfpage{53}--\blpage{72}
(\byear{2003})
\end{barticle}
\endbibitem

\bibitem{Dunbar_1993}
\begin{barticle}
\bauthor{\bsnm{Dunbar}, \binits{R.I.}}:
\batitle{Coevolution of neocortical size, group size and language in humans}.
\bjtitle{Behavioral and brain sciences}
\bvolume{16}(\bissue{4}),
\bfpage{681}--\blpage{694}
(\byear{1993})
\end{barticle}
\endbibitem

\bibitem{Dunbar_1998}
\begin{barticle}
\bauthor{\bsnm{Dunbar}, \binits{R.I.M.}}:
\batitle{The social brain hypothesis}.
\bjtitle{Evolutionary Anthropology: Issues, News, and Reviews: Issues, News,
  and Reviews}
\bvolume{6}(\bissue{5}),
\bfpage{178}--\blpage{190}
(\byear{1998})
\end{barticle}
\endbibitem

\bibitem{Dunbar_1992}
\begin{barticle}
\bauthor{\bsnm{Dunbar}, \binits{R.I.}}:
\batitle{Neocortex size as a constraint on group size in primates}.
\bjtitle{Journal of human evolution}
\bvolume{22}(\bissue{6}),
\bfpage{469}--\blpage{493}
(\byear{1992})
\end{barticle}
\endbibitem

\bibitem{Dunbar_2015}
\begin{barticle}
\bauthor{\bsnm{Dunbar}, \binits{R.I.}},
\bauthor{\bsnm{Arnaboldi}, \binits{V.}},
\bauthor{\bsnm{Conti}, \binits{M.}},
\bauthor{\bsnm{Passarella}, \binits{A.}}:
\batitle{The structure of online social networks mirrors those in the offline
  world}.
\bjtitle{Social networks}
\bvolume{43},
\bfpage{39}--\blpage{47}
(\byear{2015})
\end{barticle}
\endbibitem

\bibitem{Sutcliffe_2012}
\begin{barticle}
\bauthor{\bsnm{Sutcliffe}, \binits{A.}},
\bauthor{\bsnm{Dunbar}, \binits{R.}},
\bauthor{\bsnm{Binder}, \binits{J.}},
\bauthor{\bsnm{Arrow}, \binits{H.}}:
\batitle{Relationships and the social brain: integrating psychological and
  evolutionary perspectives}.
\bjtitle{British journal of psychology}
\bvolume{103}(\bissue{2}),
\bfpage{149}--\blpage{168}
(\byear{2012})
\end{barticle}
\endbibitem

\bibitem{Gilbert_2009}
\begin{bchapter}
\bauthor{\bsnm{Gilbert}, \binits{E.}},
\bauthor{\bsnm{Karahalios}, \binits{K.}}:
\bctitle{Predicting tie strength with social media}.
In: \bbtitle{Proceedings of the CHI},
pp. \bfpage{211}--\blpage{220}
(\byear{2009})
\end{bchapter}
\endbibitem

\bibitem{Esmailian_2015}
\begin{barticle}
\bauthor{\bsnm{Esmailian}, \binits{P.}},
\bauthor{\bsnm{Jalili}, \binits{M.}}:
\batitle{Community detection in signed networks: the role of negative ties in
  different scales}.
\bjtitle{Scientific reports}
\bvolume{5}(\bissue{1}),
\bfpage{1}--\blpage{17}
(\byear{2015})
\end{barticle}
\endbibitem

\bibitem{Shi_2016}
\begin{barticle}
\bauthor{\bsnm{Shi}, \binits{G.}},
\bauthor{\bsnm{Proutiere}, \binits{A.}},
\bauthor{\bsnm{Johansson}, \binits{M.}},
\bauthor{\bsnm{Baras}, \binits{J.S.}},
\bauthor{\bsnm{Johansson}, \binits{K.H.}}:
\batitle{The evolution of beliefs over signed social networks}.
\bjtitle{Operations Research}
\bvolume{64}(\bissue{3}),
\bfpage{585}--\blpage{604}
(\byear{2016})
\end{barticle}
\endbibitem

\bibitem{Leskovec_2010_b_Predicting}
\begin{bchapter}
\bauthor{\bsnm{Leskovec}, \binits{J.}},
\bauthor{\bsnm{Huttenlocher}, \binits{D.}},
\bauthor{\bsnm{Kleinberg}, \binits{J.}}:
\bctitle{Predicting positive and negative links in online social networks}.
In: \bbtitle{Proceedings of WWW},
pp. \bfpage{641}--\blpage{650}
(\byear{2010})
\end{bchapter}
\endbibitem

\bibitem{Gottman_1995}
\begin{bbook}
\bauthor{\bsnm{Gottman}, \binits{J.}}:
\bbtitle{Why Marriages Succeed or Fail: And How You Can Make Yours Last}.
\bpublisher{Simon and Schuster},
\blocation{US}
(\byear{1995})
\end{bbook}
\endbibitem

\bibitem{Leskovec_2010_a_Signed}
\begin{bchapter}
\bauthor{\bsnm{Leskovec}, \binits{J.}},
\bauthor{\bsnm{Huttenlocher}, \binits{D.}},
\bauthor{\bsnm{Kleinberg}, \binits{J.}}:
\bctitle{Signed networks in social media}.
In: \bbtitle{Proceedings of the CHI},
pp. \bfpage{1361}--\blpage{1370}
(\byear{2010})
\end{bchapter}
\endbibitem

\bibitem{Tacchi_2022}
\begin{botherref}
\oauthor{\bsnm{Tacchi}, \binits{J.}},
\oauthor{\bsnm{Boldrini}, \binits{C.}},
\oauthor{\bsnm{Passarella}, \binits{A.}},
\oauthor{\bsnm{Conti}, \binits{M.}}:
Signed ego network model and its application to twitter.
IEEE BigData 2022
(2022)
\end{botherref}
\endbibitem

\bibitem{Tacchi_2023}
\begin{bchapter}
\bauthor{\bsnm{Tacchi}, \binits{J.}},
\bauthor{\bsnm{Boldrini}, \binits{C.}},
\bauthor{\bsnm{Passarella}, \binits{A.}},
\bauthor{\bsnm{Conti}, \binits{M.}}:
\bctitle{Cultural differences in signed ego networks on twitter: An
  investigatory analysis}.
In: \bbtitle{Companion Proceedings of the ACM Web Conference 2023},
pp. \bfpage{1039}--\blpage{1049}
(\byear{2023})
\end{bchapter}
\endbibitem

\bibitem{Granovetter_1973}
\begin{barticle}
\bauthor{\bsnm{Granovetter}, \binits{M.S.}}:
\batitle{The strength of weak ties}.
\bjtitle{American journal of sociology}
\bvolume{78}(\bissue{6}),
\bfpage{1360}--\blpage{1380}
(\byear{1973})
\end{barticle}
\endbibitem

\bibitem{Toprak_2021_b_Region-based}
\begin{botherref}
\oauthor{\bsnm{Toprak}, \binits{M.}},
\oauthor{\bsnm{Boldrini}, \binits{C.}},
\oauthor{\bsnm{Passarella}, \binits{A.}},
\oauthor{\bsnm{Conti}, \binits{M.}}:
Structural models of human social interactions in online smart communities: the
  case of region-based journalists on twitter.
Online Social Networks and Media
\textbf{30}
(2021)
\end{botherref}
\endbibitem

\bibitem{Tang_2016}
\begin{barticle}
\bauthor{\bsnm{Tang}, \binits{J.}},
\bauthor{\bsnm{Chang}, \binits{Y.}},
\bauthor{\bsnm{Aggarwal}, \binits{C.}},
\bauthor{\bsnm{Liu}, \binits{H.}}:
\batitle{A survey of signed network mining in social media}.
\bjtitle{ACM Computing Surveys (CSUR)}
\bvolume{49}(\bissue{3}),
\bfpage{1}--\blpage{37}
(\byear{2016})
\end{barticle}
\endbibitem

\bibitem{Maniu_2011}
\begin{bchapter}
\bauthor{\bsnm{Maniu}, \binits{S.}},
\bauthor{\bsnm{Abdessalem}, \binits{T.}},
\bauthor{\bsnm{Cautis}, \binits{B.}}:
\bctitle{Casting a web of trust over wikipedia: an interaction-based approach}.
In: \bbtitle{Comp. Proceedings of WWW},
pp. \bfpage{87}--\blpage{88}
(\byear{2011})
\end{bchapter}
\endbibitem

\bibitem{Traag_2009}
\begin{botherref}
\oauthor{\bsnm{Traag}, \binits{V.A.}},
\oauthor{\bsnm{Bruggeman}, \binits{J.}}:
{Community detection in networks with positive and negative links}.
Physical Review E
\textbf{80}(3)
(2009)
\end{botherref}
\endbibitem

\bibitem{Ferrara_2015}
\begin{barticle}
\bauthor{\bsnm{Ferrara}, \binits{E.}},
\bauthor{\bsnm{Yang}, \binits{Z.}}:
\batitle{Quantifying the effect of sentiment on information diffusion in social
  media}.
\bjtitle{PeerJ Computer Science}
\bvolume{1},
\bfpage{26}
(\byear{2015})
\end{barticle}
\endbibitem

\bibitem{Coleman_1988}
\begin{barticle}
\bauthor{\bsnm{Coleman}, \binits{J.S.}}:
\batitle{Social capital in the creation of human capital}.
\bjtitle{American journal of sociology}
\bvolume{94},
\bfpage{95}--\blpage{120}
(\byear{1988})
\end{barticle}
\endbibitem

\bibitem{Javari_2014}
\begin{barticle}
\bauthor{\bsnm{Javari}, \binits{A.}},
\bauthor{\bsnm{Jalili}, \binits{M.}}:
\batitle{Cluster-based collaborative filtering for sign prediction in social
  networks with positive and negative links}.
\bjtitle{ACM TIST}
\bvolume{5}(\bissue{2}),
\bfpage{1}--\blpage{19}
(\byear{2014})
\end{barticle}
\endbibitem

\bibitem{Ye_2013}
\begin{bchapter}
\bauthor{\bsnm{Ye}, \binits{J.}},
\bauthor{\bsnm{Cheng}, \binits{H.}},
\bauthor{\bsnm{Zhu}, \binits{Z.}},
\bauthor{\bsnm{Chen}, \binits{M.}}:
\bctitle{{Predicting positive and negative links in signed social networks by
  transfer learning}}.
In: \bbtitle{WWW 2013 - Proceedings of the 22nd International Conference on
  World Wide Web}
(\byear{2013}).
doi:\doiurl{10.1145/2488388.2488517}
\end{bchapter}
\endbibitem

\bibitem{Liu_2012}
\begin{barticle}
\bauthor{\bsnm{Liu}, \binits{B.}}:
\batitle{Sentiment analysis and opinion mining}.
\bjtitle{Synthesis lectures on human language technologies}
\bvolume{5}(\bissue{1}),
\bfpage{1}--\blpage{167}
(\byear{2012})
\end{barticle}
\endbibitem

\bibitem{Hassan_2012}
\begin{bchapter}
\bauthor{\bsnm{Hassan}, \binits{A.}},
\bauthor{\bsnm{Abu-Jbara}, \binits{A.}},
\bauthor{\bsnm{Radev}, \binits{D.}}:
\bctitle{Extracting signed social networks from text}.
In: \bbtitle{Workshop Proceedings of TextGraphs-7},
pp. \bfpage{6}--\blpage{14}
(\byear{2012})
\end{bchapter}
\endbibitem

\bibitem{Heider_1946}
\begin{barticle}
\bauthor{\bsnm{Heider}, \binits{F.}}:
\batitle{Attitudes and cognitive organization}.
\bjtitle{The Journal of psychology}
\bvolume{21}(\bissue{1}),
\bfpage{107}--\blpage{112}
(\byear{1946})
\end{barticle}
\endbibitem

\bibitem{Cartwright_1956}
\begin{barticle}
\bauthor{\bsnm{Cartwright}, \binits{D.}},
\bauthor{\bsnm{Harary}, \binits{F.}}:
\batitle{Structural balance: a generalization of heider's theory.}
\bjtitle{Psychological review}
\bvolume{63}(\bissue{5}),
\bfpage{277}
(\byear{1956})
\end{barticle}
\endbibitem

\bibitem{Davis_1967}
\begin{barticle}
\bauthor{\bsnm{Davis}, \binits{J.A.}}:
\batitle{Clustering and structural balance in graphs}.
\bjtitle{Human relations}
\bvolume{20}(\bissue{2}),
\bfpage{181}--\blpage{187}
(\byear{1967})
\end{barticle}
\endbibitem

\bibitem{Arnaboldi_2017}
\begin{barticle}
\bauthor{\bsnm{Arnaboldi}, \binits{V.}},
\bauthor{\bsnm{Conti}, \binits{M.}},
\bauthor{\bsnm{Passarella}, \binits{A.}},
\bauthor{\bsnm{Dunbar}, \binits{R.I.}}:
\batitle{Online social networks and information diffusion: The role of ego
  networks}.
\bjtitle{Online Soc. Netw. Media}
\bvolume{1},
\bfpage{44}--\blpage{55}
(\byear{2017})
\end{barticle}
\endbibitem

\bibitem{Toprak_2021_a_Harnessing}
\begin{botherref}
\oauthor{\bsnm{Toprak}, \binits{M.}},
\oauthor{\bsnm{Boldrini}, \binits{C.}},
\oauthor{\bsnm{Passarella}, \binits{A.}},
\oauthor{\bsnm{Conti}, \binits{M.}}:
Harnessing the power of ego network layers for link prediction in online social
  networks.
IEEE Trans. Comput. Soc. Sys.
(2022)
\end{botherref}
\endbibitem

\bibitem{Hart_1995}
\begin{bbook}
\bauthor{\bsnm{Hart}, \binits{B.}},
\bauthor{\bsnm{Risley}, \binits{T.R.}}:
\bbtitle{Meaningful Differences in the Everyday Experience of Young American
  Children.}
\bpublisher{Paul H Brookes Publishing},
\blocation{US}
(\byear{1995})
\end{bbook}
\endbibitem

\bibitem{Hutto_2014}
\begin{bchapter}
\bauthor{\bsnm{Hutto}, \binits{C.}},
\bauthor{\bsnm{Gilbert}, \binits{E.}}:
\bctitle{Vader: A parsimonious rule-based model for sentiment analysis of
  social media text}.
In: \bbtitle{Proceedings of ICWSM},
vol. \bseriesno{8},
pp. \bfpage{216}--\blpage{225}
(\byear{2014})
\end{bchapter}
\endbibitem

\bibitem{Nguyen_2020}
\begin{botherref}
\oauthor{\bsnm{Nguyen}, \binits{D.Q.}},
\oauthor{\bsnm{Vu}, \binits{T.}},
\oauthor{\bsnm{Nguyen}, \binits{A.T.}}:
Bertweet: A pre-trained language model for english tweets.
arXiv preprint arXiv:2005.10200
(2020)
\end{botherref}
\endbibitem

\bibitem{Devlin_2018}
\begin{botherref}
\oauthor{\bsnm{Devlin}, \binits{J.}},
\oauthor{\bsnm{Chang}, \binits{M.-W.}},
\oauthor{\bsnm{Lee}, \binits{K.}},
\oauthor{\bsnm{Toutanova}, \binits{K.}}:
Bert: Pre-training of deep bidirectional transformers for language
  understanding.
arXiv preprint arXiv:1810.04805
(2018)
\end{botherref}
\endbibitem

\bibitem{Rosenthal_2019}
\begin{botherref}
\oauthor{\bsnm{Rosenthal}, \binits{S.}},
\oauthor{\bsnm{Farra}, \binits{N.}},
\oauthor{\bsnm{Nakov}, \binits{P.}}:
Semeval-2017 task 4: Sentiment analysis in twitter.
arXiv preprint arXiv:1912.00741
(2019)
\end{botherref}
\endbibitem

\bibitem{Barbieri_2021}
\begin{botherref}
\oauthor{\bsnm{Barbieri}, \binits{F.}},
\oauthor{\bsnm{Anke}, \binits{L.E.}},
\oauthor{\bsnm{Camacho-Collados}, \binits{J.}}:
Xlm-t: A multilingual language model toolkit for twitter.
arXiv preprint arXiv:2104.12250
(2021)
\end{botherref}
\endbibitem

\bibitem{Conneau_2019}
\begin{botherref}
\oauthor{\bsnm{Conneau}, \binits{A.}},
\oauthor{\bsnm{Khandelwal}, \binits{K.}},
\oauthor{\bsnm{Goyal}, \binits{N.}},
\oauthor{\bsnm{Chaudhary}, \binits{V.}},
\oauthor{\bsnm{Wenzek}, \binits{G.}},
\oauthor{\bsnm{Guzm{\'a}n}, \binits{F.}},
\oauthor{\bsnm{Grave}, \binits{E.}},
\oauthor{\bsnm{Ott}, \binits{M.}},
\oauthor{\bsnm{Zettlemoyer}, \binits{L.}},
\oauthor{\bsnm{Stoyanov}, \binits{V.}}:
Unsupervised cross-lingual representation learning at scale.
arXiv preprint arXiv:1911.02116
(2019)
\end{botherref}
\endbibitem

\bibitem{Wenzek_2019}
\begin{botherref}
\oauthor{\bsnm{Wenzek}, \binits{G.}},
\oauthor{\bsnm{Lachaux}, \binits{M.-A.}},
\oauthor{\bsnm{Conneau}, \binits{A.}},
\oauthor{\bsnm{Chaudhary}, \binits{V.}},
\oauthor{\bsnm{Guzm{\'a}n}, \binits{F.}},
\oauthor{\bsnm{Joulin}, \binits{A.}},
\oauthor{\bsnm{Grave}, \binits{E.}}:
Ccnet: Extracting high quality monolingual datasets from web crawl data.
arXiv preprint arXiv:1911.00359
(2019)
\end{botherref}
\endbibitem

\bibitem{HuggingFace_2022}
\begin{botherref}
\oauthor{\bsnm{HuggingFace}}:
Sentiment Analysis Model.
\url{\url{https://huggingface.co/sbcBI/sentiment_analysis_model}}
Accessed 2023-03-03
\end{botherref}
\endbibitem

\bibitem{MacQueen_1967}
\begin{bchapter}
\bauthor{\bsnm{MacQueen}, \binits{J.}}:
\bctitle{Some methods for classification and analysis of multivariate
  observations}.
In: \bbtitle{Proceedings of the Fifth Berkeley Symposium on Mathematical
  Statistics and Probability},
vol. \bseriesno{14},
pp. \bfpage{281}--\blpage{297}
(\byear{1967}).
\bcomment{Oakland, CA, USA}
\end{bchapter}
\endbibitem

\bibitem{Ester_1996}
\begin{bchapter}
\bauthor{\bsnm{Ester}, \binits{M.}},
\bauthor{\bsnm{Kriegel}, \binits{H.-P.}},
\bauthor{\bsnm{Sander}, \binits{J.}},
\bauthor{\bsnm{Xu}, \binits{X.}}:
\bctitle{A density-based algorithm for discovering clusters in large spatial
  databases with noise}.
In: \bbtitle{Kdd},
vol. \bseriesno{96},
pp. \bfpage{226}--\blpage{231}
(\byear{1996})
\end{bchapter}
\endbibitem

\bibitem{Fukunaga_1975}
\begin{barticle}
\bauthor{\bsnm{Fukunaga}, \binits{K.}},
\bauthor{\bsnm{Hostetler}, \binits{L.}}:
\batitle{The estimation of the gradient of a density function, with
  applications in pattern recognition}.
\bjtitle{IEEE Trans. on Inf. Theory}
\bvolume{21}(\bissue{1}),
\bfpage{32}--\blpage{40}
(\byear{1975})
\end{barticle}
\endbibitem

\bibitem{Baumeister_2001}
\begin{barticle}
\bauthor{\bsnm{Baumeister}, \binits{R.F.}},
\bauthor{\bsnm{Bratslavsky}, \binits{E.}},
\bauthor{\bsnm{Finkenauer}, \binits{C.}},
\bauthor{\bsnm{Vohs}, \binits{K.D.}}:
\batitle{Bad is stronger than good}.
\bjtitle{Review of general psychology}
\bvolume{5}(\bissue{4}),
\bfpage{323}--\blpage{370}
(\byear{2001})
\end{barticle}
\endbibitem

\bibitem{Arnaboldi_2015}
\begin{bbook}
\bauthor{\bsnm{Arnaboldi}, \binits{V.}},
\bauthor{\bsnm{Passarella}, \binits{A.}},
\bauthor{\bsnm{Conti}, \binits{M.}},
\bauthor{\bsnm{Dunbar}, \binits{R.I.}}:
\bbtitle{Online Social Networks: Human Cognitive Constraints in Facebook and
  Twitter Personal Graphs}.
\bpublisher{Elsevier},
\blocation{The Netherlands}
(\byear{2015})
\end{bbook}
\endbibitem

\bibitem{Boldrini_2018}
\begin{bchapter}
\bauthor{\bsnm{Boldrini}, \binits{C.}},
\bauthor{\bsnm{Toprak}, \binits{M.}},
\bauthor{\bsnm{Conti}, \binits{M.}},
\bauthor{\bsnm{Passarella}, \binits{A.}}:
\bctitle{Twitter and the press: an ego-centred analysis}.
In: \bbtitle{Companion Proceedings of the The Web Conference 2018},
pp. \bfpage{1471}--\blpage{1478}
(\byear{2018})
\end{bchapter}
\endbibitem

\bibitem{Ollivier_2022}
\begin{botherref}
\oauthor{\bsnm{Ollivier}, \binits{K.}},
\oauthor{\bsnm{Boldrini}, \binits{C.}},
\oauthor{\bsnm{Passarella}, \binits{A.}},
\oauthor{\bsnm{Conti}, \binits{M.}}:
Structural invariants and semantic fingerprints in the" ego network" of words.
arXiv:2203.00588
(2022)
\end{botherref}
\endbibitem

\bibitem{UK_inbound_2020}
\begin{botherref}
{List of MP Twitter Accounts}.
\url{https://www.ukinbound.org/resources/list-of-mp-twitter-accounts/}.
Last accessed: 03 Mar 2022
\end{botherref}
\endbibitem

\bibitem{Arnaboldi_2013}
\begin{bchapter}
\bauthor{\bsnm{Arnaboldi}, \binits{V.}},
\bauthor{\bsnm{Conti}, \binits{M.}},
\bauthor{\bsnm{Passarella}, \binits{A.}},
\bauthor{\bsnm{Pezzoni}, \binits{F.}}:
\bctitle{Ego networks in twitter: an experimental analysis}.
In: \bbtitle{Proceedings IEEE INFOCOM},
pp. \bfpage{3459}--\blpage{3464}
(\byear{2013})
\end{bchapter}
\endbibitem

\bibitem{Cortes_1995}
\begin{barticle}
\bauthor{\bsnm{Cortes}, \binits{C.}},
\bauthor{\bsnm{Vapnik}, \binits{V.}}:
\batitle{Support-vector networks}.
\bjtitle{Machine learning}
\bvolume{20},
\bfpage{273}--\blpage{297}
(\byear{1995})
\end{barticle}
\endbibitem

\bibitem{Rozin_2001}
\begin{barticle}
\bauthor{\bsnm{Rozin}, \binits{P.}},
\bauthor{\bsnm{Royzman}, \binits{E.B.}}:
\batitle{Negativity bias, negativity dominance, and contagion}.
\bjtitle{Personality and social psychology review}
\bvolume{5}(\bissue{4}),
\bfpage{296}--\blpage{320}
(\byear{2001})
\end{barticle}
\endbibitem

\bibitem{Toprak_2022}
\begin{barticle}
\bauthor{\bsnm{Toprak}, \binits{M.}},
\bauthor{\bsnm{Boldrini}, \binits{C.}},
\bauthor{\bsnm{Passarella}, \binits{A.}},
\bauthor{\bsnm{Conti}, \binits{M.}}:
\batitle{Journalists’ ego networks in twitter: Invariant and distinctive
  structural features}.
\bjtitle{Online Social Networks and Media}
\bvolume{30},
\bfpage{100207}
(\byear{2022})
\end{barticle}
\endbibitem

\bibitem{Ostrom_2003}
\begin{botherref}
\oauthor{\bsnm{Ostrom}, \binits{E.}}:
Toward a behavioral theory linking trust, reciprocity, and reputation.
Trust and reciprocity: Interdisciplinary lessons from experimental research,
19--79
(2003)
\end{botherref}
\endbibitem

\end{thebibliography}

\newcommand{\BMCxmlcomment}[1]{}

\BMCxmlcomment{

<refgrp>

<bibl id="B1">
  <title><p>Social networks, support cliques, and kinship</p></title>
  <aug>
    <au><snm>Dunbar</snm><fnm>RI</fnm></au>
    <au><snm>Spoors</snm><fnm>M</fnm></au>
  </aug>
  <source>Human nature</source>
  <publisher>Springer</publisher>
  <pubdate>1995</pubdate>
  <volume>6</volume>
  <issue>3</issue>
  <fpage>273</fpage>
  <lpage>-290</lpage>
</bibl>

<bibl id="B2">
  <title><p>Discrete hierarchical organization of social group
  sizes</p></title>
  <aug>
    <au><snm>Zhou</snm><fnm>W X</fnm></au>
    <au><snm>Sornette</snm><fnm>D</fnm></au>
    <au><snm>Hill</snm><fnm>RA</fnm></au>
    <au><snm>Dunbar</snm><fnm>RI</fnm></au>
  </aug>
  <source>Proceedings of the Royal Society B: Biological Sciences</source>
  <publisher>The Royal Society</publisher>
  <pubdate>2005</pubdate>
  <volume>272</volume>
  <issue>1561</issue>
  <fpage>439</fpage>
  <lpage>-444</lpage>
</bibl>

<bibl id="B3">
  <title><p>Social network size in humans</p></title>
  <aug>
    <au><snm>Hill</snm><fnm>RA</fnm></au>
    <au><snm>Dunbar</snm><fnm>RI</fnm></au>
  </aug>
  <source>Human nature</source>
  <publisher>Springer</publisher>
  <pubdate>2003</pubdate>
  <volume>14</volume>
  <issue>1</issue>
  <fpage>53</fpage>
  <lpage>-72</lpage>
</bibl>

<bibl id="B4">
  <title><p>Coevolution of neocortical size, group size and language in
  humans</p></title>
  <aug>
    <au><snm>Dunbar</snm><fnm>RI</fnm></au>
  </aug>
  <source>Behavioral and brain sciences</source>
  <publisher>Cambridge University Press</publisher>
  <pubdate>1993</pubdate>
  <volume>16</volume>
  <issue>4</issue>
  <fpage>681</fpage>
  <lpage>-694</lpage>
</bibl>

<bibl id="B5">
  <title><p>The social brain hypothesis</p></title>
  <aug>
    <au><snm>Dunbar</snm><fnm>RIM</fnm></au>
  </aug>
  <source>Evolutionary Anthropology: Issues, News, and Reviews: Issues, News,
  and Reviews</source>
  <publisher>Wiley Online Library</publisher>
  <pubdate>1998</pubdate>
  <volume>6</volume>
  <issue>5</issue>
  <fpage>178</fpage>
  <lpage>-190</lpage>
</bibl>

<bibl id="B6">
  <title><p>Neocortex size as a constraint on group size in
  primates</p></title>
  <aug>
    <au><snm>Dunbar</snm><fnm>RI</fnm></au>
  </aug>
  <source>Journal of human evolution</source>
  <publisher>Elsevier</publisher>
  <pubdate>1992</pubdate>
  <volume>22</volume>
  <issue>6</issue>
  <fpage>469</fpage>
  <lpage>-493</lpage>
</bibl>

<bibl id="B7">
  <title><p>The structure of online social networks mirrors those in the
  offline world</p></title>
  <aug>
    <au><snm>Dunbar</snm><fnm>RI</fnm></au>
    <au><snm>Arnaboldi</snm><fnm>V</fnm></au>
    <au><snm>Conti</snm><fnm>M</fnm></au>
    <au><snm>Passarella</snm><fnm>A</fnm></au>
  </aug>
  <source>Social networks</source>
  <publisher>Elsevier</publisher>
  <pubdate>2015</pubdate>
  <volume>43</volume>
  <fpage>39</fpage>
  <lpage>-47</lpage>
</bibl>

<bibl id="B8">
  <title><p>Relationships and the social brain: integrating psychological and
  evolutionary perspectives</p></title>
  <aug>
    <au><snm>Sutcliffe</snm><fnm>A</fnm></au>
    <au><snm>Dunbar</snm><fnm>R</fnm></au>
    <au><snm>Binder</snm><fnm>J</fnm></au>
    <au><snm>Arrow</snm><fnm>H</fnm></au>
  </aug>
  <source>British journal of psychology</source>
  <publisher>Wiley Online Library</publisher>
  <pubdate>2012</pubdate>
  <volume>103</volume>
  <issue>2</issue>
  <fpage>149</fpage>
  <lpage>-168</lpage>
</bibl>

<bibl id="B9">
  <title><p>Predicting tie strength with social media</p></title>
  <aug>
    <au><snm>Gilbert</snm><fnm>E</fnm></au>
    <au><snm>Karahalios</snm><fnm>K</fnm></au>
  </aug>
  <source>Proceedings of the CHI</source>
  <pubdate>2009</pubdate>
  <fpage>211</fpage>
  <lpage>-220</lpage>
</bibl>

<bibl id="B10">
  <title><p>Community detection in signed networks: the role of negative ties
  in different scales</p></title>
  <aug>
    <au><snm>Esmailian</snm><fnm>P</fnm></au>
    <au><snm>Jalili</snm><fnm>M</fnm></au>
  </aug>
  <source>Scientific reports</source>
  <publisher>Nature Publishing Group</publisher>
  <pubdate>2015</pubdate>
  <volume>5</volume>
  <issue>1</issue>
  <fpage>1</fpage>
  <lpage>-17</lpage>
</bibl>

<bibl id="B11">
  <title><p>The evolution of beliefs over signed social networks</p></title>
  <aug>
    <au><snm>Shi</snm><fnm>G</fnm></au>
    <au><snm>Proutiere</snm><fnm>A</fnm></au>
    <au><snm>Johansson</snm><fnm>M</fnm></au>
    <au><snm>Baras</snm><fnm>JS</fnm></au>
    <au><snm>Johansson</snm><fnm>KH</fnm></au>
  </aug>
  <source>Operations Research</source>
  <publisher>INFORMS</publisher>
  <pubdate>2016</pubdate>
  <volume>64</volume>
  <issue>3</issue>
  <fpage>585</fpage>
  <lpage>-604</lpage>
</bibl>

<bibl id="B12">
  <title><p>Predicting positive and negative links in online social
  networks</p></title>
  <aug>
    <au><snm>Leskovec</snm><fnm>J</fnm></au>
    <au><snm>Huttenlocher</snm><fnm>D</fnm></au>
    <au><snm>Kleinberg</snm><fnm>J</fnm></au>
  </aug>
  <source>Proceedings of WWW</source>
  <pubdate>2010</pubdate>
  <fpage>641</fpage>
  <lpage>-650</lpage>
</bibl>

<bibl id="B13">
  <title><p>Why marriages succeed or fail: And how you can make yours
  last</p></title>
  <aug>
    <au><snm>Gottman</snm><fnm>J</fnm></au>
  </aug>
  <publisher>US: Simon and Schuster</publisher>
  <pubdate>1995</pubdate>
</bibl>

<bibl id="B14">
  <title><p>Signed networks in social media</p></title>
  <aug>
    <au><snm>Leskovec</snm><fnm>J</fnm></au>
    <au><snm>Huttenlocher</snm><fnm>D</fnm></au>
    <au><snm>Kleinberg</snm><fnm>J</fnm></au>
  </aug>
  <source>Proceedings of the CHI</source>
  <pubdate>2010</pubdate>
  <fpage>1361</fpage>
  <lpage>-1370</lpage>
</bibl>

<bibl id="B15">
  <title><p>Signed ego network model and its application to Twitter</p></title>
  <aug>
    <au><snm>Tacchi</snm><fnm>J</fnm></au>
    <au><snm>Boldrini</snm><fnm>C</fnm></au>
    <au><snm>Passarella</snm><fnm>A</fnm></au>
    <au><snm>Conti</snm><fnm>M</fnm></au>
  </aug>
  <source>IEEE BigData 2022</source>
  <pubdate>2022</pubdate>
</bibl>

<bibl id="B16">
  <title><p>Cultural Differences in Signed Ego Networks on Twitter: An
  Investigatory Analysis</p></title>
  <aug>
    <au><snm>Tacchi</snm><fnm>J</fnm></au>
    <au><snm>Boldrini</snm><fnm>C</fnm></au>
    <au><snm>Passarella</snm><fnm>A</fnm></au>
    <au><snm>Conti</snm><fnm>M</fnm></au>
  </aug>
  <source>Companion Proceedings of the ACM Web Conference 2023</source>
  <pubdate>2023</pubdate>
  <fpage>1039</fpage>
  <lpage>-1049</lpage>
</bibl>

<bibl id="B17">
  <title><p>The strength of weak ties</p></title>
  <aug>
    <au><snm>Granovetter</snm><fnm>MS</fnm></au>
  </aug>
  <source>American journal of sociology</source>
  <publisher>University of Chicago Press</publisher>
  <pubdate>1973</pubdate>
  <volume>78</volume>
  <issue>6</issue>
  <fpage>1360</fpage>
  <lpage>-1380</lpage>
</bibl>

<bibl id="B18">
  <title><p>Structural Models of Human Social Interactions in Online Smart
  Communities: the Case of Region-based Journalists on Twitter</p></title>
  <aug>
    <au><snm>Toprak</snm><fnm>M</fnm></au>
    <au><snm>Boldrini</snm><fnm>C</fnm></au>
    <au><snm>Passarella</snm><fnm>A</fnm></au>
    <au><snm>Conti</snm><fnm>M</fnm></au>
  </aug>
  <source>Online Social Networks and Media</source>
  <pubdate>2021</pubdate>
  <volume>30</volume>
</bibl>

<bibl id="B19">
  <title><p>A survey of signed network mining in social media</p></title>
  <aug>
    <au><snm>Tang</snm><fnm>J</fnm></au>
    <au><snm>Chang</snm><fnm>Y</fnm></au>
    <au><snm>Aggarwal</snm><fnm>C</fnm></au>
    <au><snm>Liu</snm><fnm>H</fnm></au>
  </aug>
  <source>ACM Computing Surveys (CSUR)</source>
  <publisher>ACM New York, NY, USA</publisher>
  <pubdate>2016</pubdate>
  <volume>49</volume>
  <issue>3</issue>
  <fpage>1</fpage>
  <lpage>-37</lpage>
</bibl>

<bibl id="B20">
  <title><p>Casting a web of trust over wikipedia: an interaction-based
  approach</p></title>
  <aug>
    <au><snm>Maniu</snm><fnm>S</fnm></au>
    <au><snm>Abdessalem</snm><fnm>T</fnm></au>
    <au><snm>Cautis</snm><fnm>B</fnm></au>
  </aug>
  <source>Comp. proceedings of WWW</source>
  <pubdate>2011</pubdate>
  <fpage>87</fpage>
  <lpage>-88</lpage>
</bibl>

<bibl id="B21">
  <title><p>{Community detection in networks with positive and negative
  links}</p></title>
  <aug>
    <au><snm>Traag</snm><fnm>V. A.</fnm></au>
    <au><snm>Bruggeman</snm><fnm>J</fnm></au>
  </aug>
  <source>Physical Review E</source>
  <pubdate>2009</pubdate>
  <volume>80</volume>
  <issue>3</issue>
</bibl>

<bibl id="B22">
  <title><p>Quantifying the effect of sentiment on information diffusion in
  social media</p></title>
  <aug>
    <au><snm>Ferrara</snm><fnm>E</fnm></au>
    <au><snm>Yang</snm><fnm>Z</fnm></au>
  </aug>
  <source>PeerJ Computer Science</source>
  <publisher>PeerJ Inc.</publisher>
  <pubdate>2015</pubdate>
  <volume>1</volume>
  <fpage>e26</fpage>
</bibl>

<bibl id="B23">
  <title><p>Social capital in the creation of human capital</p></title>
  <aug>
    <au><snm>Coleman</snm><fnm>JS</fnm></au>
  </aug>
  <source>American journal of sociology</source>
  <publisher>University of Chicago Press</publisher>
  <pubdate>1988</pubdate>
  <volume>94</volume>
  <fpage>S95</fpage>
  <lpage>-S120</lpage>
</bibl>

<bibl id="B24">
  <title><p>Cluster-based collaborative filtering for sign prediction in social
  networks with positive and negative links</p></title>
  <aug>
    <au><snm>Javari</snm><fnm>A</fnm></au>
    <au><snm>Jalili</snm><fnm>M</fnm></au>
  </aug>
  <source>ACM TIST</source>
  <publisher>ACM New York, NY, USA</publisher>
  <pubdate>2014</pubdate>
  <volume>5</volume>
  <issue>2</issue>
  <fpage>1</fpage>
  <lpage>-19</lpage>
</bibl>

<bibl id="B25">
  <title><p>{Predicting positive and negative links in signed social networks
  by transfer learning}</p></title>
  <aug>
    <au><snm>Ye</snm><fnm>J</fnm></au>
    <au><snm>Cheng</snm><fnm>H</fnm></au>
    <au><snm>Zhu</snm><fnm>Z</fnm></au>
    <au><snm>Chen</snm><fnm>M</fnm></au>
  </aug>
  <source>WWW 2013 - Proceedings of the 22nd International Conference on World
  Wide Web</source>
  <pubdate>2013</pubdate>
</bibl>

<bibl id="B26">
  <title><p>Sentiment analysis and opinion mining</p></title>
  <aug>
    <au><snm>Liu</snm><fnm>B</fnm></au>
  </aug>
  <source>Synthesis lectures on human language technologies</source>
  <publisher>Morgan \& Claypool Publishers</publisher>
  <pubdate>2012</pubdate>
  <volume>5</volume>
  <issue>1</issue>
  <fpage>1</fpage>
  <lpage>-167</lpage>
</bibl>

<bibl id="B27">
  <title><p>Extracting signed social networks from text</p></title>
  <aug>
    <au><snm>Hassan</snm><fnm>A</fnm></au>
    <au><snm>Abu Jbara</snm><fnm>A</fnm></au>
    <au><snm>Radev</snm><fnm>D</fnm></au>
  </aug>
  <source>Workshop Proceedings of TextGraphs-7</source>
  <pubdate>2012</pubdate>
  <fpage>6</fpage>
  <lpage>-14</lpage>
</bibl>

<bibl id="B28">
  <title><p>Attitudes and cognitive organization</p></title>
  <aug>
    <au><snm>Heider</snm><fnm>F</fnm></au>
  </aug>
  <source>The Journal of psychology</source>
  <publisher>Taylor \& Francis</publisher>
  <pubdate>1946</pubdate>
  <volume>21</volume>
  <issue>1</issue>
  <fpage>107</fpage>
  <lpage>-112</lpage>
</bibl>

<bibl id="B29">
  <title><p>Structural balance: a generalization of Heider's
  theory.</p></title>
  <aug>
    <au><snm>Cartwright</snm><fnm>D</fnm></au>
    <au><snm>Harary</snm><fnm>F</fnm></au>
  </aug>
  <source>Psychological review</source>
  <publisher>American Psychological Association</publisher>
  <pubdate>1956</pubdate>
  <volume>63</volume>
  <issue>5</issue>
  <fpage>277</fpage>
</bibl>

<bibl id="B30">
  <title><p>Clustering and structural balance in graphs</p></title>
  <aug>
    <au><snm>Davis</snm><fnm>JA</fnm></au>
  </aug>
  <source>Human relations</source>
  <publisher>Sage Publications Sage CA: Thousand Oaks, CA</publisher>
  <pubdate>1967</pubdate>
  <volume>20</volume>
  <issue>2</issue>
  <fpage>181</fpage>
  <lpage>-187</lpage>
</bibl>

<bibl id="B31">
  <title><p>Online social networks and information diffusion: The role of ego
  networks</p></title>
  <aug>
    <au><snm>Arnaboldi</snm><fnm>V</fnm></au>
    <au><snm>Conti</snm><fnm>M</fnm></au>
    <au><snm>Passarella</snm><fnm>A</fnm></au>
    <au><snm>Dunbar</snm><fnm>RI</fnm></au>
  </aug>
  <source>Online Soc. Netw. Media</source>
  <publisher>Elsevier</publisher>
  <pubdate>2017</pubdate>
  <volume>1</volume>
  <fpage>44</fpage>
  <lpage>-55</lpage>
</bibl>

<bibl id="B32">
  <title><p>Harnessing the Power of Ego Network Layers for Link Prediction in
  Online Social Networks</p></title>
  <aug>
    <au><snm>Toprak</snm><fnm>M</fnm></au>
    <au><snm>Boldrini</snm><fnm>C</fnm></au>
    <au><snm>Passarella</snm><fnm>A</fnm></au>
    <au><snm>Conti</snm><fnm>M</fnm></au>
  </aug>
  <source>IEEE Trans. Comput. Soc. Sys.</source>
  <pubdate>2022</pubdate>
</bibl>

<bibl id="B33">
  <title><p>Meaningful differences in the everyday experience of young American
  children.</p></title>
  <aug>
    <au><snm>Hart</snm><fnm>B</fnm></au>
    <au><snm>Risley</snm><fnm>TR</fnm></au>
  </aug>
  <publisher>US: Paul H Brookes Publishing</publisher>
  <pubdate>1995</pubdate>
</bibl>

<bibl id="B34">
  <title><p>Vader: A parsimonious rule-based model for sentiment analysis of
  social media text</p></title>
  <aug>
    <au><snm>Hutto</snm><fnm>C</fnm></au>
    <au><snm>Gilbert</snm><fnm>E</fnm></au>
  </aug>
  <source>Proceedings of ICWSM</source>
  <pubdate>2014</pubdate>
  <volume>8</volume>
  <fpage>216</fpage>
  <lpage>-225</lpage>
</bibl>

<bibl id="B35">
  <title><p>BERTweet: A pre-trained language model for English
  Tweets</p></title>
  <aug>
    <au><snm>Nguyen</snm><fnm>DQ</fnm></au>
    <au><snm>Vu</snm><fnm>T</fnm></au>
    <au><snm>Nguyen</snm><fnm>AT</fnm></au>
  </aug>
  <source>arXiv preprint arXiv:2005.10200</source>
  <pubdate>2020</pubdate>
</bibl>

<bibl id="B36">
  <title><p>Bert: Pre-training of deep bidirectional transformers for language
  understanding</p></title>
  <aug>
    <au><snm>Devlin</snm><fnm>J</fnm></au>
    <au><snm>Chang</snm><fnm>MW</fnm></au>
    <au><snm>Lee</snm><fnm>K</fnm></au>
    <au><snm>Toutanova</snm><fnm>K</fnm></au>
  </aug>
  <source>arXiv preprint arXiv:1810.04805</source>
  <pubdate>2018</pubdate>
</bibl>

<bibl id="B37">
  <title><p>SemEval-2017 task 4: Sentiment analysis in Twitter</p></title>
  <aug>
    <au><snm>Rosenthal</snm><fnm>S</fnm></au>
    <au><snm>Farra</snm><fnm>N</fnm></au>
    <au><snm>Nakov</snm><fnm>P</fnm></au>
  </aug>
  <source>arXiv preprint arXiv:1912.00741</source>
  <pubdate>2019</pubdate>
</bibl>

<bibl id="B38">
  <title><p>Xlm-t: A multilingual language model toolkit for
  twitter</p></title>
  <aug>
    <au><snm>Barbieri</snm><fnm>F</fnm></au>
    <au><snm>Anke</snm><fnm>LE</fnm></au>
    <au><snm>Camacho Collados</snm><fnm>J</fnm></au>
  </aug>
  <source>arXiv preprint arXiv:2104.12250</source>
  <pubdate>2021</pubdate>
</bibl>

<bibl id="B39">
  <title><p>Unsupervised cross-lingual representation learning at
  scale</p></title>
  <aug>
    <au><snm>Conneau</snm><fnm>A</fnm></au>
    <au><snm>Khandelwal</snm><fnm>K</fnm></au>
    <au><snm>Goyal</snm><fnm>N</fnm></au>
    <au><snm>Chaudhary</snm><fnm>V</fnm></au>
    <au><snm>Wenzek</snm><fnm>G</fnm></au>
    <au><snm>Guzm{\'a}n</snm><fnm>F</fnm></au>
    <au><snm>Grave</snm><fnm>E</fnm></au>
    <au><snm>Ott</snm><fnm>M</fnm></au>
    <au><snm>Zettlemoyer</snm><fnm>L</fnm></au>
    <au><snm>Stoyanov</snm><fnm>V</fnm></au>
  </aug>
  <source>arXiv preprint arXiv:1911.02116</source>
  <pubdate>2019</pubdate>
</bibl>

<bibl id="B40">
  <title><p>CCNet: Extracting high quality monolingual datasets from web crawl
  data</p></title>
  <aug>
    <au><snm>Wenzek</snm><fnm>G</fnm></au>
    <au><snm>Lachaux</snm><fnm>MA</fnm></au>
    <au><snm>Conneau</snm><fnm>A</fnm></au>
    <au><snm>Chaudhary</snm><fnm>V</fnm></au>
    <au><snm>Guzm{\'a}n</snm><fnm>F</fnm></au>
    <au><snm>Joulin</snm><fnm>A</fnm></au>
    <au><snm>Grave</snm><fnm>E</fnm></au>
  </aug>
  <source>arXiv preprint arXiv:1911.00359</source>
  <pubdate>2019</pubdate>
</bibl>

<bibl id="B41">
  <title><p>Sentiment Analysis Model</p></title>
  <aug>
    <au><cnm>HuggingFace</cnm></au>
  </aug>
  <pubdate>2022</pubdate>
  <url>\url{https://huggingface.co/sbcBI/sentiment_analysis_model}</url>
</bibl>

<bibl id="B42">
  <title><p>Some methods for classification and analysis of multivariate
  observations</p></title>
  <aug>
    <au><snm>MacQueen</snm><fnm>J</fnm></au>
  </aug>
  <source>Proceedings of the fifth Berkeley symposium on mathematical
  statistics and probability</source>
  <pubdate>1967</pubdate>
  <volume>14</volume>
  <fpage>281</fpage>
  <lpage>-297</lpage>
</bibl>

<bibl id="B43">
  <title><p>A density-based algorithm for discovering clusters in large spatial
  databases with noise</p></title>
  <aug>
    <au><snm>Ester</snm><fnm>M</fnm></au>
    <au><snm>Kriegel</snm><fnm>HP</fnm></au>
    <au><snm>Sander</snm><fnm>J</fnm></au>
    <au><snm>Xu</snm><fnm>X</fnm></au>
  </aug>
  <source>kdd</source>
  <pubdate>1996</pubdate>
  <volume>96</volume>
  <fpage>226</fpage>
  <lpage>-231</lpage>
</bibl>

<bibl id="B44">
  <title><p>The estimation of the gradient of a density function, with
  applications in pattern recognition</p></title>
  <aug>
    <au><snm>Fukunaga</snm><fnm>K</fnm></au>
    <au><snm>Hostetler</snm><fnm>L</fnm></au>
  </aug>
  <source>IEEE Trans. on Inf. Theory</source>
  <publisher>IEEE</publisher>
  <pubdate>1975</pubdate>
  <volume>21</volume>
  <issue>1</issue>
  <fpage>32</fpage>
  <lpage>-40</lpage>
</bibl>

<bibl id="B45">
  <title><p>Bad is stronger than good</p></title>
  <aug>
    <au><snm>Baumeister</snm><fnm>RF</fnm></au>
    <au><snm>Bratslavsky</snm><fnm>E</fnm></au>
    <au><snm>Finkenauer</snm><fnm>C</fnm></au>
    <au><snm>Vohs</snm><fnm>KD</fnm></au>
  </aug>
  <source>Review of general psychology</source>
  <publisher>SAGE Publications Sage CA: Los Angeles, CA</publisher>
  <pubdate>2001</pubdate>
  <volume>5</volume>
  <issue>4</issue>
  <fpage>323</fpage>
  <lpage>-370</lpage>
</bibl>

<bibl id="B46">
  <title><p>Online social networks: human cognitive constraints in Facebook and
  Twitter personal graphs</p></title>
  <aug>
    <au><snm>Arnaboldi</snm><fnm>V</fnm></au>
    <au><snm>Passarella</snm><fnm>A</fnm></au>
    <au><snm>Conti</snm><fnm>M</fnm></au>
    <au><snm>Dunbar</snm><fnm>RI</fnm></au>
  </aug>
  <publisher>The Netherlands: Elsevier</publisher>
  <pubdate>2015</pubdate>
</bibl>

<bibl id="B47">
  <title><p>Twitter and the press: an ego-centred analysis</p></title>
  <aug>
    <au><snm>Boldrini</snm><fnm>C</fnm></au>
    <au><snm>Toprak</snm><fnm>M</fnm></au>
    <au><snm>Conti</snm><fnm>M</fnm></au>
    <au><snm>Passarella</snm><fnm>A</fnm></au>
  </aug>
  <source>Companion Proceedings of the The Web Conference 2018</source>
  <pubdate>2018</pubdate>
  <fpage>1471</fpage>
  <lpage>-1478</lpage>
</bibl>

<bibl id="B48">
  <title><p>Structural invariants and semantic fingerprints in the" ego
  network" of words</p></title>
  <aug>
    <au><snm>Ollivier</snm><fnm>K</fnm></au>
    <au><snm>Boldrini</snm><fnm>C</fnm></au>
    <au><snm>Passarella</snm><fnm>A</fnm></au>
    <au><snm>Conti</snm><fnm>M</fnm></au>
  </aug>
  <source>arXiv:2203.00588</source>
  <pubdate>2022</pubdate>
</bibl>

<bibl id="B49">
  <title><p>{List of MP Twitter Accounts}</p></title>
  <source>\url{https://www.ukinbound.org/resources/list-of-mp-twitter-accounts/}</source>
  <note>Last accessed: 03 Mar 2022</note>
</bibl>

<bibl id="B50">
  <title><p>Ego networks in twitter: an experimental analysis</p></title>
  <aug>
    <au><snm>Arnaboldi</snm><fnm>V</fnm></au>
    <au><snm>Conti</snm><fnm>M</fnm></au>
    <au><snm>Passarella</snm><fnm>A</fnm></au>
    <au><snm>Pezzoni</snm><fnm>F</fnm></au>
  </aug>
  <source>Proceedings IEEE INFOCOM</source>
  <pubdate>2013</pubdate>
  <fpage>3459</fpage>
  <lpage>-3464</lpage>
</bibl>

<bibl id="B51">
  <title><p>Support-vector networks</p></title>
  <aug>
    <au><snm>Cortes</snm><fnm>C</fnm></au>
    <au><snm>Vapnik</snm><fnm>V</fnm></au>
  </aug>
  <source>Machine learning</source>
  <publisher>Springer</publisher>
  <pubdate>1995</pubdate>
  <volume>20</volume>
  <fpage>273</fpage>
  <lpage>-297</lpage>
</bibl>

<bibl id="B52">
  <title><p>Negativity bias, negativity dominance, and contagion</p></title>
  <aug>
    <au><snm>Rozin</snm><fnm>P</fnm></au>
    <au><snm>Royzman</snm><fnm>EB</fnm></au>
  </aug>
  <source>Personality and social psychology review</source>
  <publisher>Sage Publications Sage CA: Los Angeles, CA</publisher>
  <pubdate>2001</pubdate>
  <volume>5</volume>
  <issue>4</issue>
  <fpage>296</fpage>
  <lpage>-320</lpage>
</bibl>

<bibl id="B53">
  <title><p>Journalists’ ego networks in Twitter: Invariant and distinctive
  structural features</p></title>
  <aug>
    <au><snm>Toprak</snm><fnm>M</fnm></au>
    <au><snm>Boldrini</snm><fnm>C</fnm></au>
    <au><snm>Passarella</snm><fnm>A</fnm></au>
    <au><snm>Conti</snm><fnm>M</fnm></au>
  </aug>
  <source>Online Social Networks and Media</source>
  <publisher>Elsevier</publisher>
  <pubdate>2022</pubdate>
  <volume>30</volume>
  <fpage>100207</fpage>
</bibl>

<bibl id="B54">
  <title><p>Toward a behavioral theory linking trust, reciprocity, and
  reputation.</p></title>
  <aug>
    <au><snm>Ostrom</snm><fnm>E</fnm></au>
  </aug>
  <source>Trust and reciprocity: Interdisciplinary lessons from experimental
  research</source>
  <publisher>Russell Sage Foundation</publisher>
  <pubdate>2003</pubdate>
  <fpage>19</fpage>
  <lpage>-79</lpage>
</bibl>

</refgrp>
} 
\end{backmatter}

\appendix

\section{Sentiment Model Disagreements}
\label{appendix:model_disagreement_plots}

\subsection{VADER}

\begin{figure*}
    \centering
    \includegraphics[height=0.7\textheight,width=0.95\textwidth]{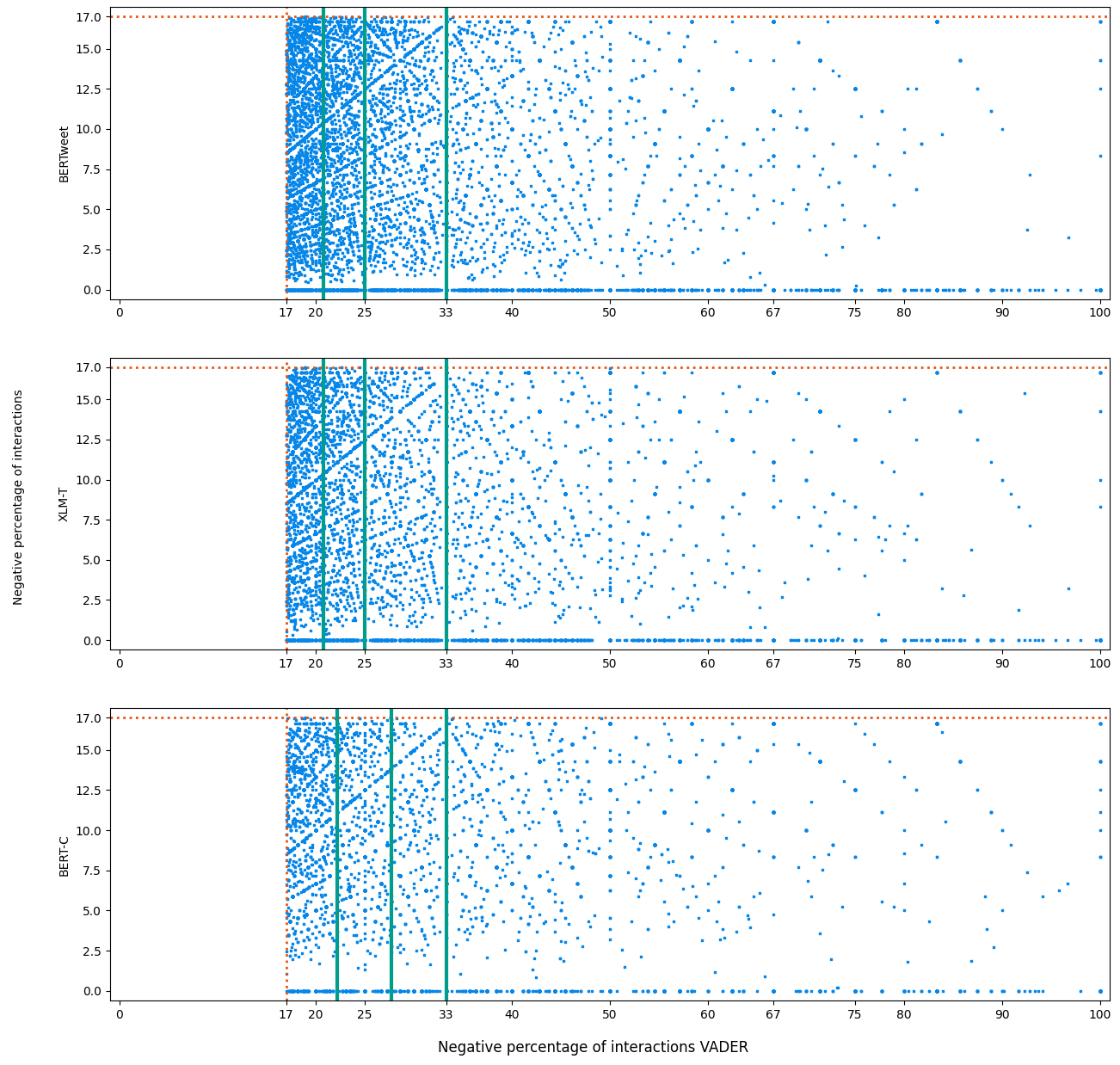}
    \caption{VADER disagreements. Each blue point represents a disagreement about the sign of a relationship, with VADER determining a negative sign and the other models determining a positive sign. The other models are, from top to bottom, BERTweet, XLM-T and BERT-C.}
    \label{fig:disagreements_plot_VADER}
\end{figure*}

A graph displaying relationship disagreements where VADER predicts a negative label for a relationship and one of the other models predicts a positive label can be seen in Figure~\ref{fig:disagreements_plot_VADER}.

\subsection{BERTweet}

\begin{figure*}
    \centering
    \includegraphics[height=0.7\textheight,width=0.95\textwidth]{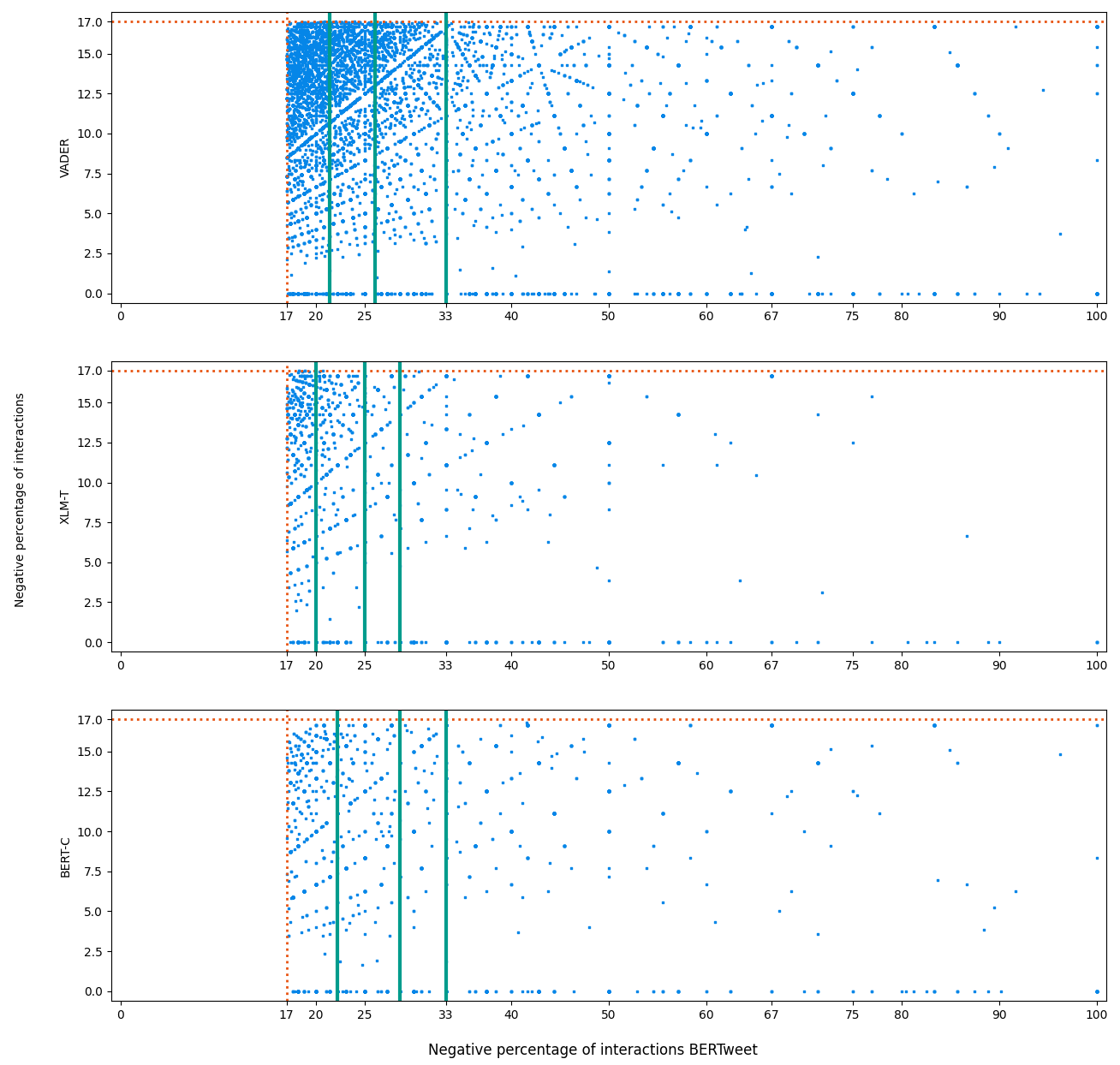}
    \caption{BERTweet disagreements. Each blue point represents a disagreement about the sign of a relationship, with BERTweet determining a negative sign and the other models determining a positive sign. The other models are, from top to bottom, VADER, XLM-T and BERT-C.}
    \label{fig:disagreements_plot_BERTweet}
\end{figure*}

A graph displaying relationship disagreements where BERTweet predicts a negative label for a relationship and one of the other models predicts a positive label can be seen in Figure~\ref{fig:disagreements_plot_BERTweet}.

\subsection{XLM-T}

\begin{figure*}
    \centering
    \includegraphics[height=0.7\textheight,width=0.95\textwidth]{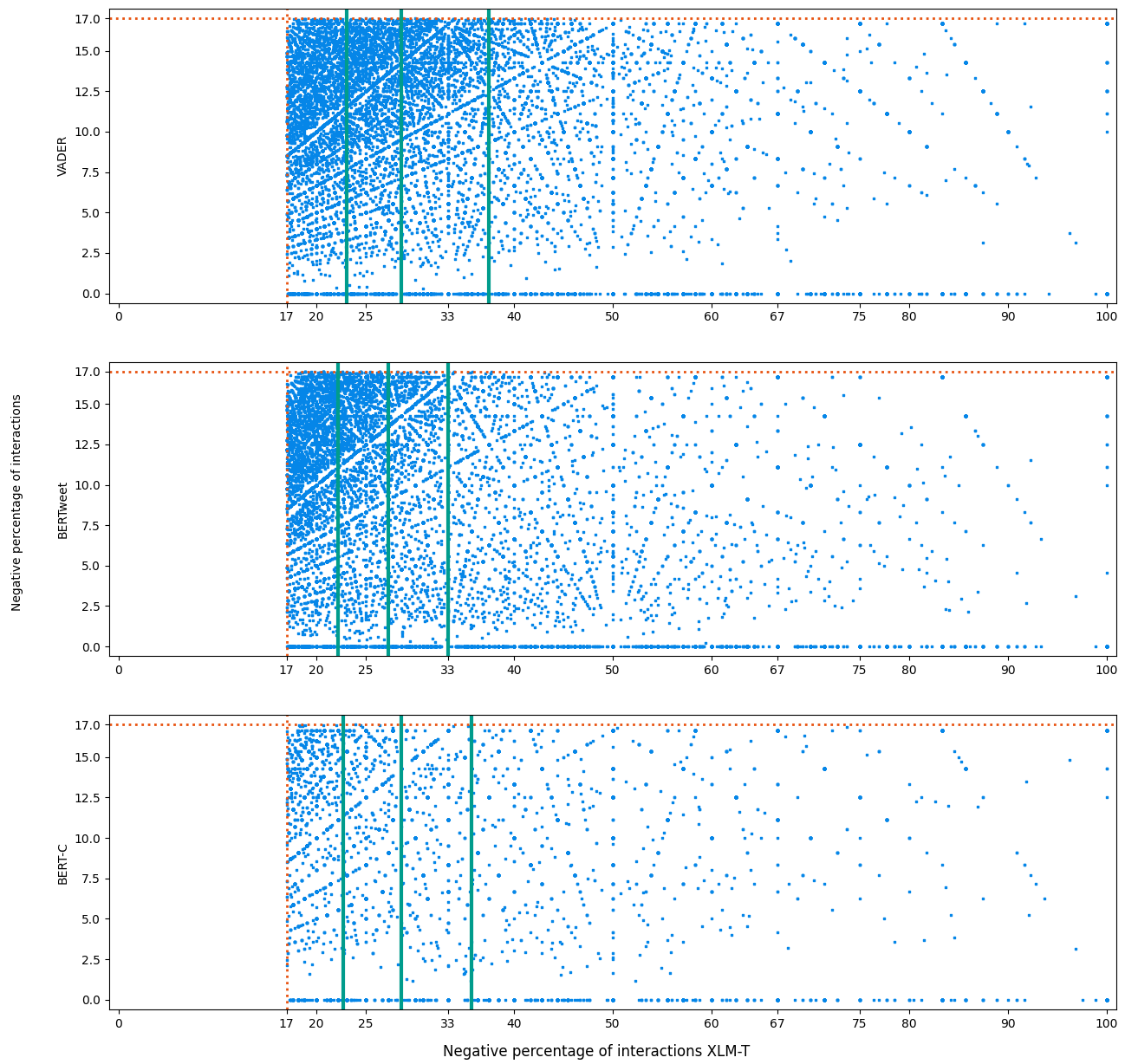}
    \caption{XLM-T disagreements. Each blue point represents a disagreement about the sign of a relationship, with XLM-T determining a negative sign and the other models determining a positive sign. The other models are, from top to bottom, VADER, BERTweet and BERT-C.}
    \label{fig:disagreements_plot_XLM}
\end{figure*}

A graph displaying relationship disagreements where XLM-T predicts a negative label for a relationship and one of the other models predicts a positive label can be seen in Figure~\ref{fig:disagreements_plot_XLM}.

\subsection{BERT-C}

\begin{figure*}
    \centering
    \includegraphics[height=0.7\textheight,width=0.95\textwidth]{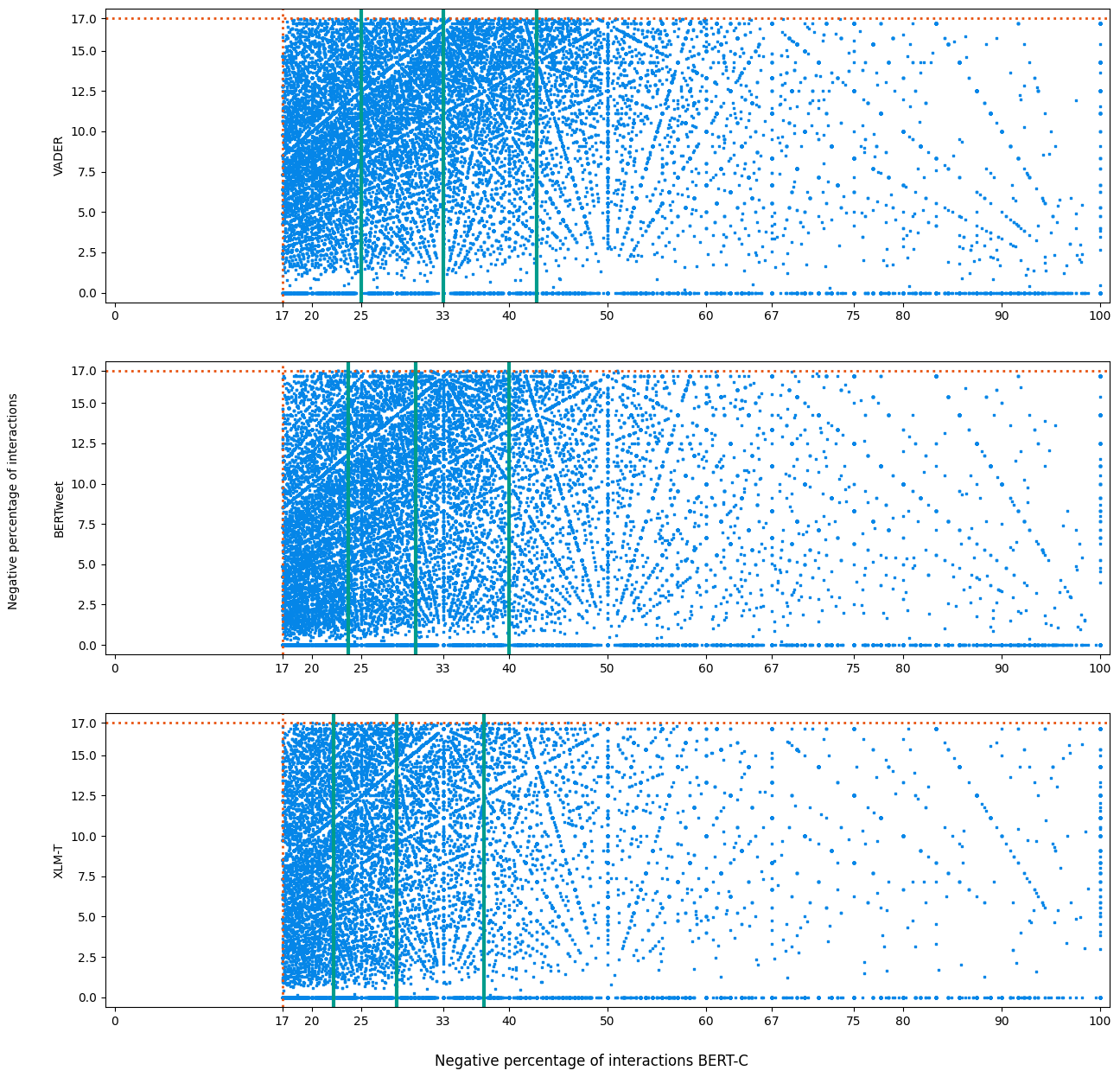}
    \caption{BERT-C disagreements. Each blue point represents a disagreement about the sign of a relationship, with BERT-C determining a negative sign and the other models determining a positive sign. The other models are, from top to bottom, VADER, BERTweet and XLM-T.}
    \label{fig:disagreements_plot_BERTC}
\end{figure*}

A graph displaying relationship disagreements where BERT-C predicts a negative label for a relationship and one of the other models predicts a positive label can be seen in Figure~\ref{fig:disagreements_plot_BERTC}.

\section{Investigation of Users' Interactions}
\label{appendix:number_interactions_per_alter_investigation}

Observing Table~\ref{num_interactions} in Section~\ref{sec:results_circles}, one may note some distinct patterns, for instance, that Egos interact most often with Alters of the inner circles. What's more, the number of interactions seems to decrease by a factor of around 2 between each circle, moving from the inside out. Furthermore, the generic users and British MPs display roughly double the number of interactions as the journalists and science writers. This is true for each level of the ENM, resulting in numbers of around 50, 25, 12, 6 and 3 for the journalists and science writers, and 100, 50, 25, 12 and 6 for the generic users and British MPs. Two exceptions to this are the innermost circles of the Australian Journalists and the Snowball dataset, both of which have numbers far higher than expected: 83.90 and 174.19 respectively.

This dichotomy between the journalists and generic users is rather unexpected because, as mentioned previously, journalists are thought to be generally more engaged with Twitter than other types of users, especially generic users and politicians \cite{Toprak_2021_b_Region-based}. To get a better understanding of these observations, we then looked at the number of Tweets and interactions each type of user generated, as well as the length of their timelines. This information is displayed in Table~\ref{interaction_investigations}\footnote{Retweets and Replies can also be Mentions if they tag different users in addition to the one being retweeted or replied to.}.

\begin{landscape}[h]
    \begin{table}[htbp]
        \centering
        \caption{Mean numbers of Tweets and percentage of Tweets that are interactions, Mentions, Retweets and Replies, as well as timeline lengths in days.}
        \label{interaction_investigations}
        \begin{tabular}{@{}lrrrrrr@{}}
            \toprule
            \textbf{Dataset} & \textbf{\# Tweets} & \textbf{\% Interactions} & \textbf{\% Mentions} & \textbf{\% Retweets} & \textbf{\% Replies} & \textbf{Timeline Length (days)}\\
            \midrule
            American Journalists & 3141.69 & 62.3 & 32.82 & 26.26 & 20.35 & 1528.38\\
            Australian Journalists & 2974.73 & 69.0 & 36.69 & 32.26 & 19.80 & 1570.55\\
            British Journalists & 3050.55 & 66.1 & 29.24 & 34.79 & 25.75 & 1544.56\\
            NYT Journalists & 3065.14 & 63.9 & 29.21 & 38.06 & 20.63 & 1455.65\\
            Science Writers & 3159.09 & 66.9 & 33.67 & 31.66 & 20.32 & 1452.46\\
            British MPs & 3135.36 & 69.7 & 34.31 & 45.17 & 13.84 & 1582.55\\
            \hdashline
            Monday Motivation & 3083.37 & 62.5 & 29.95 & 28.31 & 22.66 & 1142.65\\
            UK Users & 3025.59 & 65.9 & 29.58 & 24.91 & 32.40 & 1100.93\\
            Snowball & 3139.38 & 75.0 & 32.20 & 31.79 & 36.51 & 1426.74\\
            \bottomrule
        \end{tabular}
    \end{table}
\end{landscape}

The mean number of Tweets is close to 3,200 for all users. This shows that the majority of users, regardless of whether they are specialised or generic, in our chosen datasets are reaching the 3,200 tweet limit imposed due to the restrictions of the Twitter API. Similarly, although more surprisingly, the percentage of Tweets that are interactions is also fairly consistent across the two different categories of users; with the exception of Snowball (75.0\%), they are all very close to the 65.41\% average observed in previous work~\cite{Toprak_2022} (see Table~\ref{interaction_investigations}). However, one of the original papers that investigated the ENMs of journalists suggests that the journalists' increased level of engagement is mainly observable by the types of communications they use \cite{Toprak_2022}. Specifically, as Mentions and Retweets are generally less personal/intimate methods of interacting compared to Replies, users who rely mainly on these two methods of communicating tend to have above-average numbers of distinct peers (Alters), of presumably lower intimacy. By extension, users who mainly use Replies tend to have fewer but more intimate Alters. Given that the journalists have fewer interactions per Alter, one would expect them to have more distinct peers and, therefore, to use more Mentions and Retweets and fewer Replies than other types of users. 

Indeed, looking at the percentages of the different types of interactions in Table~\ref{interaction_investigations}, the datasets with the 4 highest percentages of both Mentions and Retweets are non-generic users and all the generic datasets have percentages of Replies that are within the top 4 highest. What's more, these observations also match the sizes of the active Ego Networks, i.e. the number of distinct peers, with the non-journalists having slightly smaller active networks (between 103.71 and 125.91) than the journalists (between 114.68 and 146.79); as displayed in the Circle 5 column of Table~\ref{circle_all}. While the two lowest percentages of Mentions do both belong to journalists (the British Journalists and NYT Journalists), these datasets have the highest percentages of Retweets, after the British MPs. Similarly, they both have percentages of Replies that are lower than the UK Users and Snowball, and the NYT Journalists' are also lower than those of the Monday Motivation dataset.

At first glance, the British MPs dataset doesn't quite seem to fit this line of reasoning. Indeed, this dataset has the highest mean number of distinct peers (146.79) as well as mean number of interactions per user that are comparable to the generic users. This would suggest that the users in this dataset simultaneously have more connections and also interact more with each connection. However, the British MPs have by far the lowest percentage of Replies, which are the most demanding way of communicating in terms of both time and cognition, and also have by far the highest percentage of Retweets, which is the least demanding method of communicating. This suggests that the British MPs employ very cognitively-efficient strategies of communicating.

\section{Negativity Metric Boxplots}
\label{appendix:negativity_metrics}

\subsection{American Journalists}

\begin{figure*}
    \centering
    \includegraphics[scale=0.39]{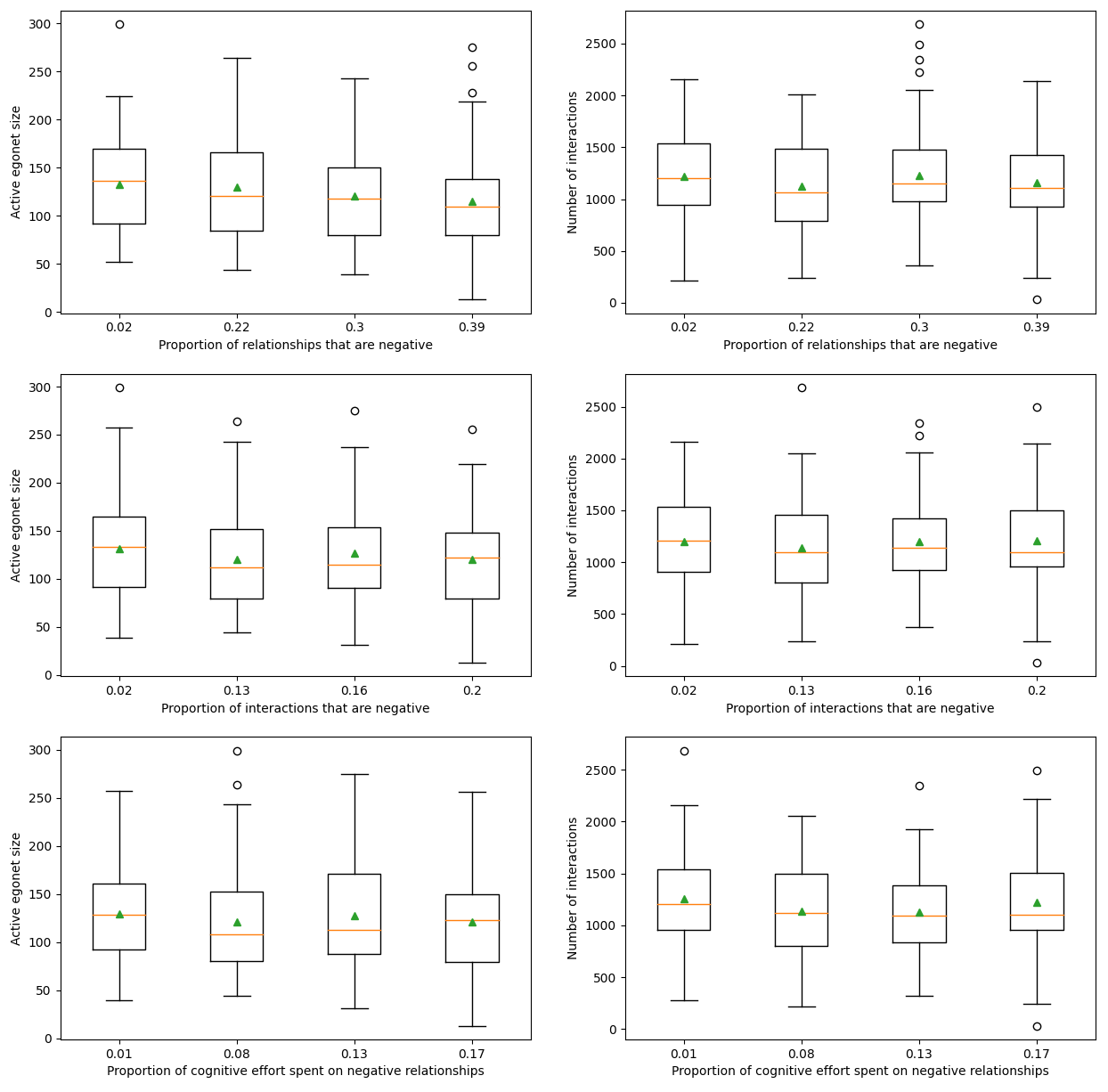}
    \caption{Boxplots for active Ego Network size (left column) and number of interactions (right column) against the 3 negativity metrics (top, middle and bottom) for the American Journalists dataset. For each group of binned Egos, the boxplots display mean (orange line), median (green triangle), first to third quartile (box), 1.5 times the interquartile range beyond the box (whiskers) and outliers (black circles).}
    \label{fig:AJ_Boxplots}
\end{figure*}

Boxplots for the American Journalists dataset, plotting active ego network size and number of interactions against each of the 3 negativity metrics, discussed in Section~\ref{sec:negativity_metrics}, can be seen in Figure~\ref{fig:AJ_Boxplots}.

\subsection{Australian Journalists}

\begin{figure*}
    \centering
    \includegraphics[scale=0.39]{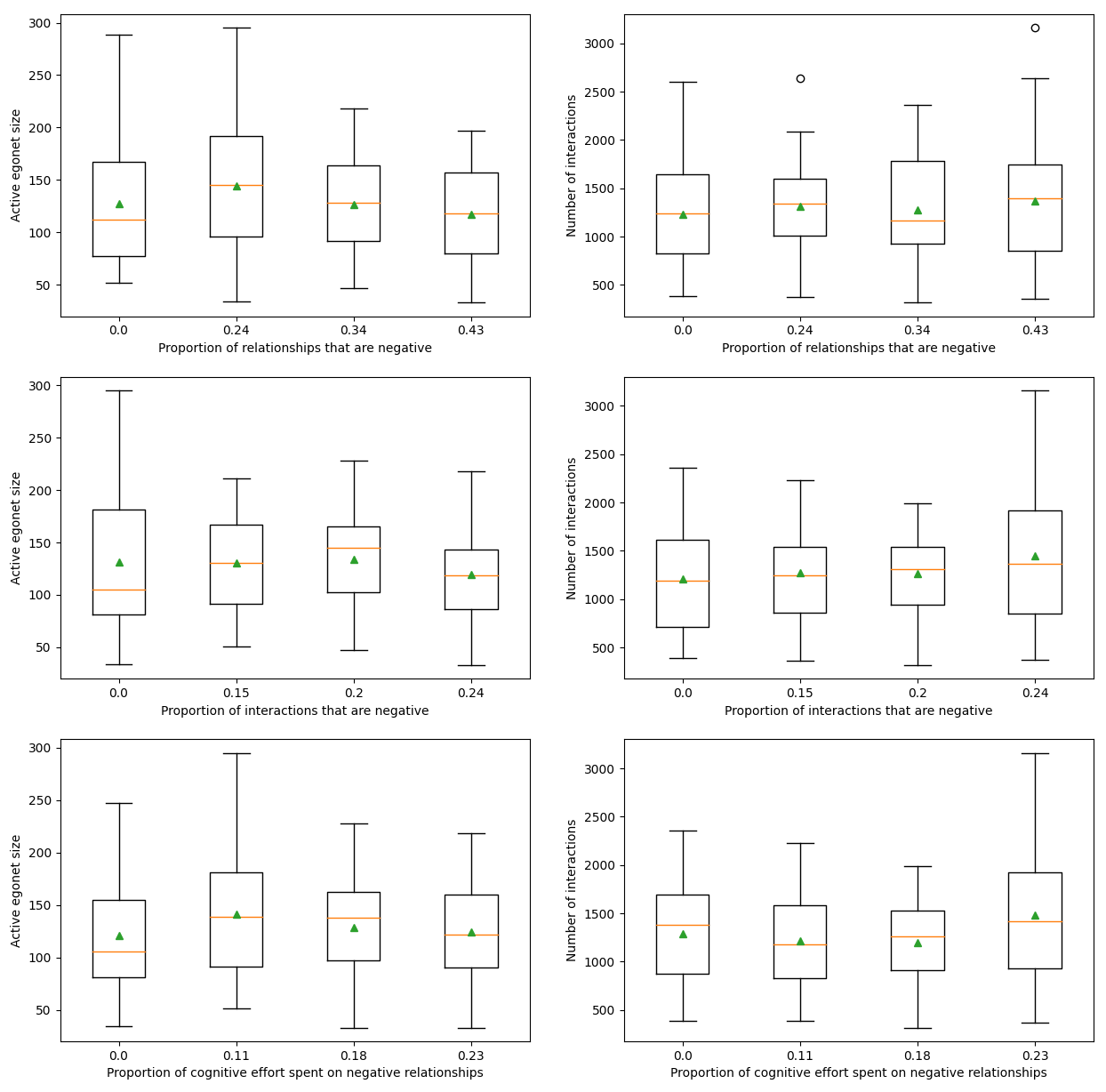}
    \caption{Boxplots for active Ego Network size (left column) and number of interactions (right column) against the 3 negativity metrics (top, middle and bottom) for the Australian Journalists dataset. For each group of binned Egos, the boxplots display mean (orange line), median (green triangle), first to third quartile (box), 1.5 times the interquartile range beyond the box (whiskers) and outliers (black circles).}
    \label{fig:AusJ_Boxplots}
\end{figure*}

Boxplots for the Australian Journalists dataset, plotting active ego network size and number of interactions against each of the 3 negativity metrics, discussed in Section~\ref{sec:negativity_metrics}, can be seen in Figure~\ref{fig:AusJ_Boxplots}.

\subsection{British Journalists}

\begin{figure*}
    \centering
    \includegraphics[scale=0.39]{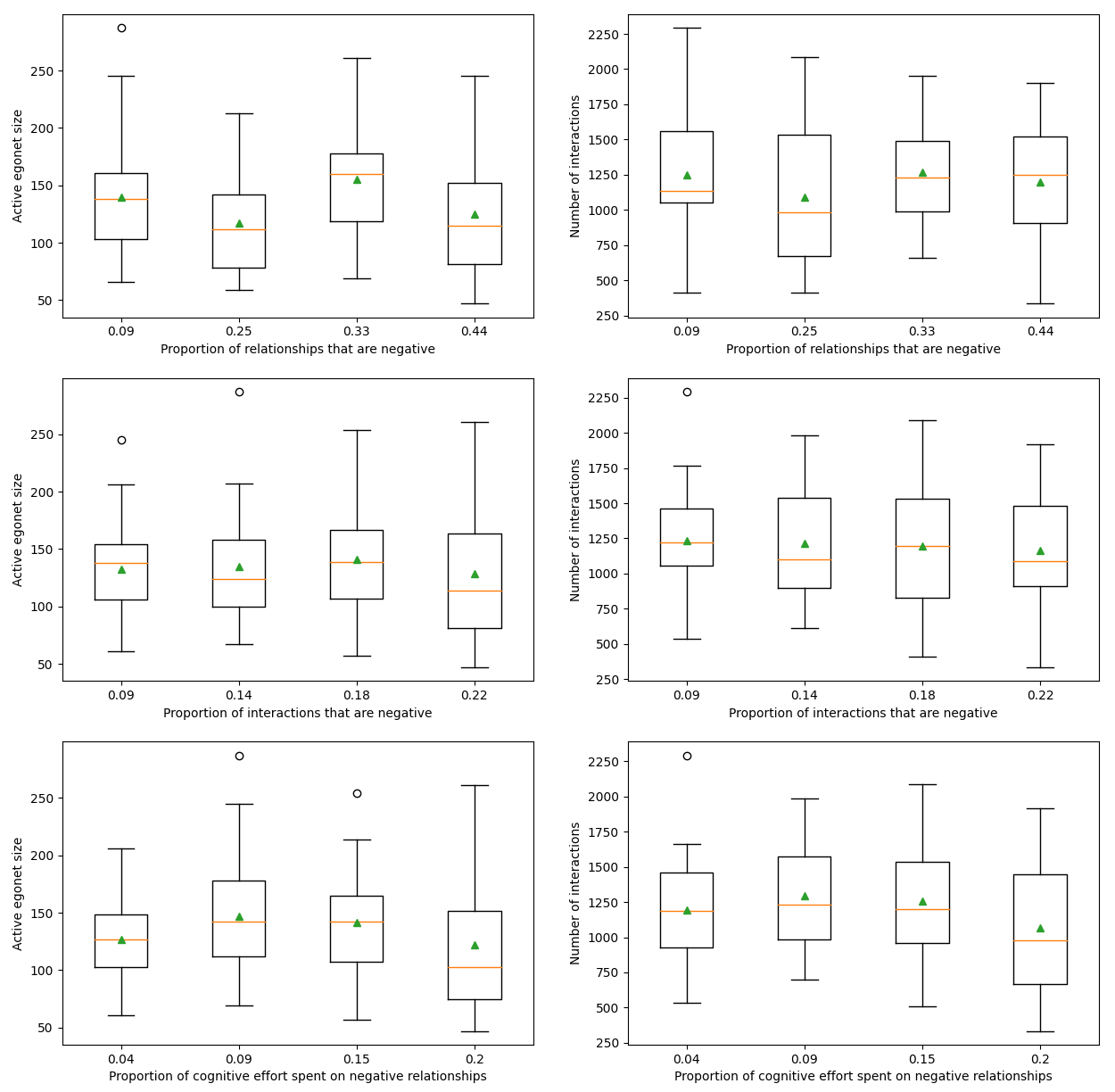}
    \caption{Boxplots for active Ego Network size (left column) and number of interactions (right column) against the 3 negativity metrics (top, middle and bottom) for the British Journalists dataset. For each group of binned Egos, the boxplots display mean (orange line), median (green triangle), first to third quartile (box), 1.5 times the interquartile range beyond the box (whiskers) and outliers (black circles).}
    \label{fig:BJ_Boxplots}
\end{figure*}

Boxplots for the British Journalists dataset, plotting active ego network size and number of interactions against each of the 3 negativity metrics, discussed in Section~\ref{sec:negativity_metrics}, can be seen in Figure~\ref{fig:BJ_Boxplots}.

\subsection{NYT Journalists}

\begin{figure*}
    \centering
    \includegraphics[scale=0.39]{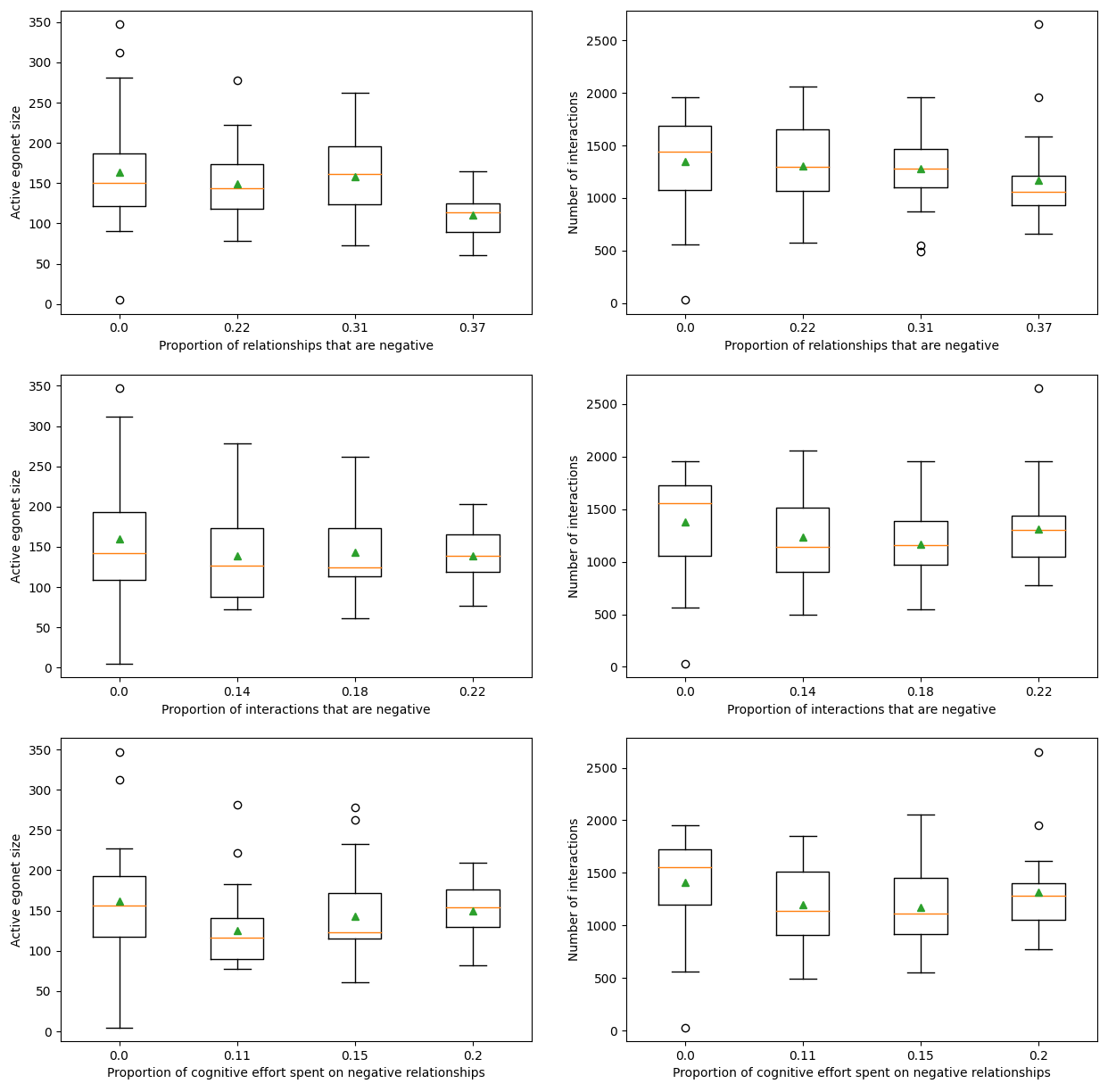}
    \caption{Boxplots for active Ego Network size (left column) and number of interactions (right column) against the 3 negativity metrics (top, middle and bottom) for the NYT Journalists dataset. For each group of binned Egos, the boxplots display mean (orange line), median (green triangle), first to third quartile (box), 1.5 times the interquartile range beyond the box (whiskers) and outliers (black circles).}
    \label{fig:NYT_Boxplots}
\end{figure*}

Boxplots for the NYT Journalists dataset, plotting active ego network size and number of interactions against each of the 3 negativity metrics, discussed in Section~\ref{sec:negativity_metrics}, can be seen in Figure~\ref{fig:NYT_Boxplots}.

\subsection{Science Writers}

\begin{figure*}
    \centering
    \includegraphics[scale=0.39]{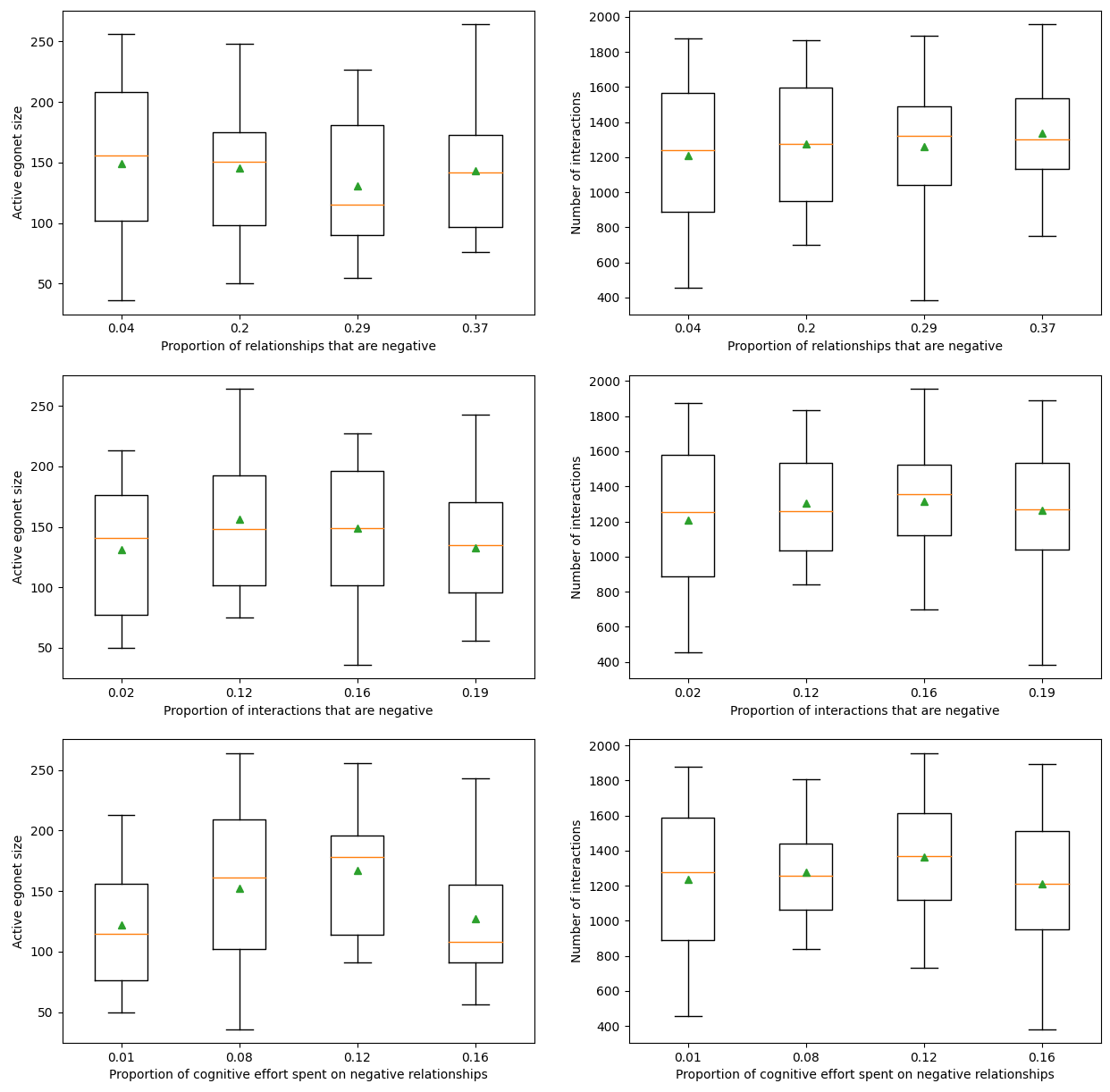}
    \caption{Boxplots for active Ego Network size (left column) and number of interactions (right column) against the 3 negativity metrics (top, middle and bottom) for the Science Writers dataset. For each group of binned Egos, the boxplots display mean (orange line), median (green triangle), first to third quartile (box), 1.5 times the interquartile range beyond the box (whiskers) and outliers (black circles).}
    \label{fig:Sci_Boxplots}
\end{figure*}

Boxplots for the Science Writers dataset, plotting active ego network size and number of interactions against each of the 3 negativity metrics, discussed in Section~\ref{sec:negativity_metrics}, can be seen in Figure~\ref{fig:Sci_Boxplots}.

\subsection{British MPs}

\begin{figure*}
    \centering
    \includegraphics[scale=0.39]{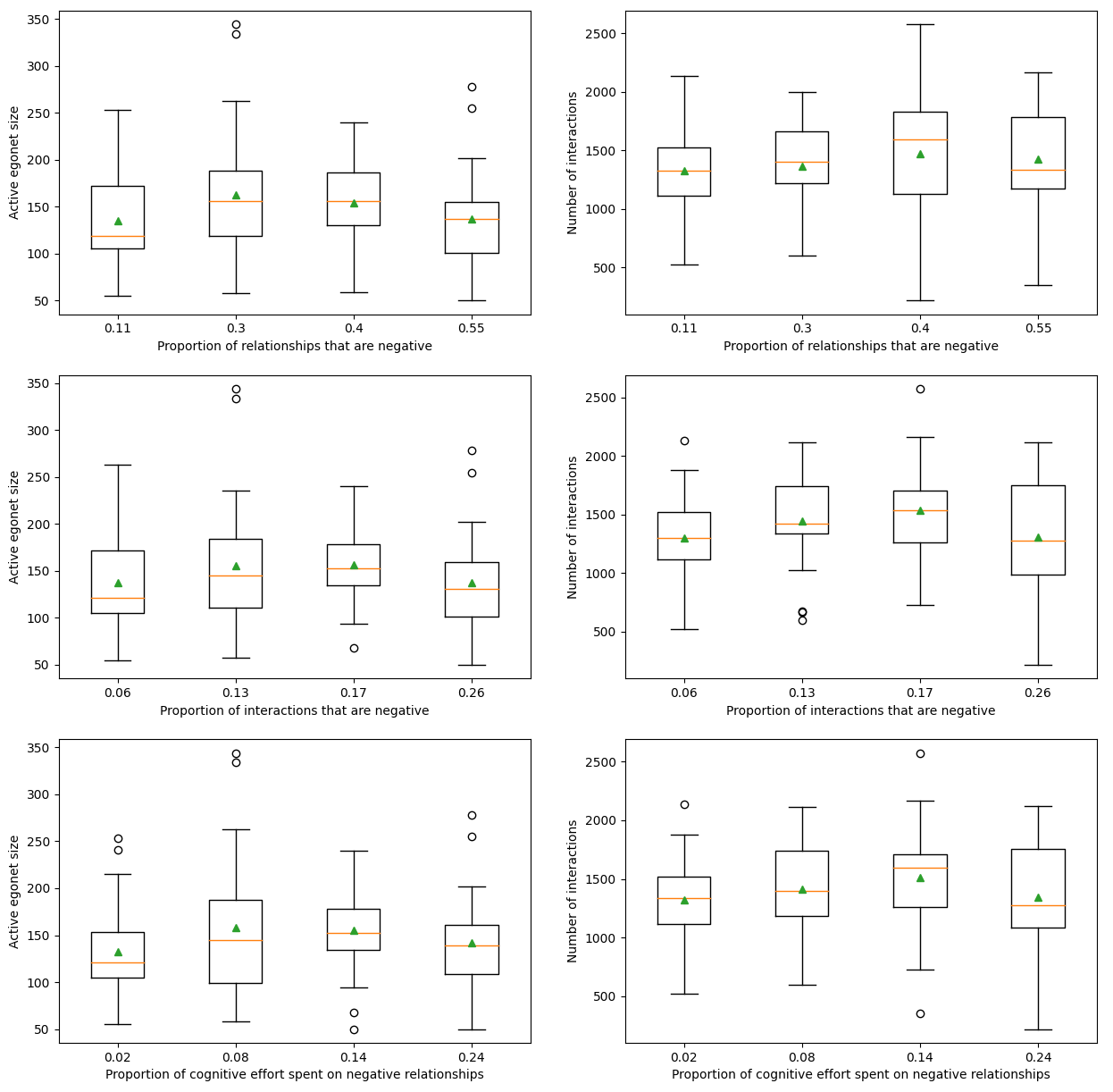}
    \caption{Boxplots for active Ego Network size (left column) and number of interactions (right column) against the 3 negativity metrics (top, middle and bottom) for the British MPs dataset. For each group of binned Egos, the boxplots display mean (orange line), median (green triangle), first to third quartile (box), 1.5 times the interquartile range beyond the box (whiskers) and outliers (black circles).}
    \label{fig:B_MPs_Boxplots}
\end{figure*}

Boxplots for the British MPs dataset, plotting active ego network size and number of interactions against each of the 3 negativity metrics, discussed in Section~\ref{sec:negativity_metrics}, can be seen in Figure~\ref{fig:B_MPs_Boxplots}.

\subsection{Monday Motivation}

\begin{figure*}
    \centering
    \includegraphics[scale=0.39]{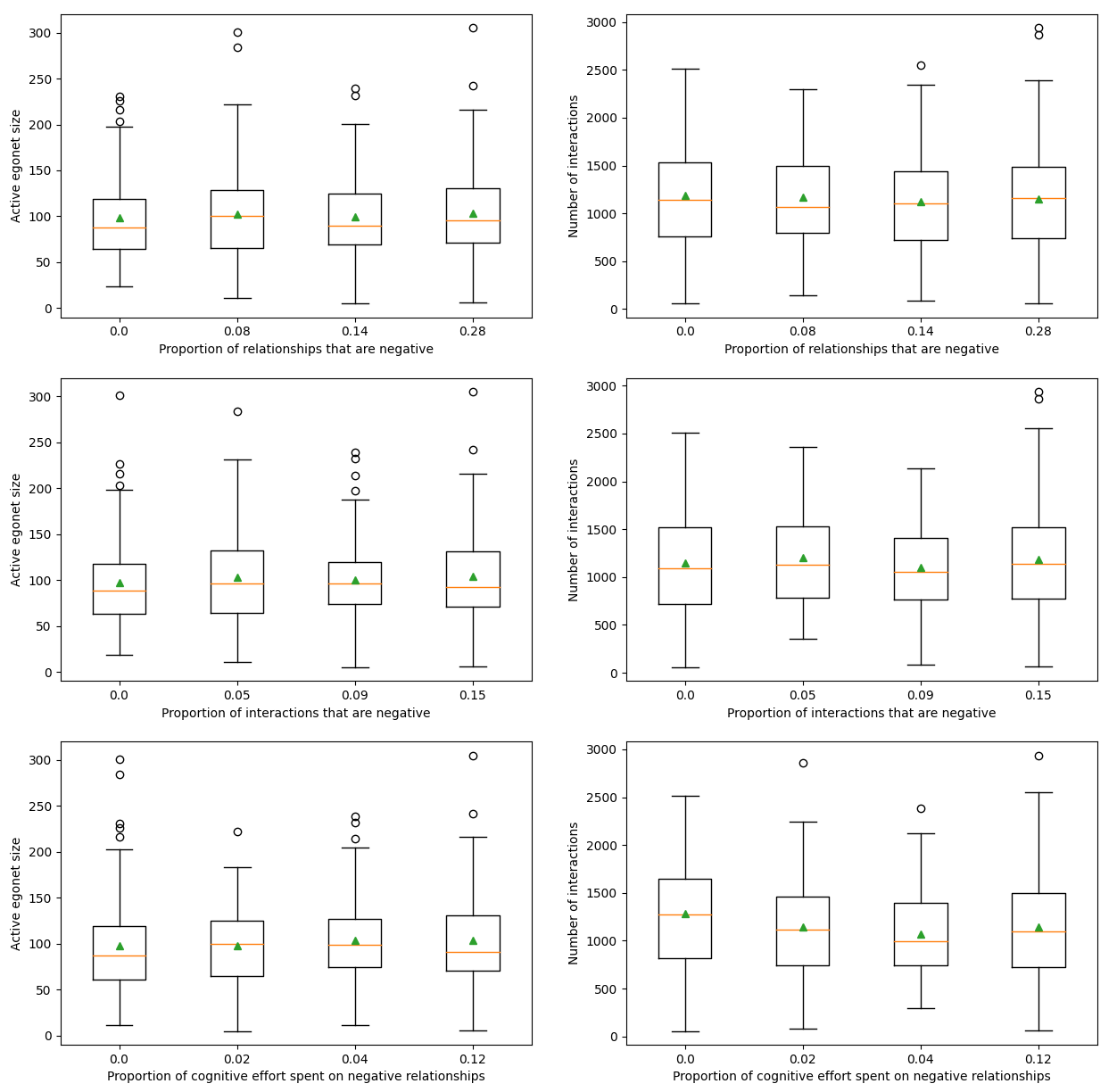}
    \caption{Boxplots for active Ego Network size (left column) and number of interactions (right column) against the 3 negativity metrics (top, middle and bottom) for the Monday Motivation dataset. For each group of binned Egos, the boxplots display mean (orange line), median (green triangle), first to third quartile (box), 1.5 times the interquartile range beyond the box (whiskers) and outliers (black circles).}
    \label{fig:R1_Boxplots}
\end{figure*}

Boxplots for the Monday Motivation dataset, plotting active ego network size and number of interactions against each of the 3 negativity metrics, discussed in Section~\ref{sec:negativity_metrics}, can be seen in Figure~\ref{fig:R1_Boxplots}.

\subsection{UK Users}

\begin{figure*}
    \centering
    \includegraphics[scale=0.39]{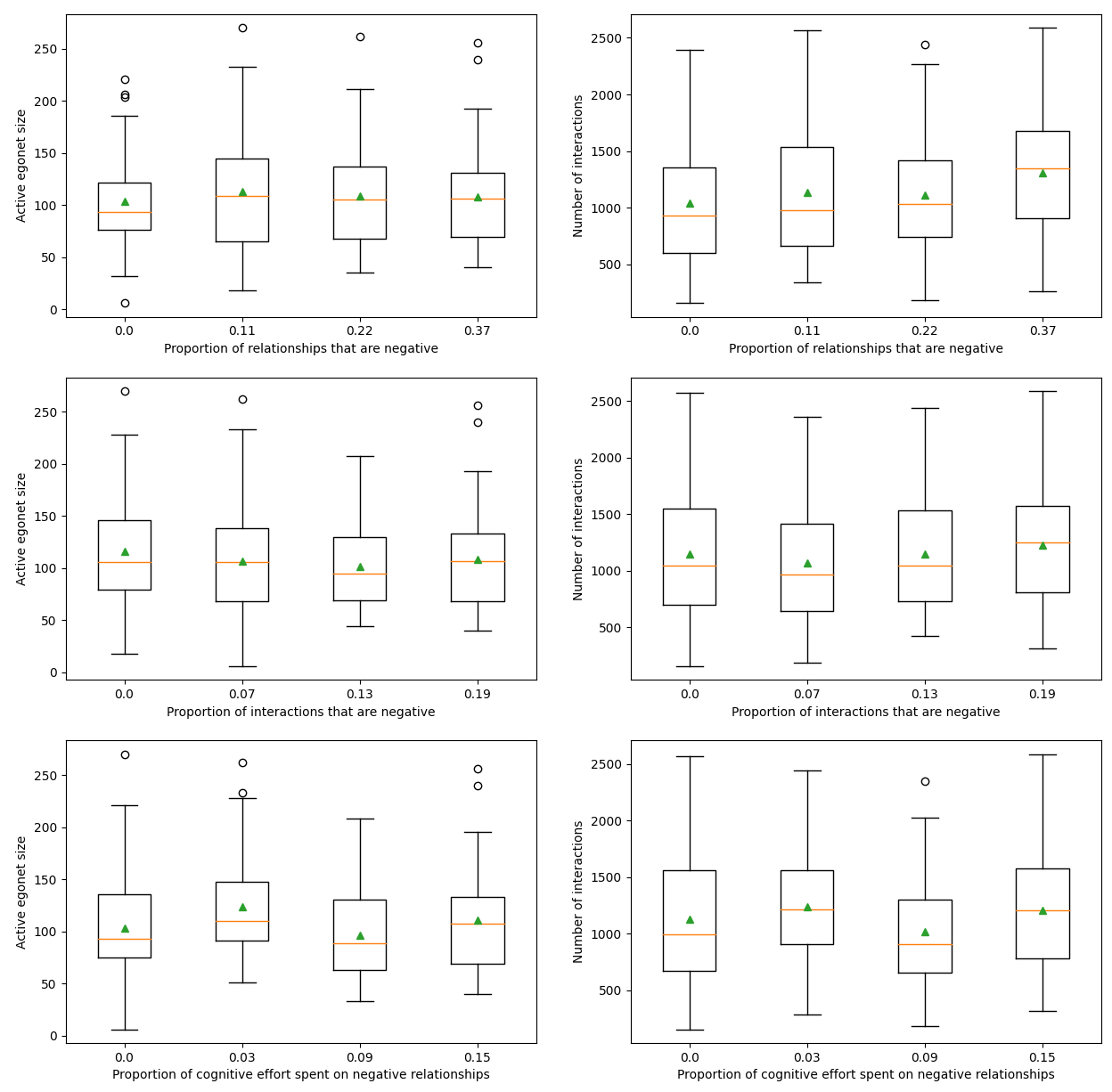}
    \caption{Boxplots for active Ego Network size (left column) and number of interactions (right column) against the 3 negativity metrics (top, middle and bottom) for the UK Users dataset. For each group of binned Egos, the boxplots display mean (orange line), median (green triangle), first to third quartile (box), 1.5 times the interquartile range beyond the box (whiskers) and outliers (black circles).}
    \label{fig:R2_Boxplots}
\end{figure*}

Boxplots for the UK Users dataset, plotting active ego network size and number of interactions against each of the 3 negativity metrics, discussed in Section~\ref{sec:negativity_metrics}, can be seen in Figure~\ref{fig:R2_Boxplots}.

\subsection{Snowball}

\begin{figure*}
    \centering
    \includegraphics[scale=0.39]{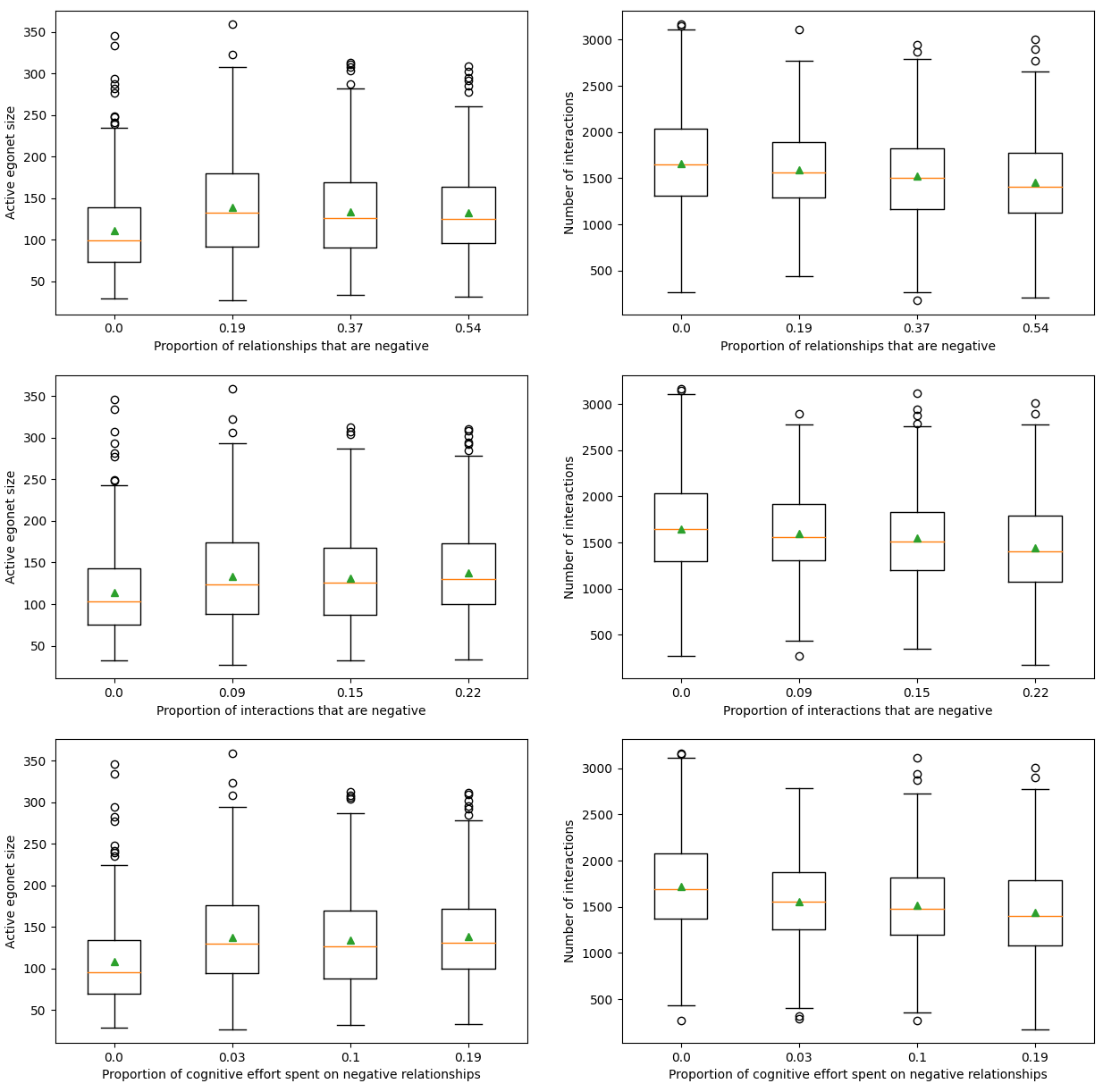}
    \caption{Boxplots for active Ego Network size (left column) and number of interactions (right column) against the 3 negativity metrics (top, middle and bottom) for the Snowball dataset. For each group of binned Egos, the boxplots display mean (orange line), median (green triangle), first to third quartile (box), 1.5 times the interquartile range beyond the box (whiskers) and outliers (black circles).}
    \label{fig:Snow_Boxplots}
\end{figure*}

Boxplots for the Snowball dataset, plotting active ego network size and number of interactions against each of the 3 negativity metrics, discussed in Section~\ref{sec:negativity_metrics}, can be seen in Figure~\ref{fig:Snow_Boxplots}.

\section{Negativity Metric t-scores}
\label{appendix:negativity_metrics_t_scores}

\subsection{Active Egonetwork Sizes}
\begin{table*}[htbp]
    \centering
    \caption{The t-scores from the pairwise comparisons between bins for Ego network sizes and negativity. Values corresponding to statistically significant p-values are displayed in bold.}
    \label{t_scores_ego_networks}
    \begin{tabular}{@{}llrrrrrrr@{}}
        \toprule
        & & \multicolumn{6}{c}{\textbf{Bin pairs}}\\
        & \textbf{Dataset} & \textbf{1-2} & \textbf{1-3} & \textbf{1-4} & \textbf{2-3} & \textbf{2-4} & \textbf{3-4}\\
        \midrule
        \multirow{9}{*}{Metric 1}
        & American Journalists & 0.328 & 1.440 & \textbf{2.198} & 1.053 & 1.790 & 0.790\\
        & Australian Journalists & -1.197 & 0.043 & 0.822 & 1.480 & \textbf{2.306} & 0.974\\
        & British Journalists & 1.426 & -0.938 & 0.886 & \textbf{-2.503} & -0.468 & 1.845\\
        & NYT Journalists & 0.794 & 0.302 & \textbf{3.361} & -0.606 & \textbf{3.479} & \textbf{4.045}\\
        & Science Writers & 0.191 & 0.884 & 0.291 & 0.770 & 0.111 & -0.660\\
        & British MPs & -1.553 & -1.332 & -0.121 & 0.514 & 1.459 & 1.214\\
        & Monday Motivation & -0.529 & -0.136 & -0.762 & 0.413 & -0.195 & -0.646\\
        & UK Users & -0.997 & -0.602 & -0.560 & 0.429 & 0.462 & 0.037\\
        & Snowball & \textbf{-5.657} & \textbf{-4.595} & \textbf{-4.471} & 1.226 & 1.456 & 0.216\\
        
        \midrule
        \multirow{9}{*}{Metric 2}
        & American Journalists & 1.205 & 0.552 & 1.397 & -0.656 & 0.105 & 0.808\\
        & Australian Journalists & 0.070 & -0.222 & 0.875 & -0.370 & 1.012 & 1.439\\
        & British Journalists & -0.162 & -0.552 & 0.247 & -0.350 & 0.360 & 0.688\\
        & NYT Journalists & 1.072 & 0.897 & 1.200 & -0.271 & -0.042 & 0.294\\
        & Science Writers & -1.193 & -0.885 & -0.051 & 0.373 & 1.219 & 0.894\\
        & British MPs & -0.962 & -1.323 & 0.029 & -0.077 & 1.004 & 1.390\\
        & Monday Motivation & -0.866 & -0.478 & -1.010 & 0.443 & -0.150 & -0.599\\
        & UK Users & 0.987 & 1.714 & 0.869 & 0.588 & -0.158 & -0.799\\
        & Snowball & \textbf{-3.926} & \textbf{-3.593} & \textbf{-4.926} & 0.431 & -0.775 & -1.249\\

        \midrule
        & American Journalists & 0.992 & 0.266 & 1.098 & -0.659 & 0.006 & 0.719\\
        & Australian Journalists & -1.481 & -0.675 & -0.280 & 0.968 & 1.366 & 0.457\\
        & British Journalists & -1.363 & -1.061 & 0.264 & 0.324 & 1.305 & 1.054\\
        & NYT Journalists & \textbf{2.061} & 1.007 & 0.771 & -1.134 & -1.977 & -0.476\\
        & Science Writers & -1.506 & \textbf{-2.459} & -0.297 & -0.710 & 1.288 & \textbf{2.257}\\
        & British MPs & -1.396 & -1.669 & -0.617 & 0.132 & 0.892 & 1.002\\
        & Monday Motivation & 0.004 & -0.910 & -0.862 & -1.083 & -1.007 & 0.003\\
        & UK Users & \textbf{-2.127} & 0.755 & -0.799 & \textbf{3.111} & 1.363 & -1.655\\
        & Snowball & \textbf{-6.271} & \textbf{-5.463} & \textbf{-6.524} & 0.805 & -0.028 & -0.859\\
        \bottomrule
    \end{tabular}
\end{table*}

\subsection{Number of Interactions}
\begin{table*}[htbp]
    \centering
    \caption{The t-scores from the pairwise comparisons between bins for the number of interactions and negativity. Values corresponding to statistically significant p-values are displayed in bold.}
    \label{t_scores_num_interactions}
    \begin{tabular}{@{}llrrrrrrr@{}}
        \toprule
        & & \multicolumn{6}{c}{\textbf{Bin pairs}}\\
        & \textbf{Dataset} & \textbf{1-2} & \textbf{1-3} & \textbf{1-4} & \textbf{2-3} & \textbf{2-4} & \textbf{3-4}\\
        \midrule
        \multirow{9}{*}{Metric 1}
        & American Journalists & 1.362 & -0.113 & 0.819 & -1.453 & -0.584 & 0.920\\
        & Australian Journalists & -0.665 & -0.380 & -0.993 & 0.290 & -0.444 & -0.680\\
        & British Journalists & 1.149 & -0.124 & 0.399 & -1.304 & -0.762 & 0.536\\
        & NYT Journalists & 0.384 & 0.593 & 1.385 & 0.229 & 1.112 & 0.902\\
        & Science Writers & -0.446 & -0.363 & -1.014 & 0.089 & -0.505 & -0.612\\
        & British MPs & -0.352 & -1.192 & -0.876 & -0.899 & -0.558 & 0.363\\
        & Monday Motivation & 0.241 & 0.967 & 0.509 & 0.746 & 0.279 & -0.456\\
        & UK Users & -0.908 & -0.673 & \textbf{-2.563} & 0.292 & -1.664 & \textbf{-2.062}\\
        & Snowball & 1.530 & \textbf{2.973} & \textbf{4.364} & 1.743 & \textbf{3.369} & 1.602\\
        
        \midrule
        \multirow{9}{*}{Metric 2}
        & American Journalists & 0.726 & -0.009 & -0.112 & -0.751 & -0.842 & -0.105\\
        & Australian Journalists & -0.575 & -0.518 & -1.665 & 0.090 & -1.230 & -1.341\\
        & British Journalists & 0.161 & 0.277 & 0.518 & 0.136 & 0.369 & 0.199\\
        & NYT Journalists & 1.113 & 1.751 & 0.533 & 0.636 & -0.695 & -1.417\\
        & Science Writers & -0.712 & -0.780 & -0.391 & -0.087 & 0.329 & 0.404\\
        & British MPs & -1.405 & \textbf{-2.229} & -0.120 & -0.769 & 1.077 & 1.786\\
        & Monday Motivation & -0.669 & 0.715 & -0.429 & 1.532 & 0.226 & -1.214\\
        & UK Users & 0.738 & 0.009 & -0.732 & -0.761 & -1.529 & -0.776\\
        & Snowball & 1.103 & \textbf{2.298} & \textbf{4.382} & 1.386 & \textbf{3.774} & \textbf{2.448}\\

        \midrule
        \multirow{9}{*}{Metric 3}
        & American Journalists & 1.592 & 1.713 & 0.440 & 0.063 & -1.110 & -1.210\\
        & Australian Journalists & 0.599 & 0.789 & -1.355 & 0.168 & -1.886 & \textbf{-2.082}\\
        & British Journalists & -0.811 & -0.467 & 0.946 & 0.281 & 1.696 & 1.322\\
        & NYT Journalists & 1.729 & 1.819 & 0.776 & 0.174 & -1.116 & -1.240\\
        & Science Writers & -0.308 & -0.891 & 0.184 & -0.779 & 0.576 & 1.198\\
        & British MPs & -0.908 & -1.683 & -0.247 & -0.784 & 0.580 & 1.310\\
        & Monday Motivation & 1.759 & \textbf{2.978} & 1.803 & 1.176 & 0.065 & -1.092\\
        & UK Users & -1.077 & 1.071 & -0.726 & \textbf{2.342} & 0.372 & -1.945\\
        & Snowball & \textbf{4.007} & \textbf{4.700} & \textbf{6.148} & 0.764 & \textbf{2.595} & 1.887\\
    \end{tabular}
\end{table*}

\end{document}